\documentclass[reprint,sort&compress,longbibliography,superscriptaddress,amsmath,aps,showpacs,prl]{revtex4-1}
\usepackage[british]{babel}
\usepackage{graphicx,epsfig}
\usepackage{float}
\usepackage{hyperref}
\usepackage{breqn}
\usepackage[utf8]{inputenc}
\usepackage{amsmath}
\usepackage{bbold}
\usepackage{subfigure}
\usepackage[normalem]{ulem}
\usepackage[dvipsnames]{xcolor}
\usepackage{hyperref}
\usepackage{soul}
\usepackage{braket}
\usepackage{tipa}
\usepackage[T5,T1]{fontenc}
\hypersetup{ 
     colorlinks   = true,
     citecolor    = blue,
}

\begin{document}
\title{Junctions and superconducting symmetry in twisted bilayer graphene}
\author{H\'ector Sainz-Cruz}
\email{hector.sainz@imdea.org}
\affiliation{$Imdea\ Nanoscience,\ Faraday\ 9,\ 28015\ Madrid,\ Spain$}
\author{Pierre A. Pantale\'on}
\affiliation{$Imdea\ Nanoscience,\ Faraday\ 9,\ 28015\ Madrid,\ Spain$}
\author{V\~{o} {\fontencoding{T5}\selectfont
Ti\'\ecircumflex{}n} Phong}
\affiliation{$Department\ of\ Physics\ and\ Astronomy,\ University\ of\ Pennsylvania,\ Philadelphia\ PA\ 19104$}
\author{Alejandro Jimeno-Pozo}
\affiliation{$Imdea\ Nanoscience,\ Faraday\ 9,\ 28015\ Madrid,\ Spain$}
\author{Francisco Guinea}
\affiliation{$Imdea\ Nanoscience,\ Faraday\ 9,\ 28015\ Madrid,\ Spain$}
\affiliation{$Donostia\ International\  Physics\ Center,\ Paseo\ Manuel\ de\ Lardizabal\ 4,\ 20018\ San\ Sebastian,\ Spain$}
\date{\today}
\begin{abstract} Junctions provide a wealth of information on the symmetry of the order parameter of superconductors.\ We analyze junctions between a scanning tunneling microscope (STM) tip and superconducting twisted bilayer graphene (TBG) and TBG Josephson junctions (JJs).\ We compare superconducting phases that are even or odd under valley exchange ($s$- or $f$-wave).\ The critical current in mixed ($s$- and $f$-) JJs strongly depends on the angle between the junction and the lattice.\ In STM-TBG junctions, due to Andreev reflection, $f$-wave leads to a prominent peak in subgap conductance, as seen in experiments.
\end{abstract}
\maketitle

{\it Introduction.} Graphene multilayers host a myriad of exotic correlated and topological phases \cite{cao2018correlated,polshyn2019large,sharpe2019emergent,serlin2020intrinsic,chen2020tunable,uri2020,saito2020independent,wong2020cascade,zondiner2020cascade,stepanov2020untying,xu2020correlated,choi2021correlation,rozen2021entropic,cao2021nematicity,stepanov2021competing,oh2021evidence,xie2021fractional,berdyugin2022out,turkel2022orderly,huang2022observation,barrera2022cascade,kim2022evidence,seiler2022quantum}.\ Perhaps most interesting and enigmatic among them is superconductivity, possibly with unconventional pairing symmetries and mechanisms, observed in alternating-twist  stacks of up to five layers \cite{cao2018unconventional, yankowitz2019, lu2019superconductors,park2021tunable, hao2021electric,park2022robust,zhang2022promotion} and in Bernal bilayers and rhombohedral trilayers  \cite{zhou2021superconductivity,zhou2022isospin,zhang2023spin,holleis23Ising}.\ Crucially, the observed superconductivity violates the Pauli limit for spin-singlet pairing \cite{cao2021pauli,park2022robust,zhou2021superconductivity,zhou2022isospin,zhang2023spin,holleis23Ising} and has been observed in settings that break time-reversal symmetry (TRS) \cite{lin2022zero}, strongly suggesting a spin-triplet pairing in these materials.\ However, the pairing may be a mixture of singlet and triplet~\cite{lake2022pairing}, and the exact symmetries involved ($s$-, $p$-, $d$- and/or $f$-) are still unknown despite intense theoretical and experimental efforts to uncover them.\ 

Recently, several experiments have studied these unconventional superconducting states using transport measurements: either with a scanning tunneling microscope (STM) tip~\cite{oh2021evidence, kim2022evidence} or with Josephson junctions (JJs)~\cite{de2021gate,rodan2021highly,diez2021magnetic,xie2022valley,hu2022valley} and Superconducting Quantum Interference Devices (SQUIDs)~\cite{portoles2022tunable}.\ In the former setup, by comparing the transmission between the STM tip and the superconducting surface in the weak and strong-coupling regimes, one can gain important insights about the symmetry of the order parameter.\ For instance, the experimental observations, such as the peak in the subgap conductance~\cite{oh2021evidence,kim2022evidence}, seem inconsistent with $s$-, $p$- and $d$-wave pairings~\cite{lake2022pairing,sukhachov2022andreev}.\ In the latter setups, the overlap of the superconductors' wavefunctions at the junction's link gives rise to a zero-frequency supercurrent whose magnitude and superconducting phase-dependence carry characteristics of the pairing symmetry~\cite{josephson1962possible,sigrist1991phenomenological, golubov2004current}.

Building on these experimental insights, we argue in this Letter that transport measurements in junctions are ideal probes of the pairing symmetry in twisted graphene superconductors, similar to the elucidation of $d$-wave pairing in cuprate superconductors~\cite{Tetal94,TK00}, and that existing STM data~\cite{oh2021evidence,kim2022evidence} are consistent with $f$-wave pairing.\ The Fermi surface of these graphene-based systems contains two valleys.\ We consider superconducting order parameters that are either even or odd under valley exchange, which in the absence of spin-orbit coupling correspond to spin-singlet $s$-wave superconductivity or spin-triplet $f$-wave superconductivity respectively.\ In `mixed' Josephson junctions connecting a $s$-wave to a $f$-wave superconductor, we observe that the critical current dramatically depends on the angle between the junction and the graphene lattice axis.\ Therefore, Josephson junctions are useful for determining whether two superconducting phases differ in their valley exchange parity.\ 

In the STM-superconductor junction, we find that the subgap conductance shows a prominent zero-bias peak for $f$-wave pairing only, due to enhanced Andreev reflection.\ This peak has been observed in experiments on both twisted bilayer~\cite{oh2021evidence} and twisted trilayer graphene~\cite{kim2022evidence}.\ This result puts forward $f$-wave pairing as a leading candidate for the superconducting symmetry of twisted bilayer graphene, which is also consistent with previous theoretical models based on Coulomb-interaction-mediated Cooper pairing~\cite{crepel2022unconventional,cea21Coulomb}.

\begin{figure*}[t]
    \centering{\includegraphics[width=17.5cm]{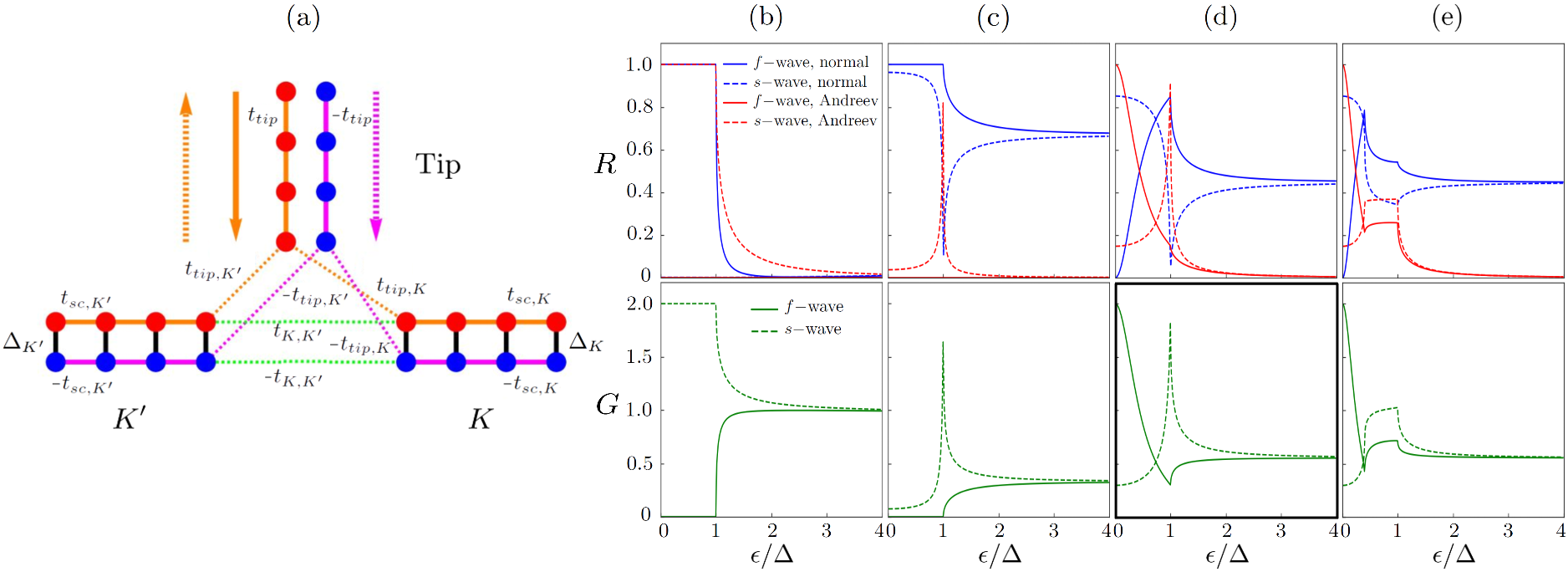}}%
    \caption{STM tip-superconducting TBG junction.\ (a) Sketch of the model for the junction used to calculate its transport properties.\ See text for details.\ (b-e): normal and Andreev reflections (top) and total conductance (bottom) for junctions with spin-singlet $s$-wave (dashed) and spin-triplet $f$-wave pairing (solid), in the perfect contact limit.\ (b) With equal Fermi velocities in all channels, $t_{tip} = t_{sc,K} = t_{sc,K'} = 1, \, t_{tip,K} = t_{tip,K'} = 1 / \sqrt{2}, \, t_{K,K'} = 0, \, \Delta_K = 0.05 , \Delta_{K'} = \pm 0.05$.\ (c) With a large Fermi velocity mismatch in the normal and superconducting channels: $t_{tip} = 10, \, t_{tip,K} = t_{tip,K'} = 10 / \sqrt{2}$, others as in (b).\ (d) With Fermi velocity mismatch and intervalley scattering $t_{K,K'} = 1$, others as in (c).\ (e) With Fermi velocity mismatch, intervalley scattering and spin-orbit coupling: $\Delta_K = 0.05 , \Delta_{K'} = \pm 0.02$, others as in (d).}
    \label{fig:1}
\end{figure*}

{\it STM tip-superconducting TBG junction.\ The model.} 
General features of transport in normal-superconductor junctions are described in Ref.~\cite{blonder1982transition}.\
The coupling between the two electrodes is given by a scattering matrix, determined by a dimensionless transmission amplitude, $T$.\ The model has been extended in~\cite{lake2022pairing,sukhachov2022andreev}.\ As in Ref.~\cite{blonder1982transition}, the normal metal tip and the superconducting electrode are described in terms of incoming and outgoing single channels.\ On the superconductor, the states in the channel are defined as suitable averages in momentum space of the quasiparticles.\ The momentum dependence of the gap leads to a momentum dependence of the mixing between electron and hole-like states in the superconductor, and it modifies the transmission of the junction, both in the tunneling and in the contact regimes.\ The Blonder-Tinkham-Klapwijk (BTK) model~\cite{blonder1982transition} has also been extended to strongly coupled superconductors, where the chemical potential can be below the bottom of the band~\cite{setiawan2022analytic,lewandowski2023andreev}.

We describe the metal$-$superconductor junction as one ingoing normal channel, which represents the tip, and two outgoing superconducting channels, which represent the two valleys in TBG.\ The signs of the gaps in these channels can be equal, describing a spin-singlet $s$-wave superconductor, or opposite, describing a spin-triplet $f$-wave superconductor~\cite{aMiscspin}.\ The model can also be applied to an Ising superconductor~\cite{holleis23Ising} in a system with strong spin-orbit coupling, characterized by spin-valley locked Cooper pairs of the type $| K , \uparrow ; K' , \downarrow \rangle$.

The three-channel model described above is discretized as a tight-binding model, see Fig.\ \ref{fig:1}(a).\ The normal channel is described by nearest-neighbor hopping $t_{tip}$, which determines its Fermi velocity and density of states.\ The superconducting channels are described by two nearest neighbor hoppings, $t_{sc,K}$ and $t_{sc,K'}$, and two gaps, $\Delta_{K}$ and $\Delta_{K'}$.\ The coupling between the normal channel and the two superconducting channels is described by the hoppings $t_{tip,K}$ and $t_{tip,K'}$.\ Without loss of generality, we assume that the Fermi energy is $\epsilon_F = 0$, so that each channel has exact electron-hole symmetry.\ Finally, we consider that the tip is a local perturbation which can induce intervalley scattering, parametrized by another hopping, $t_{K,K'}$.

We solve the transmission of the junction by matching incoming and outgoing waves in the three channels.\ If the energy $\epsilon$ is within the superconducting gaps, we use evanescent waves in the superconducting channels.\ For each energy, there are four propagating or evanescent waves in each channel.\ We assume that there is an incoming wave of electron character and amplitude 1 in the tip channel.\ In the same channel, there can be one electron and one hole outgoing channels, describing normal and Andreev reflection, with amplitudes $R_N$ and $R_A$, respectively.\ In each of the two superconducting channels there can be two decaying evanescent waves, when the energy is within the gap, or two outgoing propagating waves.\ We describe the four amplitudes as $T_{i,j}$, where $i=K,K'$ stands for the channel, and $j=1,2$ stands for the wavefunction within each channel.\ The transport properties of the junction are determined by these six amplitudes.\ The conductance of the junction is $G = 1 - | R_N |^2 + | R_A |^2$.\ The matching conditions involve the amplitudes of the wavefunctions at the three sites which describe the junction.\ The equations can be found in Ref.~\cite{SM}.

{\it STM tip-superconducting TBG junction.\ Results.} When the Fermi velocities in all channels are equal, the tip channel merges smoothly into the even combination of the $K$ and $K'$ channels and the junction behaves as described by the BTK theory in the regime of perfect contact, see Fig.~\ref{fig:1}(b).\ For $s$-wave pairing, and at zero voltage, Andreev scattering leads to a conductance twice as large as a single normal channel \cite{blonder1982transition}.\ For $f$-wave pairing, negative interference between the two hole channels cancels Andreev reflection.\ This cancellation can be expected whenever the order parameter has a sign change between states related by TRS \cite{sukhachov2022andreev}.\ At high voltages the conductance reduces to the conductance of a single channel in both cases.\ 

The bandwidth and Fermi velocity in TBG are considerably smaller than in a normal metal.\ This Fermi velocity mismatch induces elastic back-scattering in the normal phase, which reduces the conductance above the gap, see Fig.~\ref{fig:1}(c).\ Subgap Andreev reflection for $s$-wave superconductivity is strongly suppressed, and it remains zero for the $f$-wave phase, for a detailed explanation see Ref.~\cite{SM}.\ The tip can also induce a perturbation on the superconductor, on scales comparable to the atomic spacing.\ Such a perturbation will induce intervalley scattering.\ Fig.~\ref{fig:1}(d) shows results obtained for an intervalley coupling comparable to the bandwidth of the superconductor.\ This perturbation can be considered as disorder which does not violate TRS.\ The presence of intervalley scattering does not change significantly the conductance of the junction in an $s$-wave superconductor, in agreement with Anderson's theorem~\cite{anderson1959theory}.\ On the other hand, it is a pair breaking perturbation in an $f$-wave superconductor, which induces subgap states, see Ref.~\cite{SM}.\ These states allow for subgap Andreev reflection.\ As a result, the subgap conductance of the junction is strongly enhanced by intervalley scattering in a $f$-wave superconductor, leading to a zero bias peak, highlighted in Fig.~\ref{fig:1}(d), that has been seen in the experiments of Refs.~\cite{oh2021evidence,kim2022evidence}.

Recent transport experiments \cite{zhang2023spin,holleis23Ising} reveal that proximity induced spin-orbit coupling promotes the superconducting properties of Bernal bilayer graphene.\ An effect of spin-orbit coupling is to break the equivalence between the Cooper pairs $| K , \uparrow ; K' , \downarrow \rangle$ and $| K , \downarrow ; K' \uparrow \rangle$.\ In the model studied here, the spin-orbit coupling makes the two channels inequivalent.\ Results are shown in Fig.~\ref{fig:1}(e).\

\begin{figure}[t]
    \centering
    {\includegraphics[width=8cm]{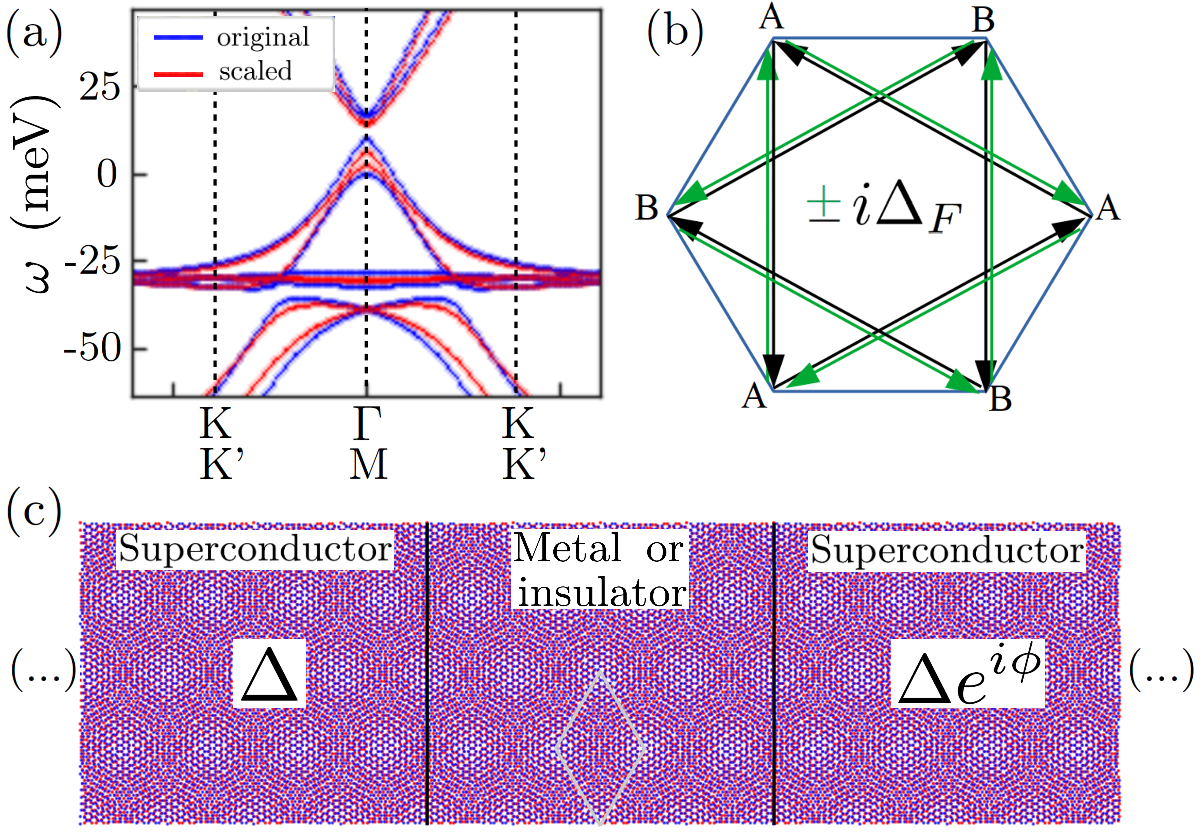}}%
    \caption{(a) Low-energy bandstructure of TBG at $\theta\approx1.08^{\circ}$ and filling $n=-2.4$, with and without scaling.\ (b) Hoppings inducing $f$-wave superconducting pairing.\ (c) Central part of the lattice of the TBG Josephson junction \cite{Misc1,gonccalves2021incommensurability}.\ The electrodes are superconductors with a phase difference of $\phi$ and the link region, with a length of four moiré periods, is metallic or insulating.\ The rhombus is a unit cell of TBG.}
    \label{fig:2}
\end{figure}

{\it Josephson junctions.\ The model.} 
For the study of JJs, our setup consists of a TBG crystal, in which the electrodes are superconducting and the weak link is in a normal metal or band insulating phase, as shown in Fig.\ \ref{fig:2}(c).\ We start from a tight-binding, non-interacting Hamiltonian ${\cal H}_0$ \cite{lin2018minimum} that includes Hartree electron-electron interactions through an electrostatic potential~\cite{GuineaNiels2018,RademakerHartree2019}.\ The parameters in the tight binding model are scaled, such that the central bands of a TBG with twist angle $\theta$ are approximated by the central bands of an equivalent lattice with twist angle $\lambda \theta$, with $\lambda >1$~\cite{gonzalez2017,vahedi2021magnetism,sainzcruz21high}, see Fig.~\ref{fig:2}(a).

The critical current comes from second order perturbation theory and is the derivative of the free energy $E$ with respect to the superconducting phase difference $\phi$:

\begin{equation}
\mathcal{I}=\frac{e}{h}\frac{\partial E}{\partial \phi}\, .
\label{eq:1}
\end{equation}
To obtain the energies of the TBG junction, we diagonalize the Bogoliubov-de Gennes Hamiltonian, 
\begin{gather}
{\cal H}_{BdG}\ket{\Psi}=
\begin{pmatrix} {\cal H}_0-\epsilon_F & f(\Delta)\\
f^{\dagger}(\Delta) & \epsilon_F-{\cal H}_0\\
\end{pmatrix}
\begin{pmatrix}  \Psi_e \\
\Psi_h\\
\end{pmatrix}
= E
\begin{pmatrix}  \Psi_e \\
\Psi_h\\ 
\end{pmatrix}\, ,
\label{eq:2}
\end{gather}
where $\epsilon_F$ is the Fermi energy.\ Again, we compare $s$-wave pairing, which we model with a on-site attractive Hubbard term $f(\Delta)=-\Delta_S \mathbb{1}$, and $f$-wave pairing, which results from Haldane-like hoppings \cite{haldane1988model,PairingAmplitudes} that allow an electron excitation to convert to a hole excitation via second nearest-neighbor imaginary intralayer hoppings, see Fig.\ \ref{fig:2}(b).

\begin{figure}[t]
   \centering{\includegraphics[scale = 0.48]{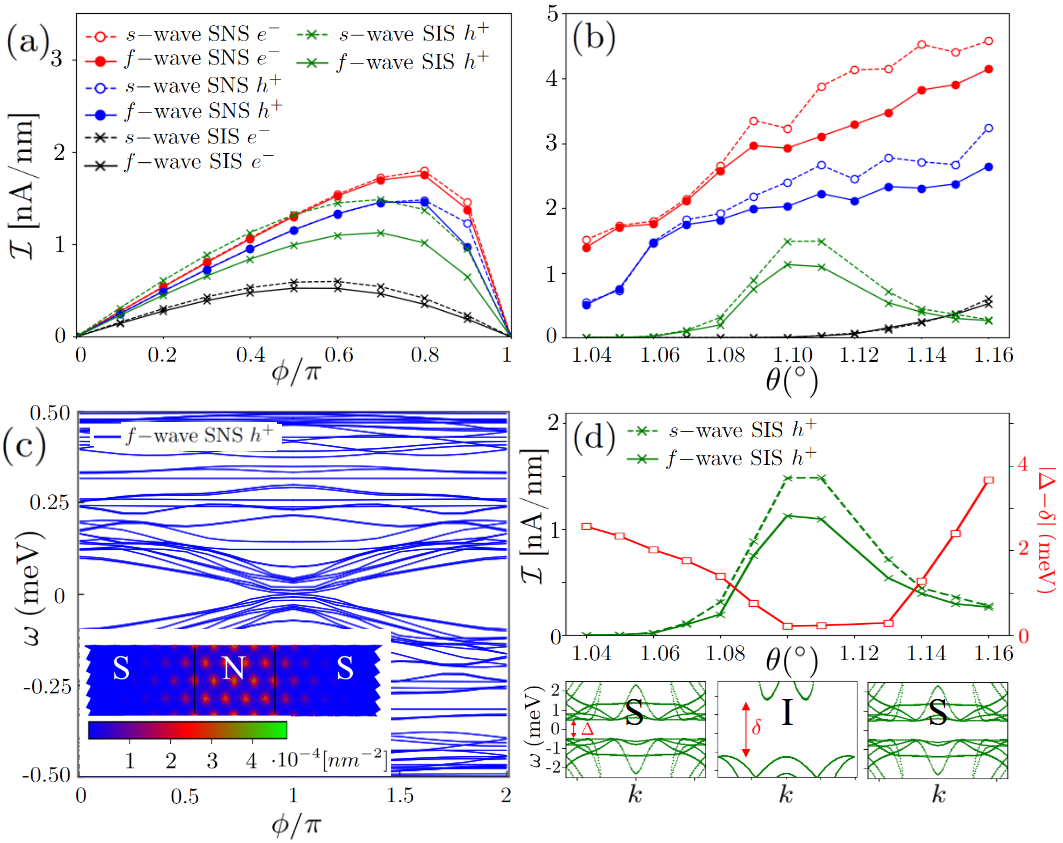}}%
    \caption{TBG Josephson junctions with equal electrodes.\ (a) Current-phase relations for near magic-angle junctions with different pairing symmetry: spin-singlet $s$-wave (dashed) or spin-triplet $f$-wave (solid), for electron and hole superconducting domes (fillings $n=\pm2.4$), with a metallic (SNS) or insulating link (SIS) \cite{bMiscSIS} We set the superconducting gap to 1 meV~\cite{oh2021evidence}.\ $\theta=1.06^{\circ}$ for SNS; $1.1^{\circ}$ for SIS $h^+$ and $1.16^{\circ}$ for SIS $e^-$.\ Units:\ nanoampere per nanometer junction width.\ (b) Critical current versus twist angle for all configurations.\ (c) Andreev spectrum at $1.06^{\circ}$.\ Inset:\ charge map of an Andreev bound state.\ (d) Critical current in SIS JJs compared to the difference between the superconducting and insulating gaps, as a function of twist angle, and a sketch of the bands in the different regions of a SIS junction.}
    \label{fig:3}
\end{figure}

{\it Josephson junctions.\ Results.} Figure~\ref{fig:3}(a) compares the current-phase relations (CPRs) of TBG JJs with $s$- and $f$-wave pairings in multiple configurations.\ CPRs can be measured with a SQUID geometry \cite{portoles2022tunable}.\ The main message of Fig.~\ref{fig:3} is that the type of pairing, $s$-wave or $f$-wave, plays a minor role when both electrodes are equal, compare dashed and solid lines in Fig.~\ref{fig:3}(a-b).\

In SNS JJs the CPR is skewed, due to high transmission of Andreev bound states, which carry over $80\%$ of the current in these junctions and are mostly localized in $AA$ stacking regions, see Fig.~\ref{fig:3}(c).\ In contrast, in SIS junctions the current comes from tunnelling states, so the CPR is sinusoidal~\cite{ambegaokar1963tunneling}.\ An exception occurs when the insulating gap in the link is comparable to the superconducting gap, resulting in skewness and large currents.\ The current in SIS junctions exponentially depends on the similitude between both gaps, see Fig.\ \ref{fig:3}(d).\ We note that the authors of Ref.~\cite{portoles2022tunable} report a sinusoidal CPR in TBG, without skewness, despite having a SNS JJ.\ This may be due to low transmission in the junction~\cite{golubov2004current}.\ Figure~\ref{fig:3}(b) shows the critical current for all JJs as a function of twist angle.\ For a comparison to experiments, see Ref.~\cite{SM}.\ The current in SNS JJs increases with twist angle, suggesting that larger Fermi velocities compensate the reduced density of states.\ Electron-hole asymmetry is very notable, e.g.\ near $\theta=1.1^{\circ}$, the current in SIS junctions with fillings $-2.4/4/-2.4$ is over two orders of magnitude larger than with $2.4/-4/2.4$ due to the asymmetry in the size of the gaps between narrow bands and electron- or hole-like remote bands.

\begin{figure}[t!]
   \centering{\includegraphics[scale = 0.32]{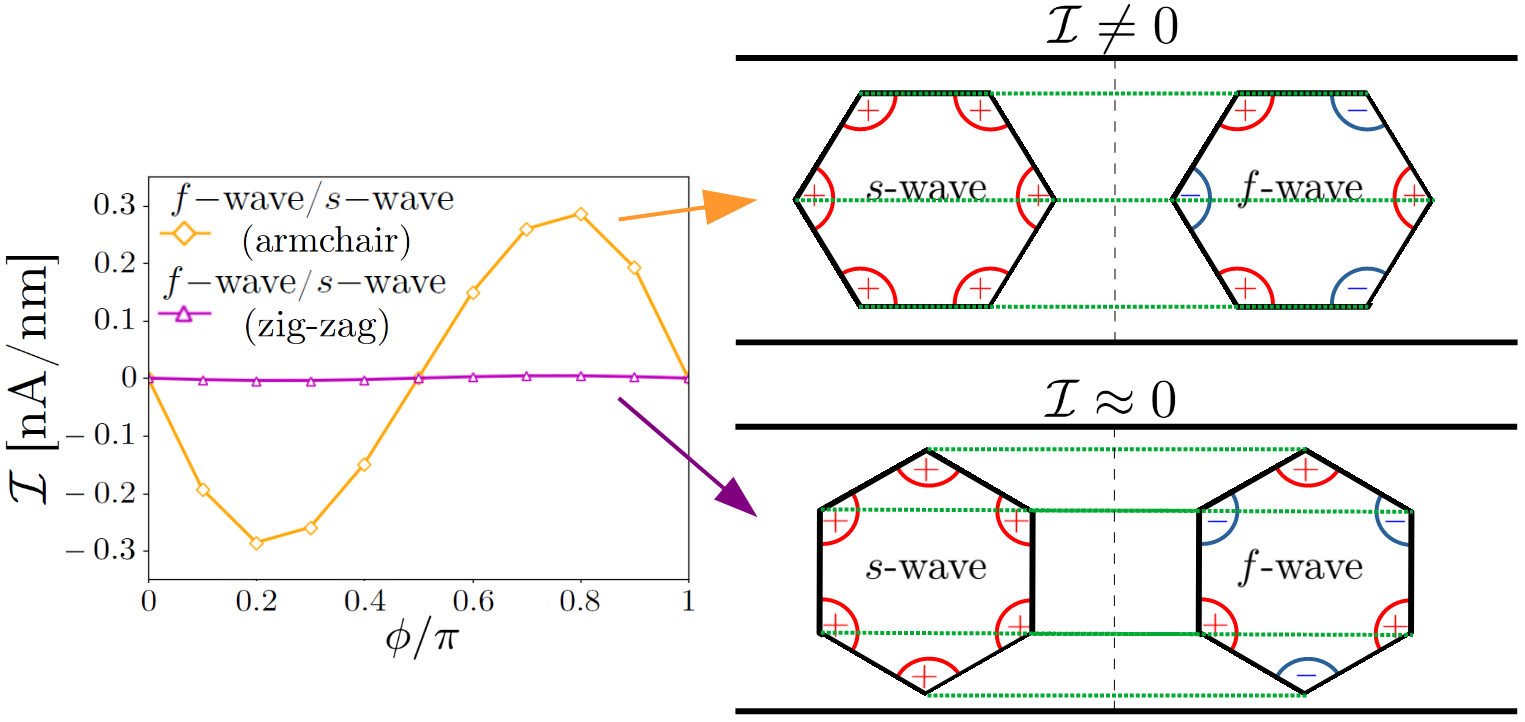}}%
    \caption{Current-phase relation in mixed $f$-wave and $s$-wave TBG Josephson junctions, nearly parallel to the graphene armchair, as in Fig.\ \ref{fig:2}(c), or zigzag directions.\ The critical current is $\sim100$ times larger for armchair junctions.}
    \label{fig:4}
\end{figure}

Ref.~\cite{de2021gate} reports a significant length dependence of the critical current in JJs prepared in mixed configurations, e.g.\ with the electrodes doped near one superconducting dome and the link near the other.\ This indicates that the superconducting pairing symmetry in the electron and hole domes may differ.\ The results in Fig.\ \ref{fig:4} for mixed $f$-wave and $s$-wave TBG JJs propose an experiment that could verify the hypothesis.\ For these JJs, the critical current dramatically depends on the angle between the junction and the lattice.\ A similar result in non-superconducting junctions was found in Ref.\ \cite{alvarado2021transport}.\ The critical current is sizeable when the junction axis is nearly parallel to the graphene armchair direction, but close to zero when parallel to the zigzag direction.\ As long as the perpendicular momentum is conserved, the zigzag JJ suffers destructive interference of the superconducting pockets along the green lines drawn in Fig.\ \ref{fig:4}.\ Also, the CPRs have a period of $\pi$, half the one of standard JJs.\ The origin of this effect is the existence of two sets of energy levels, due to coupling of the $s$-wave pocket to the two $f$-wave pockets, which have an intrinsic phase difference of $\pi$~\cite{sigrist1991phenomenological,zagoskin1997half}.\ Furthermore, the CPR shows a $\pi$-junction behaviour, i.e.\ it is first negative \cite{bulaevskii77superconducting, golubov2004current}.\ A requisite for these phenomena is that the triplet electrode is spin unpolarized, otherwise the current is zero due to spin conservation.\ The same occurs in a one-dimensional toy model~\cite{SM, zazunov2012supercurrent}.\\

{\it Discussion.} We have studied the role of the superconducting order parameter in transport through superconducting TBG junctions.\ We focus on $s$- and $f$- wave pairing (even and odd valley combinations), as these two choices are equally favored by long range interactions, either attractive or repulsive~\cite{SS20}.

We have calculated the critical current, and the current-phase relation for different types of Josephson junctions.\ JJs in which both electrodes are either $s$- or $f$-wave superconductors show similar features (unlike the $s$- and $p$- cases considered in Ref.~\cite{linder2009josephson}).\ On the other hand, the critical current in mixed ($s$- and $f$-) junctions depends strongly on the orientation of the junction with respect to the graphene lattice axes, with maxima for armchair junctions, and zeroes for zigzag junctions.\ Hence, mixed junctions are useful for determining whether two superconducting phases differ in their valley exchange parity.\ Such junctions can exist in various setups: i) different superconducting regions in the phase diagram of TBG show different order parameters~\cite{de2021gate}, ii) the superconducting state changes locally because of the spin-orbit coupling induced by a substrate~\cite{zhang2023spin}, iii)  superconducting TBG is combined with $s$-wave proximitized graphene~\cite{Heersche_2007,rui2020superconductivity}.\

For a junction between a normal STM tip and superconducting TBG, we find a prominent peak in subgap conductance for an $f$-wave order parameter, due to Andreev states induced by the tip, in agreement with the experiments of Refs.~\cite{oh2021evidence,kim2022evidence}.\ $f$-wave is also consistent with the U- and V-shaped densities of states measured in the weak coupling regime, as shown in Refs.~\cite{poduval2023vestigial,SM}.\ The agreement between the experiments~\cite{oh2021evidence,kim2022evidence} and the results presented here puts forward $f$-wave pairing as a leading candidate for the pairing symmetry of twisted graphene superconductors.\

We note that the results for the STM-superconductor junction apply equally well to all graphene superconductors~\cite{park2021tunable, hao2021electric,park2022robust,zhang2022promotion,zhou2021superconductivity,zhou2022isospin,zhang2023spin,holleis23Ising}.\ Extending the Josephson junction calculations to non-twisted graphene superconductors~\cite{zhou2021superconductivity,zhou2022isospin,zhang2023spin,holleis23Ising} is a promising direction for future research.\\

\textit{Acknowledgements} We thank Tommaso Cea, Shuichi Iwakiri, Klaus Ensslin and Fernando de Juan for fruitful discussions.\ This work was supported by funding from the European Commission, within the Graphene Flagship, Core 3, grant number 881603;\ the Severo Ochoa programme for centres of excellence in R\&D (CEX2020-001039-S/AEI/10.13039/501100011033) and grants SprQuMat (Ministerio de Ciencia e Innovación, Spain) and NMAT2D (Comunidad de Madrid, Spain).\ V.T.P. acknowledges support from  the Department of Energy under grant DE-FG02-84ER45118, the NSF Graduate Research Fellowships Program, and the P.D.\ Soros Fellowship for New Americans.

\onecolumngrid

\section*{Supplementary information}
\setcounter{equation}{0}
\setcounter{figure}{0}
\setcounter{table}{0}
\makeatletter
\renewcommand{\theequation}{S\arabic{equation}}
\renewcommand{\thefigure}{S\arabic{figure}}

\section*{S1. STM tip-superconductor junctions}
\subsection{Equations}
There are six equations relating the scattering amplitudes, as each node leads to an equation for the electron part of the wavefunction, and to another equation for the hole part, see Fig.\ \ref{fig:1}(a).\ These equations are:

\begin{align}
    \epsilon ( 1 + R_N ) &= t_{tip} ( e_N^* + R_N e_N ) + t_{tip,K} \left( \frac{T_{K,1}}{N_{K,1}} + \frac{T_{K,2}}{N_{K,2}} \right) + t_{tip,K'} \left( \frac{T_{K',1}}{N_{K',1}} + \frac{T_{K',2}}{N_{K',2}} \right) 
    \nonumber \\
    \epsilon \left( \frac{T_{K,1}}{N_{K,1}} + \frac{T_{K,2}}{N_{K,2}} \right) &= t^*_{tip,K}  ( 1 + R_N ) + t_{sc,K} \left( \frac{T_{K,1}}{N_{K,1}} e_{K,1} + \frac{T_{K,2}}{N_{K,2}} e_{K,2} \right) +t_{K,K'} \left( \frac{T_{K',1}}{N_{K',1}} + \frac{T_{K',2}}{N_{K',2}} \right)
    \nonumber \\
    \epsilon \left( \frac{T_{K',1}}{N_{K',1}} + \frac{T_{K',2}}{N_{K',2}} \right) &= t^*_{tip,K'}  ( 1 + R_N ) + t_{sc,K'} \left( \frac{T_{K',1}}{N_{K',1}} e_{K',1} + \frac{T_{K',2}}{N_{K',2}} e_{K',2}   \right)+ t_{K,K'}^* \left( \frac{T_{K,1}}{N_{K,1}} + \frac{T_{K,2}}{N_{K,2}} \right)
    \nonumber \\
    \epsilon R_A &= - t_{tip} R_A e_N^* - t_{tip,K} \left( \frac{T_{K,1} A_{K,1}}{N_{K,1}} + \frac{T_{K,2} A_{K,2}}{N_{K,2}} \right) - t_{tip,K'} \left( \frac{T_{K',1} A_{K',1}}{N_{K',1}} + \frac{T_{K',2} A_{K',2}}{N_{K',2}} \right)
    \nonumber \\
    \epsilon \left( \frac{T_{K,1} A_{K,1}}{N_{K,1}} + \frac{T_{K,2} A_{K,2}}{N_{K,2}} \right) &= - t_{tip,K}^* R_A - t_{sc,K} \left( \frac{T_{K,1} A_{K,1}}{N_{K,1}} e_{K,1} + \frac{T_{K,2} A_{K,2}}{N_{K,2}} e_{K,2} \right) 
    \nonumber \\
    &- t_{K,K'} \left( \frac{T_{K',1} A_{K',1}}{N_{K',1}} + \frac{T_{K',2} A_{K',2}}{N_{K',2}} \right)
    \nonumber \\
 \epsilon \left( \frac{T_{K',1} A_{K',1}}{N_{K',1}} + \frac{T_{K',2} A_{K',2}}{N_{K',2}} \right) &= - t_{tip,K'}^* R_A - t_{sc,K'} \left( \frac{T_{K',1} A_{K',1}}{N_{K',1}} e_{K',1} + \frac{T_{K',2} A_{K',2}}{N_{K',2}} e_{K',2} \right)  
    \nonumber \\
    &- t_{K,K'}^* \left( \frac{T_{K,1} A_{K,1}}{N_{K,1}}  + \frac{T_{K,2} A_{K,2}}{N_{K,2}} \right)
    \label{equations}
\end{align}

where $R_N , R_A, T_{K,1}, T_{K,2} , T_{K'_1} , T_{K',2}$ are the normal reflection coefficient, the Andreev reflection coefficient (associated to the tip electron and hole channels, respectively), and the transmission coefficients for the electron and hole channels in each valley, $\epsilon$ is the energy, and:
\begin{align}
    e_N &= - \frac{\epsilon}{2 t_{tip}} + i \sqrt{1 - \frac{\epsilon^2}{4 t_{tip}^2}}
    \nonumber \\
    e_{K,1} &= \left\{ \begin{array}{cc} 
    - i \sqrt{1 - \frac{\Delta_K^2 - \epsilon^2}{4 t_{sc,K}^2}} + i \sqrt{\frac{\Delta_K^2 - \epsilon^2}{4 t_{sc,K}^2}} &| \epsilon | \le \Delta _K \\
     - i \sqrt{1 + \frac{\epsilon^2 - \Delta_K^2}{4 t_{sc,K}^2}} + \sqrt{\frac{\epsilon^2 - \Delta_K^2}{4 t_{sc,K}^2}} &| \epsilon | > \Delta _K
    \end{array} \right.
       &
    e_{K,2} &= \left\{ \begin{array}{cc} 
     i \sqrt{1 - \frac{\Delta_K^2  - \epsilon^2}{4 t_{sc,K}^2}} - i \sqrt{\frac{\Delta_K^2 - \epsilon^2}{4 t_{sc,K}^2}} &| \epsilon | \le \Delta _K \\
      i \sqrt{1 - \frac{\epsilon^2 - \Delta_K^2}{4 t_{sc,K}^2}} + \sqrt{\frac{\epsilon^2 - \Delta_K^2}{4 t_{sc,K}^2}} &| \epsilon | > \Delta_K
    \end{array} \right.
     \nonumber \\
   A_{K,1} &= \left\{ \begin{array}{cc}
   \frac{- \epsilon - i  \sqrt{\Delta_{K}^2-\epsilon^2}}{\Delta_{K}} &  |\epsilon | \le \Delta _K \\
   \frac{- \epsilon - \sqrt{\epsilon^2 - \Delta_{K}^2}}{\Delta_{K}} &  |\epsilon | > \Delta _K
   \end{array} \right.
   &
    A_{K,2} &= \left\{ \begin{array}{cc}
   \frac{- \epsilon + i  \sqrt{\Delta_{K}^2-\epsilon^2}}{\Delta_{K}} &  |\epsilon | \le \Delta _K \\
   \frac{- \epsilon + \sqrt{\epsilon^2 - \Delta_{K}^2}}{\Delta_{K}} &  |\epsilon | > \Delta _K
   \end{array} \right.
   \nonumber \\
   N_{K,1} &= \sqrt{1 + | A_{K,1} |^2}
  &
    N_{K,2} &= \sqrt{1 + | A_{K,2} |^2}
\end{align}
and similar expressions for $K'$.

The model can be extended to intravalley superconducting gaps with angular dependence, like the cases discussed in Ref.\ \cite{sukhachov2022andreev}.\ Each individual scattering angle can be treated as an independent channel.\ Reflection and transmission coefficients need to be defined for each angle, and the total currents will be given by integrals over all angles.

\subsection{Subgap Andreev conductance.}
The enhancement of the conductance for voltages below the superconducting gap discussed in Ref.\ \cite{blonder1982transition} arises from processes where an incoming electron is reflected as a hole, or vice versa.\  In an $s$-wave superconductor, and in the limit of perfect transmission, these processes lead to a conductance which is twice the conductance in the normal state, as shown in the BTK theory \cite{blonder1982transition}.\ In an $s$-wave superconductor, an electron is injected into the superconductor as a coherent sum of even combinations of plane waves with momenta $\vec{k}$ and $- \vec{k}$, because due to time reversal symmetry, the normal-superconductor hopping elements satisfy $t_{tip,\vec{k}} = t_{tip,-\vec{k}}$.\ This even combination becomes coupled to another even combination of hole states, which can move back into the normal electrode.\ In a non $s$-wave superconductor, the superconducting gap changes sign.\ If there are pairs of momenta such that $\Delta_{\vec{k}} = - \Delta_{\vec{k'}}$ and $t_{tip,\vec{k}} = t_{tip,\vec{k'}}$, the amplitudes of the injected electron will be equal for $\vec{k}$ and $\vec{k'}$, but, inside the superconductor, it will be coupled to holes with amplitudes of opposite signs.\ Such a hole state cannot tunnel back into the normal electrode, and subgap Andreev conductance will be fully suppressed.\ Local tunneling processes, as expected in an STM experiment, imply momentum independent hopping elements, $t_{tip,\vec{k}} = t$.\ Hence, we can expect that the subgap Andreev conductance will be suppressed when a normal tip is coupled to generic $p$- and $d$-wave superconductors \cite{sukhachov2022andreev}, and also in the case of the $f$-wave, two valley superconductor considered here.\ The situation changes when there is intervalley scattering, see below.

\subsection{Tip induced Andreev states.}
We can understand the formation of subgap Andreev states by the intervalley elastic scattering induced by the tip by using the model shown in Fig.\ \ref{fig:1}(a).\ The model reduces to a simple tight binding model:
\begin{align}
    {\cal H} &= {\cal H}_{sc1} + {\cal H}_{sc2} + {\cal H}_{tip} \nonumber \\
    {\cal H}_{sc1} &= t_{sc1} \sum_{n=-\infty}^0 \left( c_{e,n}^\dagger c_{e,n-1} - c_{h,n}^\dagger c_{h,n-1} \right) + \Delta_{sc1} \sum_{n=-\infty}^0 c_{e,n}^\dagger c_{h,n} + h. c. \nonumber \\
    {\cal H}_{sc2} &= t_{sc2} \sum_{n=1}^{n=\infty} \left( c_{e,n}^\dagger c_{e,n+1} - c_{h,n}^\dagger c_{h,n+1} \right) + \Delta_{sc2} \sum_{n=1}^{n=\infty} c_{e,n}^\dagger c_{h,n} + h. c. \nonumber \\
    {\cal H}_{tip} &= t_{K,K'} \left( c_{e,0}^\dagger c_{e,1} - c_{h,0}^\dagger c_{h,1} \right) + h. c.
\end{align}
The Green's function at sites $n=0$ and $n=1$ of the system can be obtained from the Green's functions at the same sites in the absence of intervalley coupling:
\begin{align}
      \left(  \begin{array}{cc} G_{0,0} ( \omega )  &G_{0,1} ( \omega ) \\ G_{1,0} ( \omega ) &G_{1,1} ( \omega ) \end{array} \right) &= \left(  \begin{array}{cc} \bar{G}_{0,0}^{-1} ( \omega ) &t_{K,K'} {\cal I}_2 \\ t_{K,K'} {\cal I}_2 &\bar{G}_{1,1}^{-1} ( \omega ) \end{array} \right)^{-1}
\end{align}
where $\bar{G}_{0,0} ( \omega )$ and $\bar{G}_{1,1} ( \omega )$ are surface Green's functions associated to ${\cal H}_{sc1}$ and ${\cal H}_{sc2}$, and ${\cal I}_2$ is a $2 \times 2$ identity matrix.\ By changing to a basis defined by $c_{e,n} \pm c_{h,n}$, these matrix functions are:
\begin{align}
    \bar{G}_{0,0}^{-1} ( \omega ) &= \left( \begin{array}{cc}
    \frac{\omega - \Delta_{sc1}}{2} + \frac{\sqrt{( \omega^2 - \Delta_{sc1}^2) ( \omega^2 - \Delta_{sc1}^2 - 4 t_{sc1}^2)}}{2 ( \omega + \Delta_{sc1} )} &0\\
    0 &\frac{\omega + \Delta_{sc1}}{2} + \frac{\sqrt{( \omega^2 - \Delta_{sc1}^2) ( \omega^2 - \Delta_{sc1}^2 - 4 t_{sc1}^2)}}{2 ( \omega - \Delta_{sc1} )}
    \end{array} \right)
\end{align}
and an equivalent expression for $G_{1,1}^{-1} ( \omega )$.\ Finally, for $t_{sc1} = t_{sc2} = t$ and $\Delta_{sc1} = \Delta_{sc2} = \Delta$, the Andreev states are defined by the equations:
\begin{align}
     \frac{\omega \mp \Delta}{2}  + \frac{\sqrt{( \omega^2 - \Delta^2) ( \omega^2 - \Delta^2 - 4 t^2)}}{2 ( \omega \pm \Delta )} &= \pm t_{K,K'}
\end{align}
For $t_{K,K'} \ll \Delta , t$ this equation gives Andreev states near the edge of the superconducting gap, $\omega = \pm \Delta$, and for $t_{K,K'} = t$ the Andreev states move to the center of the gap, $\omega = 0$.\ The parameter $t$ describes a high energy cutoff of the order of the bandwidth.\ For TBG near a magic angle, it is reasonable to expect that the perturbation due to the tip in the contact regime is such that $t_{K,K'} \gtrsim t$, so that, in an $f$-wave superconductor, Andreev states near the center of the gap  will exist.

\subsection{Weak coupling conductance}

\begin{figure}[h]
   \centering{\includegraphics[width=9.5cm]{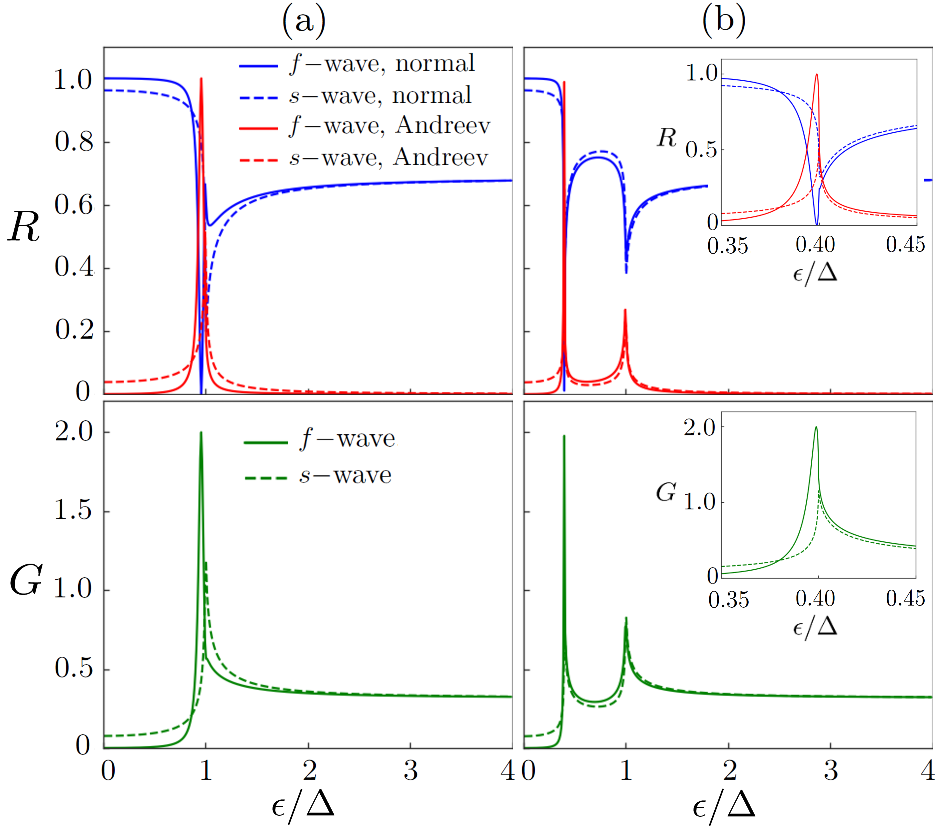}}
    \caption{Normal and Andreev reflection (top) and total conductance (bottom) in a STM tip-superconducting TBG junction in the weak coupling regime.\ (a) With Fermi velocity mismatch and intervalley scattering.\ The parameters used are $t_{tip} = 10, \, t_{sc,K} = t_{sc,K'} = 1, \, t_{tip,K} = t_{tip,K'} = 1 / \sqrt{2}, \, t_{K,K'} = 0.2, \, \Delta_K = 0.05, \, \Delta_{K'} = \pm 0.05$.\ (b) With Fermi velocity mismatch, intervalley scattering and spin-orbit coupling.\ $\Delta_K = 0.05, \, \Delta_{K'} = \pm 0.02$, others as in (a).\ The insets show a zoom near the edge of the smaller gap.} 
    \label{fig:s1}
\end{figure}

We show in Fig.\ \ref{fig:s1}(a) results in the regime where the normal transmission of the junction is small.\ The intervalley coupling has also been reduced.\ The Andreev reflection for the $s$-wave superconductor is notably suppressed.\ On the other hand, the intervalley coupling still induces subgap states near the edges of the gap of the $f$-wave superconductor.\ As a result, Andreev reflection persists, and leads to a peak in the junction conductance.\ Still, the conductance above and below the gap are similar for $s$-wave and $f$-wave superconductor.\ Therefore, in this weak coupling regime, it would be difficult for an experiment to tell apart the two pairings.\ As shown in Fig.\ \ref{fig:s1}(b), the transport characteristics of $f$-wave and $s$-wave superconductors remain similar to each other when spin-orbit coupling is included.

\section*{S2. Josephson junctions}
\subsection{Critical current: comparison to experiments}
In this section, we compare the critical currents obtained with our microscopic simulations to the ones measured in the experiments of Ref.\ \cite{de2021gate} In the main text we have used a superconducting gap of 1 meV, inferred from STM measurements \cite{oh2021evidence}.\ However the transport data of Ref.\ \cite{de2021gate} suggests a gap of $\sim0.1$ meV instead, so we set $\Delta=0.1$ meV in the following calculations.\ The experimental twist angle is 1.06$^{\circ}\pm0.04^{\circ}$ and the junction width is 1200 nm.\ Simulating a junction of this width is numerically prohibitive, so we extrapolate the results for a junction of width $\sim 50$ nm.\ We focus on the setup with a link of length $L_j\approx100$ nm.\ Note that, since the gate-defined potential profile is expected to be smooth, the effective link is most likely shorter than 100 nm.

In this setup, when the electrodes are tuned to the optimal filling for superconductivity ($n\approx-2.4$) and the link is slightly detuned from that ($n\approx-2.4+\delta n$), the junction is in a SNS configuration and the measured critical current is $\mathcal{I}_c\approx 250$ nA.\ When the link is doped instead to the gap between the narrow bands and the hole-like remote bands ($n=-4$), the junction is in a SIS configuration and the critical current is $\mathcal{I}_c\approx 40$ nA.

\begin{table}[h]
\vspace{+0.5cm}
\begin{tabular}{|p{4cm}|p{3cm}||p{3cm}|p{3cm}|}
 \hline
 \multicolumn{4}{|c|}{Experiment \cite{de2021gate}} \\
 \hline
 Twist angle & Link length & SNS $\mathcal{I}_c$ &SIS $\mathcal{I}_c$\\
 \hline
 1.06$^{\circ}\pm0.04^{\circ}$  & 100 nm   & \textbf{250 nA}    & \textbf{40 nA}\\
 \hline
 \hline
 \multicolumn{4}{|c|}{Theory} \\
 \hline
 Twist angle and pairing & Link length & SNS $\mathcal{I}_c$ &SIS $\mathcal{I}_c$\\
 \hline
 1.06$^{\circ}$ and $s$-wave  & 80 nm   & 420 nA    & 10$^{-4}$ nA\\
 \hline
  1.06$^{\circ}$ and $f$-wave  & 80 nm   & 240 nA    & 10$^{-4}$ nA\\
 \hline
  1.06$^{\circ}$ and $s$-wave  & 40 nm   & 490 nA    & 0.1 nA\\
 \hline
  1.12$^{\circ}$ and $s$-wave  & 80 nm   & 350 nA    & 3 nA\\
 \hline
  1.12$^{\circ}$ and $s$-wave  & 40 nm   & \textbf{420 nA}    & \textbf{40 nA}\\
 \hline
  1.12$^{\circ}$ and $f$-wave  & 40 nm   & \textbf{325 nA}   & \textbf{16 nA}\\
 \hline
\end{tabular}
 \caption{Critical current in TBG Josephson junctions:\ comparison between experiment and theory.}
 \label{table1}
\end{table}

In Table \ref{table1}, we present the results of the calculations in the two configurations.\ We consider two possibilities for the twist angle (1.06$^{\circ}$ and 1.12$^{\circ}$), as well as a longer (80 nm) and a shorter (40 nm) link.\ In some cases we also compare $s$- to $f$-wave pairing.\ We set the dielectric constant to $\epsilon_r=7$.\ The results are in very good agreement with the experiments.\ In particular, the critical current in the SNS configuration always matches the order magnitude of the experimental one, regardless of details such as the twist angle or the effective length of the link.\ In the case of the SIS junction, the current depends exponentially on the length of the link and the insulating gap between narrow and remote bands.\ Crucially, this insulating gap is 8 meV for 1.06$^{\circ}$ and diminishes to 4 meV at 1.12$^{\circ}$.\ At 1.12$^{\circ}$ and with a 40 nm link, the SIS current matches the experimental value.\ Therefore, a junction with this angle and link length reproduces simultaneously the SNS and SIS critical currents measured in Ref.\ \cite{de2021gate}.\ Note that an angle of 1.12$^{\circ}$ is well within the combined error bar of the experiment plus the uncertainty in the tight binding parameters.\ The results suggest that the top gate of 100 nm used in the experiment induces a rather smooth potential profile, leading to an effective link around half that length.\ A complementary or even alternative explanation is that the insulating phase has some metallic character due to defects.\ $s$- and $f$-wave lead to similar critical currents.

\subsection{Details of the model}
The junction lattice is built following the same procedure as in Ref.\ \cite{sainzcruz21high}.\ To avoid border transport, we impose periodic boundary conditions from top to bottom, which leads to a folding of the Brillouin zone.\ The folded bandstructure has more than two flat bands, e.g.\ four in Fig.\ \ref{fig:2}(a).\ Following the notation of \cite{sainzcruz21high}, we build a TBG nanotube with chiral vectors (44,2)@(-44,-2), which has a twist angle of 4.41$^{\circ}$ and 2704 sites in its unit cell.

We use a tight binding Hamiltonian given by \cite{lin2018minimum}
\begin{dmath}
{{\cal H}_0=-\sum_{i\neq j,m}\gamma_{ij}^{mm}(c^{\dagger}_{i,m}c_{j,m}+h.c.)}-\sum_{i, j,m}\gamma_{ij}^{m,m+1}(c^{\dagger}_{i,m}c_{j,m+1}+h.c.)+\sum_{i,m}V_H(n)c^{\dagger}_{i,m}c_{i,m}\, ,
\label{eq:s3}
\end{dmath}
where $i,j$ run over the lattice sites and $m$ is the layer index.\ $H_0$ includes intralayer hopping to nearest-neighbors only $\gamma_{ij}^{mm}=t_{\parallel}$ and interlayer hopping that decays exponentially away from the vertical direction, $\gamma_{ij}^{m,m+1}=t_{\perp}e^{-(\sqrt{r^2+d^2}-d)/\lambda_\perp}\frac{d^2}{r^2+d^2}$, where $d=0.335$ nm is the distance between layers, $t_{\parallel}=3.09$ eV and $t_{\perp}=0.39$ eV are the intralayer and interlayer hopping amplitudes and $\lambda_\perp=0.027$ nm is a cutoff for the interlayer hopping \cite{lin2018minimum}.\

We perform a scaling approximation, based on the fact that, within the continuum model, the bands of TBG depend, to first order, on a dimensionless parameter \cite{bistritzer2011moire},
\begin{equation}
\alpha = \frac{at_{\perp}}{2\hbar v_F \sin ( \theta / 2 )} \propto \frac{t_{\perp}}{t_{\parallel}\theta}\, .
\label{eq:s4}
\end{equation}
where $a$ is the lattice constant, $v_F$ is the Fermi velocity.\ Thus, a small angle $\theta$ can be simulated with a larger one $\theta^{\prime}$, doing the following transformations:\ $t_{\parallel}\rightarrow\frac{1}{\lambda} t_{\parallel}$, $a\rightarrow\lambda a$, $d\rightarrow\lambda d$, with $\lambda=\sin(\frac{\theta^{\prime}}{2})/\sin(\frac{\theta}{2})$ \cite{gonzalez2017,vahedi2021magnetism,sainzcruz21high}.\ This approximation reproduces well the low-energy bandstructure, as shown in Fig.\ \ref{fig:2}(a) in the main text.\ It is worth noting that scaling leads to a rigid blueshift of the bandstructure, of up to $\sim 20$ meV, which we have removed in the figure.\ We use scaling factors $\lambda \sim$ 4.

\begin{figure}[h]
   \centering{\includegraphics[width=12cm]{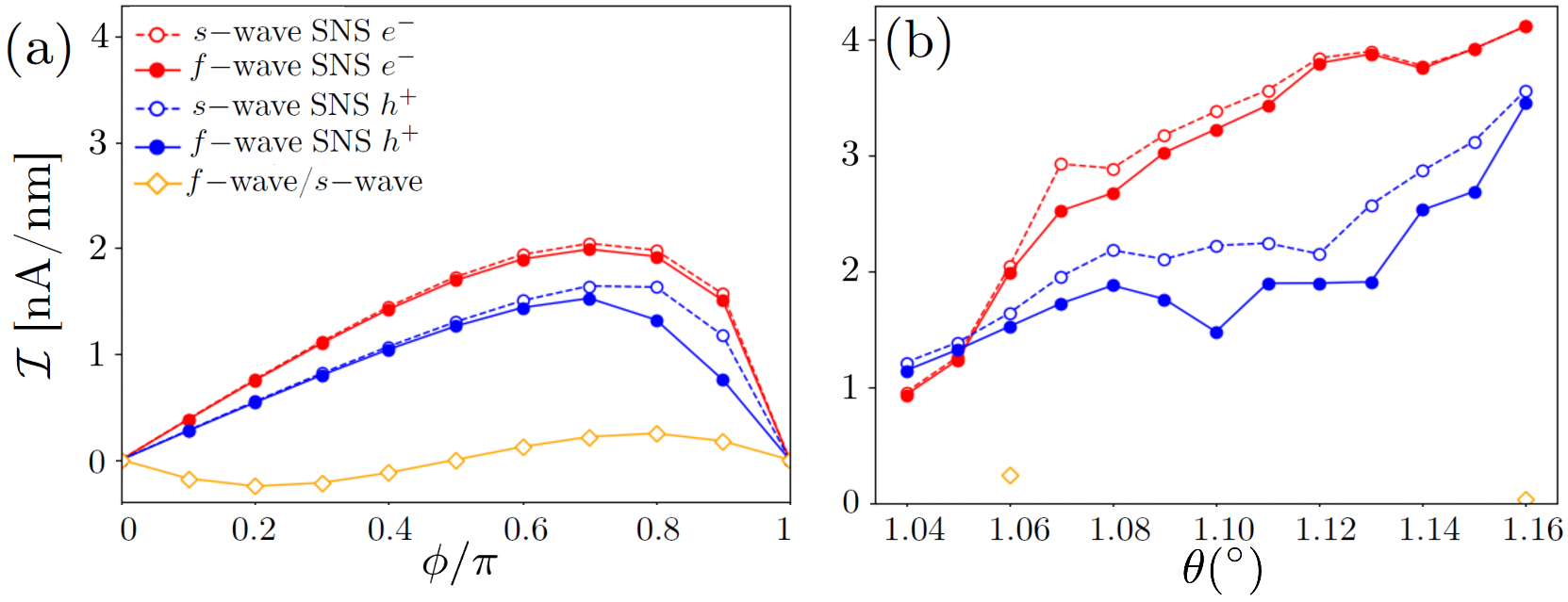}}%
    \caption{(a) Current-phase relation in TBG SNS and mixed JJs.\ (b) Critical current versus twist angle for all configurations.\ Everything equal to Fig.\ \ref{fig:3}(a-b) in the main text, but with dielectric constant $\epsilon_r=7$ instead of 4.}
    \label{fig:s2}
\end{figure}

In Eq.\ (\ref{eq:s3}), the Hartree term is $V_H(n)=\frac{2 \rho(n)}{\epsilon_r L_ M}\sum^3_{i=1}\cos({G_i}\cdot{r})$, where $G_i$ are the reciprocal lattice vectors, $r$ the position, $L_M$ the moiré period, $\epsilon_r$ the dielectric constant due to hBN encapsulation and $\rho(n)$ a filling dependent parameter, listed in Table \ref{table2}.\ Realistic values for the dielectric constant of hBN-encapsulated twisted bilayer graphene are usually in the range $\epsilon_r=4-10$, and our original choice was $\epsilon_r=4$.\ However, this value leads to band distortions so severe that, for some twist angles, the gaps between the central narrow bands and the remote dispersive bands are closed.\ Therefore it is not possible to form a proper SIS junction in these cases, because the bandgap for the insulating `I' region is not available.\ For this reason, the results for SIS junctions in Fig.\ \ref{fig:3} in the main text are calculated with $\epsilon_r=7$, so the gaps are preserved, while the results for SNS and mixed junctions are calculated with $\epsilon_r=4$.\ However, it is worth noting that, in contrast to SIS JJs, the current in SNS and mixed JJs depends only weakly on $\epsilon$.\ We show this explicitly in Fig.\ \ref{fig:s2}, which includes CPRs and critical currents for these JJs, for $\epsilon_r=7$.\ The results are qualitatively the same as those in Fig.\ \ref{fig:3} and \ref{fig:4} in the main text.

\begin{figure}[h]
   \centering{\includegraphics[width=11cm]{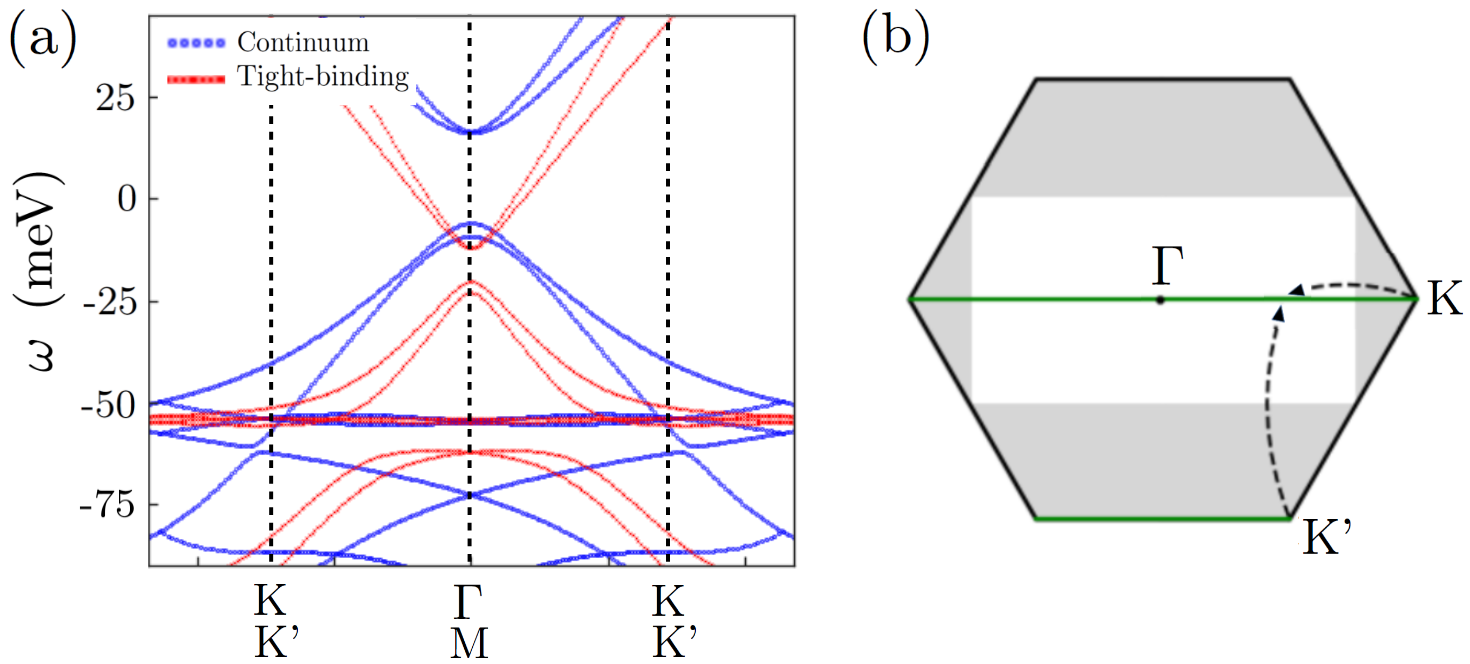}}%
    \caption{(a) Low energy TBG bands for a tight-binding model (red) and a continuum model, both including Hartree interactions.\ Filling $n=-2.4$, $\theta=1.06^{\circ}$ and $\epsilon_r=4$.\ (b) There are four bands because the system is technically a nanotube, in this case with a unit cell twice the size that of TBG, so its Brillouin zone results from the folding depicted in the figure.\ The green lines are the momentum values allowed by the periodic boundary condition which closes the nanotube.\ For a detailed discussion, see Ref.\ \cite{sainzcruz21high}.}
    \label{fig:s3}
\end{figure}

To obtain $\rho(n)$, we fit the bandstructure of the tight-binding model to the continuum model of Ref.\ \cite{moon2014optical} and do a self-consistent calculation.\ Fig.\ \ref{fig:s3} compares the resulting bandstructures obtained with the continuum and tight-binding Hamiltonians including the Hartree term.\ The bands are in fair agreement.\ In particular, note that the very narrow bands near -55 meV, which are similar in both cases, set the Fermi level, and over $95\%$ of the critical current comes from states in a small window around the gap which opens at the Fermi level.\

\begin{table}[h]
\begin{tabular}{|p{2.5cm}||p{1cm}|p{1cm}|p{1cm}|p{1cm}|p{1cm}|p{1cm}|p{1cm}|p{1cm}|p{1cm}|p{1cm}|p{1cm}|p{1cm}|p{1cm}|}
 \hline
 \multicolumn{14}{|c|}{Values of $\rho$ for the Hartree term} \\
 \hline
 & 1.04$^{\circ}$ & 1.05$^{\circ}$ & 1.06$^{\circ}$ & 1.07$^{\circ}$ & 1.08$^{\circ}$& 1.09$^{\circ}$& 1.10$^{\circ}$& 1.11$^{\circ}$& 1.12$^{\circ}$& 1.13$^{\circ}$& 1.14$^{\circ}$& 1.15$^{\circ}$& 1.16$^{\circ}$\\
 \hline
 $n=-2.4$; $\epsilon=4$ & -0.802 & -0.795 & -0.768 & -0.778 & -0.776 & -0.776 & -0.77 &-0.776 & -0.77 &-0.773 & -0.766 & -0.766 & -0.769 \\
 \hline
 $n=+2.4$; $\epsilon=4$  & 0.802 & 0.812 & 0.768 & 0.76 & 0.78 & 0.795 & 0.787 & 0.779 & 0.785 & 0.793 & 0.775 & 0.785 & 0.781 \\
 \hline
 $n=-2.4$; $\epsilon=7$  & -0.962 & -0.961 & -0.948 & -0.946 & -0.942 & -0.934 & -0.927 & -0.921 & -0.911 & -0.9 & -0.896 & -0.887 & -0.876 \\
 \hline
 $n=+2.4$; $\epsilon=7$  & 0.97 & 0.971 & 0.971 & 0.966 & 0.961 & 0.956 & 0.944 & 0.936 & 0.93 & 0.917 & 0.907 & 0.896 & 0.888 \\
 \hline
\end{tabular}
\label{tab:table_isingSOC}
\caption{Calculated values of the constant $\rho$ as a function of twist angle, filling and dielectric constant.}
\label{table2}
\end{table}

\begin{figure}[h]
   \centering{\includegraphics[width=6.5cm]{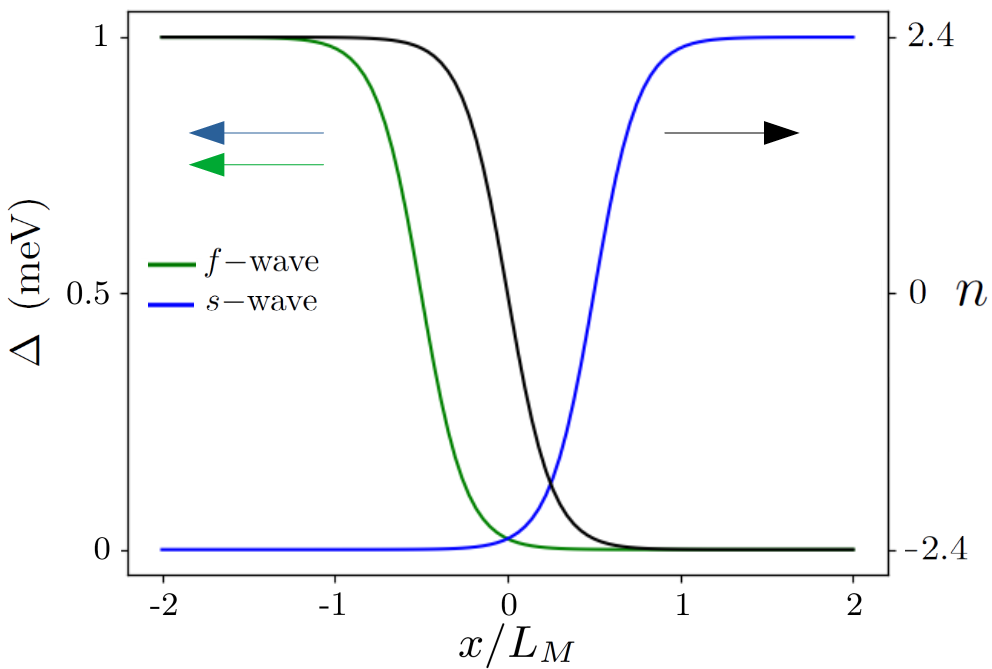}}%
    \caption{Superconducting gaps and filling at the interface between $f$-wave and $s$-wave electrodes in the mixed junction, versus position in units of moiré period, measured from the center of the junction.}
    \label{fig:s4}
\end{figure}

To obtain the results in Fig.\ \ref{fig:3} and \ref{fig:4}, we have exploited the fact that most of the critical current comes from states near the superconducting gap $\Delta=1$ meV, in fact, we have observed that states in the window $[-2\Delta, 2\Delta]$ carry over 95$\%$ of the current.\ Therefore, we have approximated Eq.\ \ref{eq:1} in the main text as:
\begin{equation}
\mathcal{I}\approx\frac{e}{h}\frac{\partial}{\partial\phi}\sum_{\mid\epsilon_i\mid<2\Delta}\epsilon_i   
\end{equation}
For the calculation of the current in the mixed junction, spin is explicitly included in the model, so the Hamiltonian in Eq.\ \ref{eq:2} is doubled.\ The $f$-wave electrode is at filling $n=2.4$ and the $s$-wave electrode at $n=-2.4$.\ The interface region interpolates smoothly between these two fillings.\ The gaps are similarly smoothed, see Fig.\ \ref{fig:s4}, in contrast to SNS and SIS junctions, for which we use hard boundary conditions.

With regards to the lenght of the system, we have found that 24 moiré unit cells per electrode are enough to reach convergent results, except in the mixed armchair junction, where 41 cells per electrode are needed.\ The low-energy spectrum was obtain with the library ARPACK.\ The complexity was approximately $\mathcal{O}(N^2)$ with N the number of sites in the system.\ To verify the algorithm was working as intended, we first reproduced some of the results in one-dimensional chains obtained with a Green's functions technique in Ref.\ \cite{perfetto2009equilibrium}.

\subsection{Toy model junction}
\begin{figure}[h]
   \centering{\includegraphics[width=12.5cm]{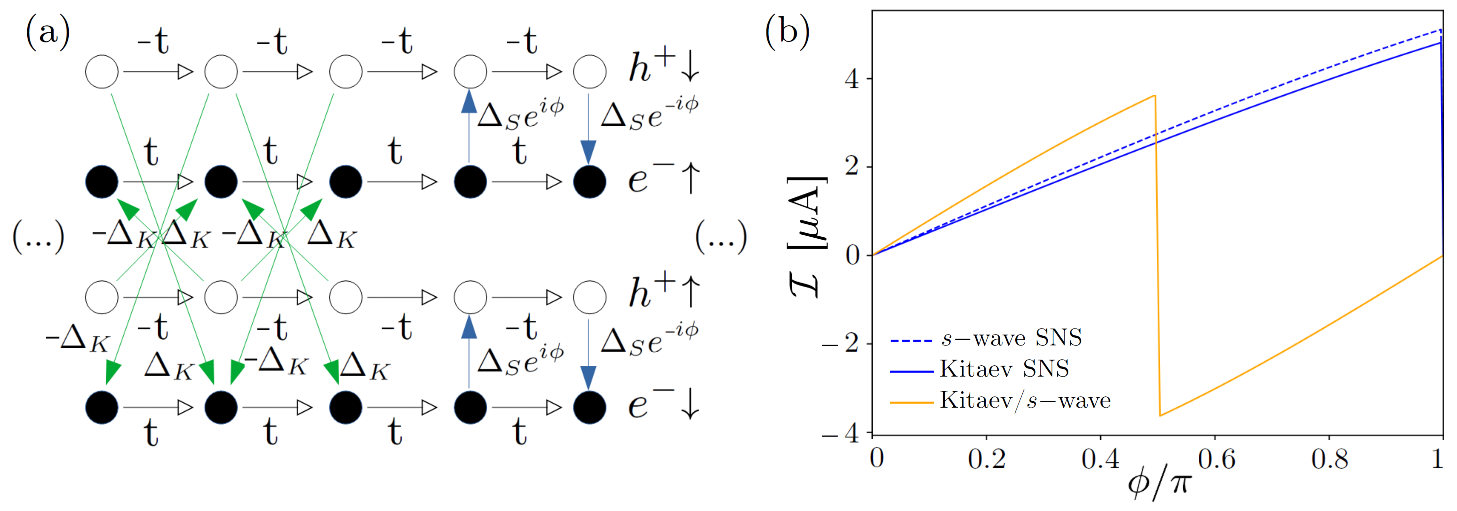}}%
    \caption{(a) Schematic of a mixed one-dimensional toy model junction, with Kitaev pairing on one side and $s$-wave pairing on the other.\
    (b) CPR of the mixed junction, compared to SNS junctions.\ The parameters used are $t=1$, $\Delta_K=0.1$ and $\Delta_s=0.2$.\ SNS junctions have five metallic atoms in the link.}
    \label{fig:s5}
\end{figure}

In the main text we showed that the critical current in mixed $f$-wave/$s$-wave TBG junctions has a phase periodicity of $\pi$, half of conventional junctions.\ This phenomenon was also found in Ref.\ \cite{zazunov2012supercurrent} in one-dimensional mixed chains.\ Fig.\ \ref{fig:s5} depicts a toy model which reproduces the result:\ in a chain of atoms with spin-unpolarized Kitaev ($p$-wave) pairing \cite{kitaev2001unpaired} on one side and $s$-wave pairing on the other, the current is $\pi$-periodic.\ Note that the saw-tooth profiles are a consequence of fully ballistic transport \cite{ishii1970josephson, golubov2004current}.

\subsection{Andreev spectra}

\begin{figure}[h]
   \centering{\includegraphics[width=16cm]{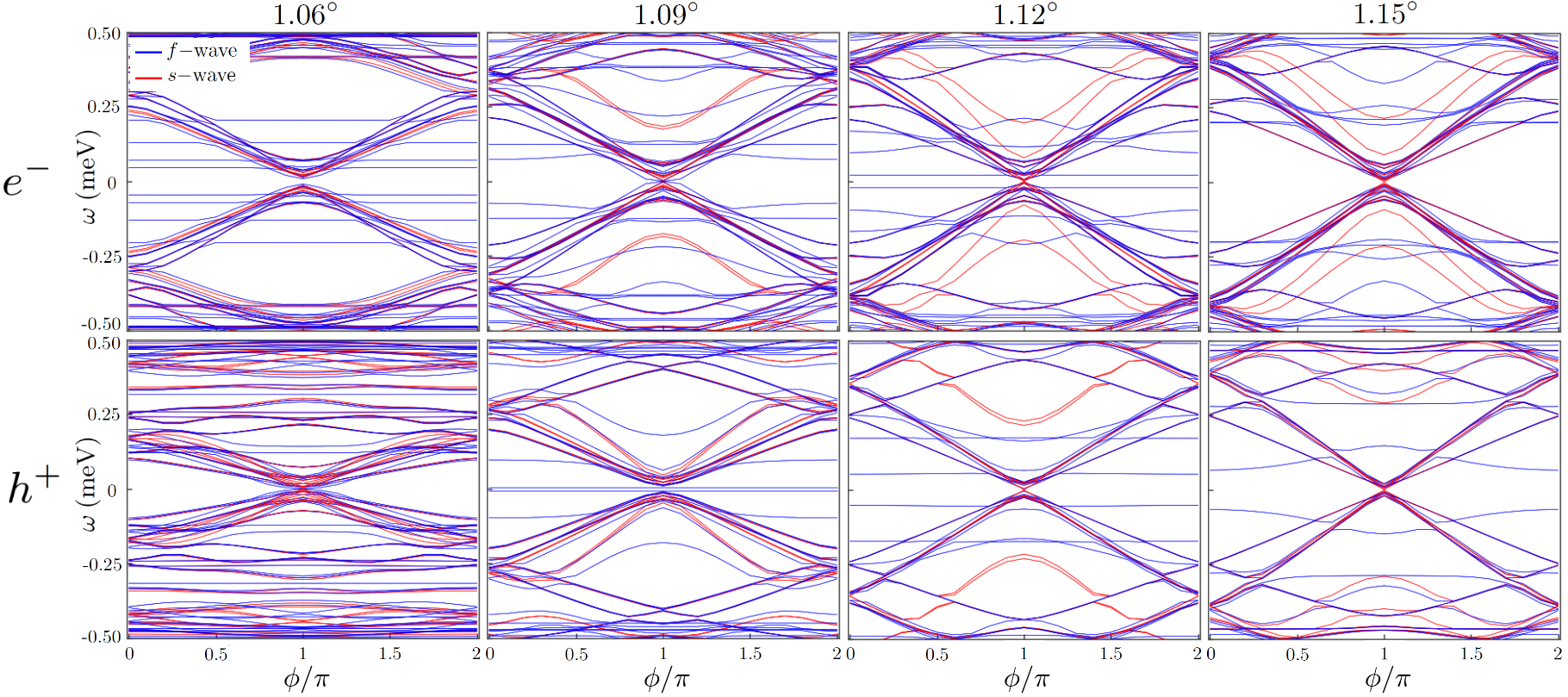}}%
    \caption{Subgap Andreev spectra in TBG SNS junctions, as a function of the superconducting phase difference, for $f$-wave and $s$-wave pairings and for electron (top row) and hole (bottom) domes, at different twist angles.\ $\epsilon_r=4$.}
    \label{fig:s6}
\end{figure}

\begin{figure}[h]
   \centering{\includegraphics[width=4.5cm]{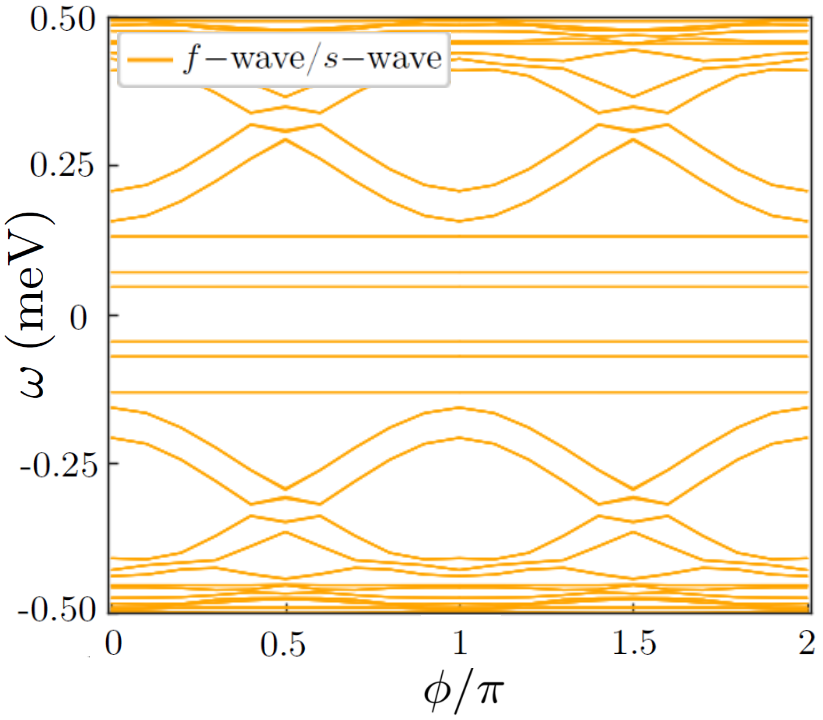}}%
    \caption{Subgap Andreev spectra in a mixed $f$-wave/$s$-wave TBG Josephson junction.\ $\theta=1.06^{\circ}$.\ $\epsilon_r=4$.}
    \label{fig:s7}
\end{figure}

Figure \ref{fig:s6} shows the subgap Andreev spectra in TBG junctions in the SNS configuration, for $f$-wave and $s$-wave parings at different fillings and twist angles.\ As stated in the main text, these states carry most of the current in SNS junctions.\ $f$-wave and $s$-wave have similar spectra, except for the quasi-flat levels that appear for $f$-wave pairing.\ These states, which are localized near the edges of the sample, are precursors of Majorana modes, which will be analyzed in a forthcoming publication.\ The spectrum changes fast near the magic angle, compare 1.06$^{\circ}$ and 1.09$^{\circ}$.\ The fact that the critical current increases with twist angle in these junctions is seen here as the growth of the slope of the Andreev levels with angle.\ There is marked electron-hole asymmetry.\ Finally, Fig.\ \ref{fig:s7} shows the subgap spectrum of a mixed $f$-wave/$s$-wave junction, which has a periodicity of $\pi$.

\begin{thebibliography}{83}%
\makeatletter
\providecommand \@ifxundefined [1]{%
 \@ifx{#1\undefined}
}%
\providecommand \@ifnum [1]{%
 \ifnum #1\expandafter \@firstoftwo
 \else \expandafter \@secondoftwo
 \fi
}%
\providecommand \@ifx [1]{%
 \ifx #1\expandafter \@firstoftwo
 \else \expandafter \@secondoftwo
 \fi
}%
\providecommand \natexlab [1]{#1}%
\providecommand \enquote  [1]{``#1''}%
\providecommand \bibnamefont  [1]{#1}%
\providecommand \bibfnamefont [1]{#1}%
\providecommand \citenamefont [1]{#1}%
\providecommand \href@noop [0]{\@secondoftwo}%
\providecommand \href [0]{\begingroup \@sanitize@url \@href}%
\providecommand \@href[1]{\@@startlink{#1}\@@href}%
\providecommand \@@href[1]{\endgroup#1\@@endlink}%
\providecommand \@sanitize@url [0]{\catcode `\\12\catcode `\$12\catcode
  `\&12\catcode `\#12\catcode `\^12\catcode `\_12\catcode `\%12\relax}%
\providecommand \@@startlink[1]{}%
\providecommand \@@endlink[0]{}%
\providecommand \url  [0]{\begingroup\@sanitize@url \@url }%
\providecommand \@url [1]{\endgroup\@href {#1}{\urlprefix }}%
\providecommand \urlprefix  [0]{URL }%
\providecommand \Eprint [0]{\href }%
\providecommand \doibase [0]{http://dx.doi.org/}%
\providecommand \selectlanguage [0]{\@gobble}%
\providecommand \bibinfo  [0]{\@secondoftwo}%
\providecommand \bibfield  [0]{\@secondoftwo}%
\providecommand \translation [1]{[#1]}%
\providecommand \BibitemOpen [0]{}%
\providecommand \bibitemStop [0]{}%
\providecommand \bibitemNoStop [0]{.\EOS\space}%
\providecommand \EOS [0]{\spacefactor3000\relax}%
\providecommand \BibitemShut  [1]{\csname bibitem#1\endcsname}%
\let\auto@bib@innerbib\@empty
\bibitem [{\citenamefont {Cao}\ \emph {et~al.}(2018{\natexlab{a}})\citenamefont
  {Cao}, \citenamefont {Fatemi}, \citenamefont {Demir}, \citenamefont {Fang},
  \citenamefont {Tomarken}, \citenamefont {Luo}, \citenamefont
  {Sanchez-Yamagishi}, \citenamefont {Watanabe}, \citenamefont {Taniguchi},
  \citenamefont {Kaxiras}, \citenamefont {Ashoori},\ and\ \citenamefont
  {Jarillo-Herrero}}]{cao2018correlated}%
  \BibitemOpen
  \bibfield  {author} {\bibinfo {author} {\bibfnamefont {Yuan}\ \bibnamefont
  {Cao}}, \bibinfo {author} {\bibfnamefont {Valla}\ \bibnamefont {Fatemi}},
  \bibinfo {author} {\bibfnamefont {Ahmet}\ \bibnamefont {Demir}}, \bibinfo
  {author} {\bibfnamefont {Shiang}\ \bibnamefont {Fang}}, \bibinfo {author}
  {\bibfnamefont {Spencer~L.}\ \bibnamefont {Tomarken}}, \bibinfo {author}
  {\bibfnamefont {Jason~Y.}\ \bibnamefont {Luo}}, \bibinfo {author}
  {\bibfnamefont {Javier~D.}\ \bibnamefont {Sanchez-Yamagishi}}, \bibinfo
  {author} {\bibfnamefont {Kenji}\ \bibnamefont {Watanabe}}, \bibinfo {author}
  {\bibfnamefont {Takashi}\ \bibnamefont {Taniguchi}}, \bibinfo {author}
  {\bibfnamefont {Efthimios}\ \bibnamefont {Kaxiras}}, \bibinfo {author}
  {\bibfnamefont {Ray~C.}\ \bibnamefont {Ashoori}}, \ and\ \bibinfo {author}
  {\bibfnamefont {Pablo}\ \bibnamefont {Jarillo-Herrero}},\ }\bibfield  {title}
  {\enquote {\bibinfo {title} {Correlated insulator behaviour at half-filling
  in magic-angle graphene superlattices},}\ }\href {\doibase
  10.1038/nature26154} {\bibfield  {journal} {\bibinfo  {journal} {Nature}\
  }\textbf {\bibinfo {volume} {556}},\ \bibinfo {pages} {80--84} (\bibinfo
  {year} {2018}{\natexlab{a}})}\BibitemShut {NoStop}%
\bibitem [{\citenamefont {Polshyn}\ \emph {et~al.}(2019)\citenamefont
  {Polshyn}, \citenamefont {Yankowitz}, \citenamefont {Chen}, \citenamefont
  {Zhang}, \citenamefont {Watanabe}, \citenamefont {Taniguchi}, \citenamefont
  {Dean},\ and\ \citenamefont {Young}}]{polshyn2019large}%
  \BibitemOpen
  \bibfield  {author} {\bibinfo {author} {\bibfnamefont {Hryhoriy}\
  \bibnamefont {Polshyn}}, \bibinfo {author} {\bibfnamefont {Matthew}\
  \bibnamefont {Yankowitz}}, \bibinfo {author} {\bibfnamefont {Shaowen}\
  \bibnamefont {Chen}}, \bibinfo {author} {\bibfnamefont {Yuxuan}\ \bibnamefont
  {Zhang}}, \bibinfo {author} {\bibfnamefont {K.}~\bibnamefont {Watanabe}},
  \bibinfo {author} {\bibfnamefont {T.}~\bibnamefont {Taniguchi}}, \bibinfo
  {author} {\bibfnamefont {Cory~R.}\ \bibnamefont {Dean}}, \ and\ \bibinfo
  {author} {\bibfnamefont {Andrea~F.}\ \bibnamefont {Young}},\ }\bibfield
  {title} {\enquote {\bibinfo {title} {Large linear-in-temperature resistivity
  in twisted bilayer graphene},}\ }\href {\doibase 10.1038/s41567-019-0596-3}
  {\bibfield  {journal} {\bibinfo  {journal} {Nature Physics}\ }\textbf
  {\bibinfo {volume} {15}},\ \bibinfo {pages} {1011--1016} (\bibinfo {year}
  {2019})}\BibitemShut {NoStop}%
\bibitem [{\citenamefont {Sharpe}\ \emph {et~al.}(2019)\citenamefont {Sharpe},
  \citenamefont {Fox}, \citenamefont {Barnard}, \citenamefont {Finney},
  \citenamefont {Watanabe}, \citenamefont {Taniguchi}, \citenamefont
  {Kastner},\ and\ \citenamefont {Goldhaber-Gordon}}]{sharpe2019emergent}%
  \BibitemOpen
  \bibfield  {author} {\bibinfo {author} {\bibfnamefont {Aaron~L.}\
  \bibnamefont {Sharpe}}, \bibinfo {author} {\bibfnamefont {Eli~J.}\
  \bibnamefont {Fox}}, \bibinfo {author} {\bibfnamefont {Arthur~W.}\
  \bibnamefont {Barnard}}, \bibinfo {author} {\bibfnamefont {Joe}\ \bibnamefont
  {Finney}}, \bibinfo {author} {\bibfnamefont {Kenji}\ \bibnamefont
  {Watanabe}}, \bibinfo {author} {\bibfnamefont {Takashi}\ \bibnamefont
  {Taniguchi}}, \bibinfo {author} {\bibfnamefont {M.~A.}\ \bibnamefont
  {Kastner}}, \ and\ \bibinfo {author} {\bibfnamefont {David}\ \bibnamefont
  {Goldhaber-Gordon}},\ }\bibfield  {title} {\enquote {\bibinfo {title}
  {Emergent ferromagnetism near three-quarters filling in twisted bilayer
  graphene},}\ }\href {\doibase 10.1126/science.aaw3780} {\bibfield  {journal}
  {\bibinfo  {journal} {Science}\ }\textbf {\bibinfo {volume} {365}},\ \bibinfo
  {pages} {605--608} (\bibinfo {year} {2019})}\BibitemShut {NoStop}%
\bibitem [{\citenamefont {Serlin}\ \emph {et~al.}(2020)\citenamefont {Serlin},
  \citenamefont {Tschirhart}, \citenamefont {Polshyn}, \citenamefont {Zhang},
  \citenamefont {Zhu}, \citenamefont {Watanabe}, \citenamefont {Taniguchi},
  \citenamefont {Balents},\ and\ \citenamefont {Young}}]{serlin2020intrinsic}%
  \BibitemOpen
  \bibfield  {author} {\bibinfo {author} {\bibfnamefont {M.}~\bibnamefont
  {Serlin}}, \bibinfo {author} {\bibfnamefont {C.~L.}\ \bibnamefont
  {Tschirhart}}, \bibinfo {author} {\bibfnamefont {H.}~\bibnamefont {Polshyn}},
  \bibinfo {author} {\bibfnamefont {Y.}~\bibnamefont {Zhang}}, \bibinfo
  {author} {\bibfnamefont {J.}~\bibnamefont {Zhu}}, \bibinfo {author}
  {\bibfnamefont {K.}~\bibnamefont {Watanabe}}, \bibinfo {author}
  {\bibfnamefont {T.}~\bibnamefont {Taniguchi}}, \bibinfo {author}
  {\bibfnamefont {L.}~\bibnamefont {Balents}}, \ and\ \bibinfo {author}
  {\bibfnamefont {A.~F.}\ \bibnamefont {Young}},\ }\bibfield  {title} {\enquote
  {\bibinfo {title} {Intrinsic quantized anomalous hall effect in a moir{\'{e}}
  heterostructure},}\ }\href {\doibase 10.1126/science.aay5533} {\bibfield
  {journal} {\bibinfo  {journal} {Science}\ }\textbf {\bibinfo {volume}
  {367}},\ \bibinfo {pages} {900--903} (\bibinfo {year} {2020})}\BibitemShut
  {NoStop}%
\bibitem [{\citenamefont {Chen}\ \emph {et~al.}(2020)\citenamefont {Chen},
  \citenamefont {Sharpe}, \citenamefont {Fox}, \citenamefont {Zhang},
  \citenamefont {Wang}, \citenamefont {Jiang}, \citenamefont {Lyu},
  \citenamefont {Li}, \citenamefont {Watanabe}, \citenamefont {Taniguchi},
  \citenamefont {Shi}, \citenamefont {Senthil}, \citenamefont
  {Goldhaber-Gordon}, \citenamefont {Zhang},\ and\ \citenamefont
  {Wang}}]{chen2020tunable}%
  \BibitemOpen
  \bibfield  {author} {\bibinfo {author} {\bibfnamefont {Guorui}\ \bibnamefont
  {Chen}}, \bibinfo {author} {\bibfnamefont {Aaron~L.}\ \bibnamefont {Sharpe}},
  \bibinfo {author} {\bibfnamefont {Eli~J.}\ \bibnamefont {Fox}}, \bibinfo
  {author} {\bibfnamefont {Ya-Hui}\ \bibnamefont {Zhang}}, \bibinfo {author}
  {\bibfnamefont {Shaoxin}\ \bibnamefont {Wang}}, \bibinfo {author}
  {\bibfnamefont {Lili}\ \bibnamefont {Jiang}}, \bibinfo {author}
  {\bibfnamefont {Bosai}\ \bibnamefont {Lyu}}, \bibinfo {author} {\bibfnamefont
  {Hongyuan}\ \bibnamefont {Li}}, \bibinfo {author} {\bibfnamefont {Kenji}\
  \bibnamefont {Watanabe}}, \bibinfo {author} {\bibfnamefont {Takashi}\
  \bibnamefont {Taniguchi}}, \bibinfo {author} {\bibfnamefont {Zhiwen}\
  \bibnamefont {Shi}}, \bibinfo {author} {\bibfnamefont {T.}~\bibnamefont
  {Senthil}}, \bibinfo {author} {\bibfnamefont {David}\ \bibnamefont
  {Goldhaber-Gordon}}, \bibinfo {author} {\bibfnamefont {Yuanbo}\ \bibnamefont
  {Zhang}}, \ and\ \bibinfo {author} {\bibfnamefont {Feng}\ \bibnamefont
  {Wang}},\ }\bibfield  {title} {\enquote {\bibinfo {title} {Tunable correlated
  chern insulator and ferromagnetism in a moir{\'{e}} superlattice},}\ }\href
  {\doibase 10.1038/s41586-020-2049-7} {\bibfield  {journal} {\bibinfo
  {journal} {Nature}\ }\textbf {\bibinfo {volume} {579}},\ \bibinfo {pages}
  {56--61} (\bibinfo {year} {2020})}\BibitemShut {NoStop}%
\bibitem [{\citenamefont {Uri}\ \emph {et~al.}(2020)\citenamefont {Uri},
  \citenamefont {Grover}, \citenamefont {Cao}, \citenamefont {Crosse},
  \citenamefont {Bagani}, \citenamefont {Rodan-Legrain}, \citenamefont
  {Myasoedov}, \citenamefont {Watanabe}, \citenamefont {Taniguchi},
  \citenamefont {Moon}, \citenamefont {Koshino}, \citenamefont
  {Jarillo-Herrero},\ and\ \citenamefont {Zeldov}}]{uri2020}%
  \BibitemOpen
  \bibfield  {author} {\bibinfo {author} {\bibfnamefont {A.}~\bibnamefont
  {Uri}}, \bibinfo {author} {\bibfnamefont {S.}~\bibnamefont {Grover}},
  \bibinfo {author} {\bibfnamefont {Y.}~\bibnamefont {Cao}}, \bibinfo {author}
  {\bibfnamefont {J.~A.}\ \bibnamefont {Crosse}}, \bibinfo {author}
  {\bibfnamefont {K.}~\bibnamefont {Bagani}}, \bibinfo {author} {\bibfnamefont
  {D.}~\bibnamefont {Rodan-Legrain}}, \bibinfo {author} {\bibfnamefont
  {Y.}~\bibnamefont {Myasoedov}}, \bibinfo {author} {\bibfnamefont
  {K.}~\bibnamefont {Watanabe}}, \bibinfo {author} {\bibfnamefont
  {T.}~\bibnamefont {Taniguchi}}, \bibinfo {author} {\bibfnamefont
  {P.}~\bibnamefont {Moon}}, \bibinfo {author} {\bibfnamefont {M.}~\bibnamefont
  {Koshino}}, \bibinfo {author} {\bibfnamefont {P.}~\bibnamefont
  {Jarillo-Herrero}}, \ and\ \bibinfo {author} {\bibfnamefont {E.}~\bibnamefont
  {Zeldov}},\ }\bibfield  {title} {\enquote {\bibinfo {title} {Mapping the
  twist-angle disorder and landau levels in magic-angle graphene},}\ }\href
  {\doibase 10.1038/s41586-020-2255-3} {\bibfield  {journal} {\bibinfo
  {journal} {Nature}\ }\textbf {\bibinfo {volume} {581}},\ \bibinfo {pages}
  {47--52} (\bibinfo {year} {2020})}\BibitemShut {NoStop}%
\bibitem [{\citenamefont {Saito}\ \emph {et~al.}(2020)\citenamefont {Saito},
  \citenamefont {Ge}, \citenamefont {Watanabe}, \citenamefont {Taniguchi},\
  and\ \citenamefont {Young}}]{saito2020independent}%
  \BibitemOpen
  \bibfield  {author} {\bibinfo {author} {\bibfnamefont {Yu}~\bibnamefont
  {Saito}}, \bibinfo {author} {\bibfnamefont {Jingyuan}\ \bibnamefont {Ge}},
  \bibinfo {author} {\bibfnamefont {Kenji}\ \bibnamefont {Watanabe}}, \bibinfo
  {author} {\bibfnamefont {Takashi}\ \bibnamefont {Taniguchi}}, \ and\ \bibinfo
  {author} {\bibfnamefont {Andrea~F.}\ \bibnamefont {Young}},\ }\bibfield
  {title} {\enquote {\bibinfo {title} {Independent superconductors and
  correlated insulators in twisted bilayer graphene},}\ }\href {\doibase
  10.1038/s41567-020-0928-3} {\bibfield  {journal} {\bibinfo  {journal} {Nature
  Physics}\ }\textbf {\bibinfo {volume} {16}},\ \bibinfo {pages} {926--930}
  (\bibinfo {year} {2020})}\BibitemShut {NoStop}%
\bibitem [{\citenamefont {Wong}\ \emph {et~al.}(2020)\citenamefont {Wong},
  \citenamefont {Nuckolls}, \citenamefont {Oh}, \citenamefont {Lian},
  \citenamefont {Xie}, \citenamefont {Jeon}, \citenamefont {Watanabe},
  \citenamefont {Taniguchi}, \citenamefont {Bernevig},\ and\ \citenamefont
  {Yazdani}}]{wong2020cascade}%
  \BibitemOpen
  \bibfield  {author} {\bibinfo {author} {\bibfnamefont {Dillon}\ \bibnamefont
  {Wong}}, \bibinfo {author} {\bibfnamefont {Kevin~P.}\ \bibnamefont
  {Nuckolls}}, \bibinfo {author} {\bibfnamefont {Myungchul}\ \bibnamefont
  {Oh}}, \bibinfo {author} {\bibfnamefont {Biao}\ \bibnamefont {Lian}},
  \bibinfo {author} {\bibfnamefont {Yonglong}\ \bibnamefont {Xie}}, \bibinfo
  {author} {\bibfnamefont {Sangjun}\ \bibnamefont {Jeon}}, \bibinfo {author}
  {\bibfnamefont {Kenji}\ \bibnamefont {Watanabe}}, \bibinfo {author}
  {\bibfnamefont {Takashi}\ \bibnamefont {Taniguchi}}, \bibinfo {author}
  {\bibfnamefont {B.~Andrei}\ \bibnamefont {Bernevig}}, \ and\ \bibinfo
  {author} {\bibfnamefont {Ali}\ \bibnamefont {Yazdani}},\ }\bibfield  {title}
  {\enquote {\bibinfo {title} {Cascade of electronic transitions in magic-angle
  twisted bilayer graphene},}\ }\href {\doibase 10.1038/s41586-020-2339-0}
  {\bibfield  {journal} {\bibinfo  {journal} {Nature}\ }\textbf {\bibinfo
  {volume} {582}},\ \bibinfo {pages} {198--202} (\bibinfo {year}
  {2020})}\BibitemShut {NoStop}%
\bibitem [{\citenamefont {Zondiner}\ \emph {et~al.}(2020)\citenamefont
  {Zondiner}, \citenamefont {Rozen}, \citenamefont {Rodan-Legrain},
  \citenamefont {Cao}, \citenamefont {Queiroz}, \citenamefont {Taniguchi},
  \citenamefont {Watanabe}, \citenamefont {Oreg}, \citenamefont {von Oppen},
  \citenamefont {Stern}, \citenamefont {Berg}, \citenamefont
  {Jarillo-Herrero},\ and\ \citenamefont {Ilani}}]{zondiner2020cascade}%
  \BibitemOpen
  \bibfield  {author} {\bibinfo {author} {\bibfnamefont {U.}~\bibnamefont
  {Zondiner}}, \bibinfo {author} {\bibfnamefont {A.}~\bibnamefont {Rozen}},
  \bibinfo {author} {\bibfnamefont {D.}~\bibnamefont {Rodan-Legrain}}, \bibinfo
  {author} {\bibfnamefont {Y.}~\bibnamefont {Cao}}, \bibinfo {author}
  {\bibfnamefont {R.}~\bibnamefont {Queiroz}}, \bibinfo {author} {\bibfnamefont
  {T.}~\bibnamefont {Taniguchi}}, \bibinfo {author} {\bibfnamefont
  {K.}~\bibnamefont {Watanabe}}, \bibinfo {author} {\bibfnamefont
  {Y.}~\bibnamefont {Oreg}}, \bibinfo {author} {\bibfnamefont {F.}~\bibnamefont
  {von Oppen}}, \bibinfo {author} {\bibfnamefont {Ady}\ \bibnamefont {Stern}},
  \bibinfo {author} {\bibfnamefont {E.}~\bibnamefont {Berg}}, \bibinfo {author}
  {\bibfnamefont {P.}~\bibnamefont {Jarillo-Herrero}}, \ and\ \bibinfo {author}
  {\bibfnamefont {S.}~\bibnamefont {Ilani}},\ }\bibfield  {title} {\enquote
  {\bibinfo {title} {Cascade of phase transitions and dirac revivals in
  magic-angle graphene},}\ }\href {\doibase 10.1038/s41586-020-2373-y}
  {\bibfield  {journal} {\bibinfo  {journal} {Nature}\ }\textbf {\bibinfo
  {volume} {582}},\ \bibinfo {pages} {203--208} (\bibinfo {year}
  {2020})}\BibitemShut {NoStop}%
\bibitem [{\citenamefont {Stepanov}\ \emph {et~al.}(2020)\citenamefont
  {Stepanov}, \citenamefont {Das}, \citenamefont {Lu}, \citenamefont
  {Fahimniya}, \citenamefont {Watanabe}, \citenamefont {Taniguchi},
  \citenamefont {Koppens}, \citenamefont {Lischner}, \citenamefont {Levitov},\
  and\ \citenamefont {Efetov}}]{stepanov2020untying}%
  \BibitemOpen
  \bibfield  {author} {\bibinfo {author} {\bibfnamefont {Petr}\ \bibnamefont
  {Stepanov}}, \bibinfo {author} {\bibfnamefont {Ipsita}\ \bibnamefont {Das}},
  \bibinfo {author} {\bibfnamefont {Xiaobo}\ \bibnamefont {Lu}}, \bibinfo
  {author} {\bibfnamefont {Ali}\ \bibnamefont {Fahimniya}}, \bibinfo {author}
  {\bibfnamefont {Kenji}\ \bibnamefont {Watanabe}}, \bibinfo {author}
  {\bibfnamefont {Takashi}\ \bibnamefont {Taniguchi}}, \bibinfo {author}
  {\bibfnamefont {Frank H.~L.}\ \bibnamefont {Koppens}}, \bibinfo {author}
  {\bibfnamefont {Johannes}\ \bibnamefont {Lischner}}, \bibinfo {author}
  {\bibfnamefont {Leonid}\ \bibnamefont {Levitov}}, \ and\ \bibinfo {author}
  {\bibfnamefont {Dmitri~K.}\ \bibnamefont {Efetov}},\ }\bibfield  {title}
  {\enquote {\bibinfo {title} {Untying the insulating and superconducting
  orders in magic-angle graphene},}\ }\href {\doibase
  10.1038/s41586-020-2459-6} {\bibfield  {journal} {\bibinfo  {journal}
  {Nature}\ }\textbf {\bibinfo {volume} {583}},\ \bibinfo {pages} {375--378}
  (\bibinfo {year} {2020})}\BibitemShut {NoStop}%
\bibitem [{\citenamefont {Xu}\ \emph {et~al.}(2020)\citenamefont {Xu},
  \citenamefont {Liu}, \citenamefont {Rhodes}, \citenamefont {Watanabe},
  \citenamefont {Taniguchi}, \citenamefont {Hone}, \citenamefont {Elser},
  \citenamefont {Mak},\ and\ \citenamefont {Shan}}]{xu2020correlated}%
  \BibitemOpen
  \bibfield  {author} {\bibinfo {author} {\bibfnamefont {Yang}\ \bibnamefont
  {Xu}}, \bibinfo {author} {\bibfnamefont {Song}\ \bibnamefont {Liu}}, \bibinfo
  {author} {\bibfnamefont {Daniel~A.}\ \bibnamefont {Rhodes}}, \bibinfo
  {author} {\bibfnamefont {Kenji}\ \bibnamefont {Watanabe}}, \bibinfo {author}
  {\bibfnamefont {Takashi}\ \bibnamefont {Taniguchi}}, \bibinfo {author}
  {\bibfnamefont {James}\ \bibnamefont {Hone}}, \bibinfo {author}
  {\bibfnamefont {Veit}\ \bibnamefont {Elser}}, \bibinfo {author}
  {\bibfnamefont {Kin~Fai}\ \bibnamefont {Mak}}, \ and\ \bibinfo {author}
  {\bibfnamefont {Jie}\ \bibnamefont {Shan}},\ }\bibfield  {title} {\enquote
  {\bibinfo {title} {Correlated insulating states at fractional fillings of
  moir{\'{e}} superlattices},}\ }\href {\doibase 10.1038/s41586-020-2868-6}
  {\bibfield  {journal} {\bibinfo  {journal} {Nature}\ }\textbf {\bibinfo
  {volume} {587}},\ \bibinfo {pages} {214--218} (\bibinfo {year}
  {2020})}\BibitemShut {NoStop}%
\bibitem [{\citenamefont {Choi}\ \emph {et~al.}(2021)\citenamefont {Choi},
  \citenamefont {Kim}, \citenamefont {Peng}, \citenamefont {Thomson},
  \citenamefont {Lewandowski}, \citenamefont {Polski}, \citenamefont {Zhang},
  \citenamefont {Arora}, \citenamefont {Watanabe}, \citenamefont {Taniguchi},
  \citenamefont {Alicea},\ and\ \citenamefont
  {Nadj-Perge}}]{choi2021correlation}%
  \BibitemOpen
  \bibfield  {author} {\bibinfo {author} {\bibfnamefont {Youngjoon}\
  \bibnamefont {Choi}}, \bibinfo {author} {\bibfnamefont {Hyunjin}\
  \bibnamefont {Kim}}, \bibinfo {author} {\bibfnamefont {Yang}\ \bibnamefont
  {Peng}}, \bibinfo {author} {\bibfnamefont {Alex}\ \bibnamefont {Thomson}},
  \bibinfo {author} {\bibfnamefont {Cyprian}\ \bibnamefont {Lewandowski}},
  \bibinfo {author} {\bibfnamefont {Robert}\ \bibnamefont {Polski}}, \bibinfo
  {author} {\bibfnamefont {Yiran}\ \bibnamefont {Zhang}}, \bibinfo {author}
  {\bibfnamefont {Harpreet~Singh}\ \bibnamefont {Arora}}, \bibinfo {author}
  {\bibfnamefont {Kenji}\ \bibnamefont {Watanabe}}, \bibinfo {author}
  {\bibfnamefont {Takashi}\ \bibnamefont {Taniguchi}}, \bibinfo {author}
  {\bibfnamefont {Jason}\ \bibnamefont {Alicea}}, \ and\ \bibinfo {author}
  {\bibfnamefont {Stevan}\ \bibnamefont {Nadj-Perge}},\ }\bibfield  {title}
  {\enquote {\bibinfo {title} {Correlation-driven topological phases in
  magic-angle twisted bilayer graphene},}\ }\href {\doibase
  10.1038/s41586-020-03159-7} {\bibfield  {journal} {\bibinfo  {journal}
  {Nature}\ }\textbf {\bibinfo {volume} {589}},\ \bibinfo {pages} {536--541}
  (\bibinfo {year} {2021})}\BibitemShut {NoStop}%
\bibitem [{\citenamefont {Rozen}\ \emph {et~al.}(2021)\citenamefont {Rozen},
  \citenamefont {Park}, \citenamefont {Zondiner}, \citenamefont {Cao},
  \citenamefont {Rodan-Legrain}, \citenamefont {Taniguchi}, \citenamefont
  {Watanabe}, \citenamefont {Oreg}, \citenamefont {Stern}, \citenamefont
  {Berg}, \citenamefont {Jarillo-Herrero},\ and\ \citenamefont
  {Ilani}}]{rozen2021entropic}%
  \BibitemOpen
  \bibfield  {author} {\bibinfo {author} {\bibfnamefont {Asaf}\ \bibnamefont
  {Rozen}}, \bibinfo {author} {\bibfnamefont {Jeong~Min}\ \bibnamefont {Park}},
  \bibinfo {author} {\bibfnamefont {Uri}\ \bibnamefont {Zondiner}}, \bibinfo
  {author} {\bibfnamefont {Yuan}\ \bibnamefont {Cao}}, \bibinfo {author}
  {\bibfnamefont {Daniel}\ \bibnamefont {Rodan-Legrain}}, \bibinfo {author}
  {\bibfnamefont {Takashi}\ \bibnamefont {Taniguchi}}, \bibinfo {author}
  {\bibfnamefont {Kenji}\ \bibnamefont {Watanabe}}, \bibinfo {author}
  {\bibfnamefont {Yuval}\ \bibnamefont {Oreg}}, \bibinfo {author}
  {\bibfnamefont {Ady}\ \bibnamefont {Stern}}, \bibinfo {author} {\bibfnamefont
  {Erez}\ \bibnamefont {Berg}}, \bibinfo {author} {\bibfnamefont {Pablo}\
  \bibnamefont {Jarillo-Herrero}}, \ and\ \bibinfo {author} {\bibfnamefont
  {Shahal}\ \bibnamefont {Ilani}},\ }\bibfield  {title} {\enquote {\bibinfo
  {title} {Entropic evidence for a pomeranchuk effect in magic-angle
  graphene},}\ }\href {\doibase 10.1038/s41586-021-03319-3} {\bibfield
  {journal} {\bibinfo  {journal} {Nature}\ }\textbf {\bibinfo {volume} {592}},\
  \bibinfo {pages} {214--219} (\bibinfo {year} {2021})}\BibitemShut {NoStop}%
\bibitem [{\citenamefont {Cao}\ \emph {et~al.}(2021{\natexlab{a}})\citenamefont
  {Cao}, \citenamefont {Rodan-Legrain}, \citenamefont {Park}, \citenamefont
  {Yuan}, \citenamefont {Watanabe}, \citenamefont {Taniguchi}, \citenamefont
  {Fernandes}, \citenamefont {Fu},\ and\ \citenamefont
  {Jarillo-Herrero}}]{cao2021nematicity}%
  \BibitemOpen
  \bibfield  {author} {\bibinfo {author} {\bibfnamefont {Yuan}\ \bibnamefont
  {Cao}}, \bibinfo {author} {\bibfnamefont {Daniel}\ \bibnamefont
  {Rodan-Legrain}}, \bibinfo {author} {\bibfnamefont {Jeong~Min}\ \bibnamefont
  {Park}}, \bibinfo {author} {\bibfnamefont {Noah F.~Q.}\ \bibnamefont {Yuan}},
  \bibinfo {author} {\bibfnamefont {Kenji}\ \bibnamefont {Watanabe}}, \bibinfo
  {author} {\bibfnamefont {Takashi}\ \bibnamefont {Taniguchi}}, \bibinfo
  {author} {\bibfnamefont {Rafael~M.}\ \bibnamefont {Fernandes}}, \bibinfo
  {author} {\bibfnamefont {Liang}\ \bibnamefont {Fu}}, \ and\ \bibinfo {author}
  {\bibfnamefont {Pablo}\ \bibnamefont {Jarillo-Herrero}},\ }\bibfield  {title}
  {\enquote {\bibinfo {title} {Nematicity and competing orders in
  superconducting magic-angle graphene},}\ }\href {\doibase
  10.1126/science.abc2836} {\bibfield  {journal} {\bibinfo  {journal}
  {Science}\ }\textbf {\bibinfo {volume} {372}},\ \bibinfo {pages} {264--271}
  (\bibinfo {year} {2021}{\natexlab{a}})}\BibitemShut {NoStop}%
\bibitem [{\citenamefont {Stepanov}\ \emph {et~al.}(2021)\citenamefont
  {Stepanov}, \citenamefont {Xie}, \citenamefont {Taniguchi}, \citenamefont
  {Watanabe}, \citenamefont {Lu}, \citenamefont {MacDonald}, \citenamefont
  {Bernevig},\ and\ \citenamefont {Efetov}}]{stepanov2021competing}%
  \BibitemOpen
  \bibfield  {author} {\bibinfo {author} {\bibfnamefont {Petr}\ \bibnamefont
  {Stepanov}}, \bibinfo {author} {\bibfnamefont {Ming}\ \bibnamefont {Xie}},
  \bibinfo {author} {\bibfnamefont {Takashi}\ \bibnamefont {Taniguchi}},
  \bibinfo {author} {\bibfnamefont {Kenji}\ \bibnamefont {Watanabe}}, \bibinfo
  {author} {\bibfnamefont {Xiaobo}\ \bibnamefont {Lu}}, \bibinfo {author}
  {\bibfnamefont {Allan~H.}\ \bibnamefont {MacDonald}}, \bibinfo {author}
  {\bibfnamefont {B.~Andrei}\ \bibnamefont {Bernevig}}, \ and\ \bibinfo
  {author} {\bibfnamefont {Dmitri~K.}\ \bibnamefont {Efetov}},\ }\bibfield
  {title} {\enquote {\bibinfo {title} {Competing zero-field chern insulators in
  superconducting twisted bilayer graphene},}\ }\href {\doibase
  10.1103/PhysRevLett.127.197701} {\bibfield  {journal} {\bibinfo  {journal}
  {Phys. Rev. Lett.}\ }\textbf {\bibinfo {volume} {127}},\ \bibinfo {pages}
  {197701} (\bibinfo {year} {2021})}\BibitemShut {NoStop}%
\bibitem [{\citenamefont {Oh}\ \emph {et~al.}(2021)\citenamefont {Oh},
  \citenamefont {Nuckolls}, \citenamefont {Wong}, \citenamefont {Lee},
  \citenamefont {Liu}, \citenamefont {Watanabe}, \citenamefont {Taniguchi},\
  and\ \citenamefont {Yazdani}}]{oh2021evidence}%
  \BibitemOpen
  \bibfield  {author} {\bibinfo {author} {\bibfnamefont {Myungchul}\
  \bibnamefont {Oh}}, \bibinfo {author} {\bibfnamefont {Kevin~P.}\ \bibnamefont
  {Nuckolls}}, \bibinfo {author} {\bibfnamefont {Dillon}\ \bibnamefont {Wong}},
  \bibinfo {author} {\bibfnamefont {Ryan~L.}\ \bibnamefont {Lee}}, \bibinfo
  {author} {\bibfnamefont {Xiaomeng}\ \bibnamefont {Liu}}, \bibinfo {author}
  {\bibfnamefont {Kenji}\ \bibnamefont {Watanabe}}, \bibinfo {author}
  {\bibfnamefont {Takashi}\ \bibnamefont {Taniguchi}}, \ and\ \bibinfo {author}
  {\bibfnamefont {Ali}\ \bibnamefont {Yazdani}},\ }\bibfield  {title} {\enquote
  {\bibinfo {title} {Evidence for unconventional superconductivity in twisted
  bilayer graphene},}\ }\href {\doibase 10.1038/s41586-021-04121-x} {\bibfield
  {journal} {\bibinfo  {journal} {Nature}\ }\textbf {\bibinfo {volume} {600}},\
  \bibinfo {pages} {240--245} (\bibinfo {year} {2021})}\BibitemShut {NoStop}%
\bibitem [{\citenamefont {Xie}\ \emph {et~al.}(2021)\citenamefont {Xie},
  \citenamefont {Pierce}, \citenamefont {Park}, \citenamefont {Parker},
  \citenamefont {Khalaf}, \citenamefont {Ledwith}, \citenamefont {Cao},
  \citenamefont {Lee}, \citenamefont {Chen}, \citenamefont {Forrester},
  \citenamefont {Watanabe}, \citenamefont {Taniguchi}, \citenamefont
  {Vishwanath}, \citenamefont {Jarillo-Herrero},\ and\ \citenamefont
  {Yacoby}}]{xie2021fractional}%
  \BibitemOpen
  \bibfield  {author} {\bibinfo {author} {\bibfnamefont {Yonglong}\
  \bibnamefont {Xie}}, \bibinfo {author} {\bibfnamefont {Andrew~T.}\
  \bibnamefont {Pierce}}, \bibinfo {author} {\bibfnamefont {Jeong~Min}\
  \bibnamefont {Park}}, \bibinfo {author} {\bibfnamefont {Daniel~E.}\
  \bibnamefont {Parker}}, \bibinfo {author} {\bibfnamefont {Eslam}\
  \bibnamefont {Khalaf}}, \bibinfo {author} {\bibfnamefont {Patrick}\
  \bibnamefont {Ledwith}}, \bibinfo {author} {\bibfnamefont {Yuan}\
  \bibnamefont {Cao}}, \bibinfo {author} {\bibfnamefont {Seung~Hwan}\
  \bibnamefont {Lee}}, \bibinfo {author} {\bibfnamefont {Shaowen}\ \bibnamefont
  {Chen}}, \bibinfo {author} {\bibfnamefont {Patrick~R.}\ \bibnamefont
  {Forrester}}, \bibinfo {author} {\bibfnamefont {Kenji}\ \bibnamefont
  {Watanabe}}, \bibinfo {author} {\bibfnamefont {Takashi}\ \bibnamefont
  {Taniguchi}}, \bibinfo {author} {\bibfnamefont {Ashvin}\ \bibnamefont
  {Vishwanath}}, \bibinfo {author} {\bibfnamefont {Pablo}\ \bibnamefont
  {Jarillo-Herrero}}, \ and\ \bibinfo {author} {\bibfnamefont {Amir}\
  \bibnamefont {Yacoby}},\ }\bibfield  {title} {\enquote {\bibinfo {title}
  {Fractional chern insulators in magic-angle twisted bilayer graphene},}\
  }\href {\doibase 10.1038/s41586-021-04002-3} {\bibfield  {journal} {\bibinfo
  {journal} {Nature}\ }\textbf {\bibinfo {volume} {600}},\ \bibinfo {pages}
  {439--443} (\bibinfo {year} {2021})}\BibitemShut {NoStop}%
\bibitem [{\citenamefont {Berdyugin}\ \emph {et~al.}(2022)\citenamefont
  {Berdyugin}, \citenamefont {Xin}, \citenamefont {Gao}, \citenamefont
  {Slizovskiy}, \citenamefont {Dong}, \citenamefont {Bhattacharjee},
  \citenamefont {Kumaravadivel}, \citenamefont {Xu}, \citenamefont
  {Ponomarenko}, \citenamefont {Holwill}, \citenamefont {Bandurin},
  \citenamefont {Kim}, \citenamefont {Cao}, \citenamefont {Greenaway},
  \citenamefont {Novoselov}, \citenamefont {Grigorieva}, \citenamefont
  {Watanabe}, \citenamefont {Taniguchi}, \citenamefont {Fal'ko}, \citenamefont
  {Levitov}, \citenamefont {Kumar},\ and\ \citenamefont
  {Geim}}]{berdyugin2022out}%
  \BibitemOpen
  \bibfield  {author} {\bibinfo {author} {\bibfnamefont {Alexey~I.}\
  \bibnamefont {Berdyugin}}, \bibinfo {author} {\bibfnamefont {Na}~\bibnamefont
  {Xin}}, \bibinfo {author} {\bibfnamefont {Haoyang}\ \bibnamefont {Gao}},
  \bibinfo {author} {\bibfnamefont {Sergey}\ \bibnamefont {Slizovskiy}},
  \bibinfo {author} {\bibfnamefont {Zhiyu}\ \bibnamefont {Dong}}, \bibinfo
  {author} {\bibfnamefont {Shubhadeep}\ \bibnamefont {Bhattacharjee}}, \bibinfo
  {author} {\bibfnamefont {P.}~\bibnamefont {Kumaravadivel}}, \bibinfo {author}
  {\bibfnamefont {Shuigang}\ \bibnamefont {Xu}}, \bibinfo {author}
  {\bibfnamefont {L.~A.}\ \bibnamefont {Ponomarenko}}, \bibinfo {author}
  {\bibfnamefont {Matthew}\ \bibnamefont {Holwill}}, \bibinfo {author}
  {\bibfnamefont {D.~A.}\ \bibnamefont {Bandurin}}, \bibinfo {author}
  {\bibfnamefont {Minsoo}\ \bibnamefont {Kim}}, \bibinfo {author}
  {\bibfnamefont {Yang}\ \bibnamefont {Cao}}, \bibinfo {author} {\bibfnamefont
  {M.~T.}\ \bibnamefont {Greenaway}}, \bibinfo {author} {\bibfnamefont {K.~S.}\
  \bibnamefont {Novoselov}}, \bibinfo {author} {\bibfnamefont {I.~V.}\
  \bibnamefont {Grigorieva}}, \bibinfo {author} {\bibfnamefont
  {K.}~\bibnamefont {Watanabe}}, \bibinfo {author} {\bibfnamefont
  {T.}~\bibnamefont {Taniguchi}}, \bibinfo {author} {\bibfnamefont {V.~I.}\
  \bibnamefont {Fal'ko}}, \bibinfo {author} {\bibfnamefont {L.~S.}\
  \bibnamefont {Levitov}}, \bibinfo {author} {\bibfnamefont {Roshan~Krishna}\
  \bibnamefont {Kumar}}, \ and\ \bibinfo {author} {\bibfnamefont {A.~K.}\
  \bibnamefont {Geim}},\ }\bibfield  {title} {\enquote {\bibinfo {title}
  {Out-of-equilibrium criticalities in graphene superlattices},}\ }\href
  {\doibase 10.1126/science.abi8627} {\bibfield  {journal} {\bibinfo  {journal}
  {Science}\ }\textbf {\bibinfo {volume} {375}},\ \bibinfo {pages} {430--433}
  (\bibinfo {year} {2022})}\BibitemShut {NoStop}%
\bibitem [{\citenamefont {Turkel}\ \emph {et~al.}(2022)\citenamefont {Turkel},
  \citenamefont {Swann}, \citenamefont {Zhu}, \citenamefont {Christos},
  \citenamefont {Watanabe}, \citenamefont {Taniguchi}, \citenamefont {Sachdev},
  \citenamefont {Scheurer}, \citenamefont {Kaxiras}, \citenamefont {Dean},\
  and\ \citenamefont {Pasupathy}}]{turkel2022orderly}%
  \BibitemOpen
  \bibfield  {author} {\bibinfo {author} {\bibfnamefont {Simon}\ \bibnamefont
  {Turkel}}, \bibinfo {author} {\bibfnamefont {Joshua}\ \bibnamefont {Swann}},
  \bibinfo {author} {\bibfnamefont {Ziyan}\ \bibnamefont {Zhu}}, \bibinfo
  {author} {\bibfnamefont {Maine}\ \bibnamefont {Christos}}, \bibinfo {author}
  {\bibfnamefont {K.}~\bibnamefont {Watanabe}}, \bibinfo {author}
  {\bibfnamefont {T.}~\bibnamefont {Taniguchi}}, \bibinfo {author}
  {\bibfnamefont {Subir}\ \bibnamefont {Sachdev}}, \bibinfo {author}
  {\bibfnamefont {Mathias~S.}\ \bibnamefont {Scheurer}}, \bibinfo {author}
  {\bibfnamefont {Efthimios}\ \bibnamefont {Kaxiras}}, \bibinfo {author}
  {\bibfnamefont {Cory~R.}\ \bibnamefont {Dean}}, \ and\ \bibinfo {author}
  {\bibfnamefont {Abhay~N.}\ \bibnamefont {Pasupathy}},\ }\bibfield  {title}
  {\enquote {\bibinfo {title} {Orderly disorder in magic-angle twisted trilayer
  graphene},}\ }\href {\doibase 10.1126/science.abk1895} {\bibfield  {journal}
  {\bibinfo  {journal} {Science}\ }\textbf {\bibinfo {volume} {376}},\ \bibinfo
  {pages} {193--199} (\bibinfo {year} {2022})}\BibitemShut {NoStop}%
\bibitem [{\citenamefont {Huang}\ \emph {et~al.}(2022)\citenamefont {Huang},
  \citenamefont {Tu}, \citenamefont {Shen}, \citenamefont {Zheng},
  \citenamefont {Wang}, \citenamefont {Wang}, \citenamefont {Khaliji},
  \citenamefont {Park}, \citenamefont {Liu}, \citenamefont {Yang},
  \citenamefont {Zhang}, \citenamefont {Shao}, \citenamefont {Li},
  \citenamefont {Low}, \citenamefont {Shi},\ and\ \citenamefont
  {Wang}}]{huang2022observation}%
  \BibitemOpen
  \bibfield  {author} {\bibinfo {author} {\bibfnamefont {Tianye}\ \bibnamefont
  {Huang}}, \bibinfo {author} {\bibfnamefont {Xuecou}\ \bibnamefont {Tu}},
  \bibinfo {author} {\bibfnamefont {Changqing}\ \bibnamefont {Shen}}, \bibinfo
  {author} {\bibfnamefont {Binjie}\ \bibnamefont {Zheng}}, \bibinfo {author}
  {\bibfnamefont {Junzhuan}\ \bibnamefont {Wang}}, \bibinfo {author}
  {\bibfnamefont {Hao}\ \bibnamefont {Wang}}, \bibinfo {author} {\bibfnamefont
  {Kaveh}\ \bibnamefont {Khaliji}}, \bibinfo {author} {\bibfnamefont
  {Sang~Hyun}\ \bibnamefont {Park}}, \bibinfo {author} {\bibfnamefont
  {Zhiyong}\ \bibnamefont {Liu}}, \bibinfo {author} {\bibfnamefont {Teng}\
  \bibnamefont {Yang}}, \bibinfo {author} {\bibfnamefont {Zhidong}\
  \bibnamefont {Zhang}}, \bibinfo {author} {\bibfnamefont {Lei}\ \bibnamefont
  {Shao}}, \bibinfo {author} {\bibfnamefont {Xuesong}\ \bibnamefont {Li}},
  \bibinfo {author} {\bibfnamefont {Tony}\ \bibnamefont {Low}}, \bibinfo
  {author} {\bibfnamefont {Yi}~\bibnamefont {Shi}}, \ and\ \bibinfo {author}
  {\bibfnamefont {Xiaomu}\ \bibnamefont {Wang}},\ }\bibfield  {title} {\enquote
  {\bibinfo {title} {Observation of chiral and slow plasmons in twisted bilayer
  graphene},}\ }\href {\doibase 10.1038/s41586-022-04520-8} {\bibfield
  {journal} {\bibinfo  {journal} {Nature}\ }\textbf {\bibinfo {volume} {605}},\
  \bibinfo {pages} {63--68} (\bibinfo {year} {2022})}\BibitemShut {NoStop}%
\bibitem [{\citenamefont {de~la Barrera}\ \emph {et~al.}(2022)\citenamefont
  {de~la Barrera}, \citenamefont {Aronson}, \citenamefont {Zheng},
  \citenamefont {Watanabe}, \citenamefont {Taniguchi}, \citenamefont {Ma},
  \citenamefont {Jarillo-Herrero},\ and\ \citenamefont
  {Ashoori}}]{barrera2022cascade}%
  \BibitemOpen
  \bibfield  {author} {\bibinfo {author} {\bibfnamefont {Sergio~C.}\
  \bibnamefont {de~la Barrera}}, \bibinfo {author} {\bibfnamefont {Samuel}\
  \bibnamefont {Aronson}}, \bibinfo {author} {\bibfnamefont {Zhiren}\
  \bibnamefont {Zheng}}, \bibinfo {author} {\bibfnamefont {Kenji}\ \bibnamefont
  {Watanabe}}, \bibinfo {author} {\bibfnamefont {Takashi}\ \bibnamefont
  {Taniguchi}}, \bibinfo {author} {\bibfnamefont {Qiong}\ \bibnamefont {Ma}},
  \bibinfo {author} {\bibfnamefont {Pablo}\ \bibnamefont {Jarillo-Herrero}}, \
  and\ \bibinfo {author} {\bibfnamefont {Raymond}\ \bibnamefont {Ashoori}},\
  }\bibfield  {title} {\enquote {\bibinfo {title} {Cascade of isospin phase
  transitions in bernal-stacked bilayer graphene at zero magnetic field},}\
  }\href {\doibase 10.1038/s41567-022-01616-w} {\bibfield  {journal} {\bibinfo
  {journal} {Nature Physics}\ }\textbf {\bibinfo {volume} {18}},\ \bibinfo
  {pages} {771--775} (\bibinfo {year} {2022})}\BibitemShut {NoStop}%
\bibitem [{\citenamefont {Kim}\ \emph {et~al.}(2022)\citenamefont {Kim},
  \citenamefont {Choi}, \citenamefont {Lewandowski}, \citenamefont {Thomson},
  \citenamefont {Zhang}, \citenamefont {Polski}, \citenamefont {Watanabe},
  \citenamefont {Taniguchi}, \citenamefont {Alicea},\ and\ \citenamefont
  {Nadj-Perge}}]{kim2022evidence}%
  \BibitemOpen
  \bibfield  {author} {\bibinfo {author} {\bibfnamefont {Hyunjin}\ \bibnamefont
  {Kim}}, \bibinfo {author} {\bibfnamefont {Youngjoon}\ \bibnamefont {Choi}},
  \bibinfo {author} {\bibfnamefont {Cyprian}\ \bibnamefont {Lewandowski}},
  \bibinfo {author} {\bibfnamefont {Alex}\ \bibnamefont {Thomson}}, \bibinfo
  {author} {\bibfnamefont {Yiran}\ \bibnamefont {Zhang}}, \bibinfo {author}
  {\bibfnamefont {Robert}\ \bibnamefont {Polski}}, \bibinfo {author}
  {\bibfnamefont {Kenji}\ \bibnamefont {Watanabe}}, \bibinfo {author}
  {\bibfnamefont {Takashi}\ \bibnamefont {Taniguchi}}, \bibinfo {author}
  {\bibfnamefont {Jason}\ \bibnamefont {Alicea}}, \ and\ \bibinfo {author}
  {\bibfnamefont {Stevan}\ \bibnamefont {Nadj-Perge}},\ }\bibfield  {title}
  {\enquote {\bibinfo {title} {Evidence for unconventional superconductivity in
  twisted trilayer graphene},}\ }\href {\doibase 10.1038/s41586-022-04715-z}
  {\bibfield  {journal} {\bibinfo  {journal} {Nature}\ }\textbf {\bibinfo
  {volume} {606}},\ \bibinfo {pages} {494--500} (\bibinfo {year}
  {2022})}\BibitemShut {NoStop}%
\bibitem [{\citenamefont {Seiler}\ \emph {et~al.}(2022)\citenamefont {Seiler},
  \citenamefont {Geisenhof}, \citenamefont {Winterer}, \citenamefont
  {Watanabe}, \citenamefont {Taniguchi}, \citenamefont {Xu}, \citenamefont
  {Zhang},\ and\ \citenamefont {Weitz}}]{seiler2022quantum}%
  \BibitemOpen
  \bibfield  {author} {\bibinfo {author} {\bibfnamefont {Anna~M.}\ \bibnamefont
  {Seiler}}, \bibinfo {author} {\bibfnamefont {Fabian~R.}\ \bibnamefont
  {Geisenhof}}, \bibinfo {author} {\bibfnamefont {Felix}\ \bibnamefont
  {Winterer}}, \bibinfo {author} {\bibfnamefont {Kenji}\ \bibnamefont
  {Watanabe}}, \bibinfo {author} {\bibfnamefont {Takashi}\ \bibnamefont
  {Taniguchi}}, \bibinfo {author} {\bibfnamefont {Tianyi}\ \bibnamefont {Xu}},
  \bibinfo {author} {\bibfnamefont {Fan}\ \bibnamefont {Zhang}}, \ and\
  \bibinfo {author} {\bibfnamefont {R.~Thomas}\ \bibnamefont {Weitz}},\
  }\bibfield  {title} {\enquote {\bibinfo {title} {Quantum cascade of
  correlated phases in trigonally warped bilayer graphene},}\ }\href {\doibase
  10.1038/s41586-022-04937-1} {\bibfield  {journal} {\bibinfo  {journal}
  {Nature}\ }\textbf {\bibinfo {volume} {608}},\ \bibinfo {pages} {298--302}
  (\bibinfo {year} {2022})}\BibitemShut {NoStop}%
\bibitem [{\citenamefont {Cao}\ \emph {et~al.}(2018{\natexlab{b}})\citenamefont
  {Cao}, \citenamefont {Fatemi}, \citenamefont {Fang}, \citenamefont
  {Watanabe}, \citenamefont {Taniguchi}, \citenamefont {Kaxiras},\ and\
  \citenamefont {Jarillo-Herrero}}]{cao2018unconventional}%
  \BibitemOpen
  \bibfield  {author} {\bibinfo {author} {\bibfnamefont {Yuan}\ \bibnamefont
  {Cao}}, \bibinfo {author} {\bibfnamefont {Valla}\ \bibnamefont {Fatemi}},
  \bibinfo {author} {\bibfnamefont {Shiang}\ \bibnamefont {Fang}}, \bibinfo
  {author} {\bibfnamefont {Kenji}\ \bibnamefont {Watanabe}}, \bibinfo {author}
  {\bibfnamefont {Takashi}\ \bibnamefont {Taniguchi}}, \bibinfo {author}
  {\bibfnamefont {Efthimios}\ \bibnamefont {Kaxiras}}, \ and\ \bibinfo {author}
  {\bibfnamefont {Pablo}\ \bibnamefont {Jarillo-Herrero}},\ }\bibfield  {title}
  {\enquote {\bibinfo {title} {Unconventional superconductivity in magic-angle
  graphene superlattices},}\ }\href {\doibase 10.1038/nature26160} {\bibfield
  {journal} {\bibinfo  {journal} {Nature}\ }\textbf {\bibinfo {volume} {556}},\
  \bibinfo {pages} {43--50} (\bibinfo {year} {2018}{\natexlab{b}})}\BibitemShut
  {NoStop}%
\bibitem [{\citenamefont {Yankowitz}\ \emph {et~al.}(2019)\citenamefont
  {Yankowitz}, \citenamefont {Chen}, \citenamefont {Polshyn}, \citenamefont
  {Zhang}, \citenamefont {Watanabe}, \citenamefont {Taniguchi}, \citenamefont
  {Graf}, \citenamefont {Young},\ and\ \citenamefont {Dean}}]{yankowitz2019}%
  \BibitemOpen
  \bibfield  {author} {\bibinfo {author} {\bibfnamefont {Matthew}\ \bibnamefont
  {Yankowitz}}, \bibinfo {author} {\bibfnamefont {Shaowen}\ \bibnamefont
  {Chen}}, \bibinfo {author} {\bibfnamefont {Hryhoriy}\ \bibnamefont
  {Polshyn}}, \bibinfo {author} {\bibfnamefont {Yuxuan}\ \bibnamefont {Zhang}},
  \bibinfo {author} {\bibfnamefont {K.}~\bibnamefont {Watanabe}}, \bibinfo
  {author} {\bibfnamefont {T.}~\bibnamefont {Taniguchi}}, \bibinfo {author}
  {\bibfnamefont {David}\ \bibnamefont {Graf}}, \bibinfo {author}
  {\bibfnamefont {Andrea~F.}\ \bibnamefont {Young}}, \ and\ \bibinfo {author}
  {\bibfnamefont {Cory~R.}\ \bibnamefont {Dean}},\ }\bibfield  {title}
  {\enquote {\bibinfo {title} {Tuning superconductivity in twisted bilayer
  graphene},}\ }\href {\doibase 10.1126/science.aav1910} {\bibfield  {journal}
  {\bibinfo  {journal} {Science}\ }\textbf {\bibinfo {volume} {363}},\ \bibinfo
  {pages} {1059--1064} (\bibinfo {year} {2019})}\BibitemShut {NoStop}%
\bibitem [{\citenamefont {Lu}\ \emph {et~al.}(2019)\citenamefont {Lu},
  \citenamefont {Stepanov}, \citenamefont {Yang}, \citenamefont {Xie},
  \citenamefont {Aamir}, \citenamefont {Das}, \citenamefont {Urgell},
  \citenamefont {Watanabe}, \citenamefont {Taniguchi}, \citenamefont {Zhang},
  \citenamefont {Bachtold}, \citenamefont {MacDonald},\ and\ \citenamefont
  {Efetov}}]{lu2019superconductors}%
  \BibitemOpen
  \bibfield  {author} {\bibinfo {author} {\bibfnamefont {Xiaobo}\ \bibnamefont
  {Lu}}, \bibinfo {author} {\bibfnamefont {Petr}\ \bibnamefont {Stepanov}},
  \bibinfo {author} {\bibfnamefont {Wei}\ \bibnamefont {Yang}}, \bibinfo
  {author} {\bibfnamefont {Ming}\ \bibnamefont {Xie}}, \bibinfo {author}
  {\bibfnamefont {Mohammed~Ali}\ \bibnamefont {Aamir}}, \bibinfo {author}
  {\bibfnamefont {Ipsita}\ \bibnamefont {Das}}, \bibinfo {author}
  {\bibfnamefont {Carles}\ \bibnamefont {Urgell}}, \bibinfo {author}
  {\bibfnamefont {Kenji}\ \bibnamefont {Watanabe}}, \bibinfo {author}
  {\bibfnamefont {Takashi}\ \bibnamefont {Taniguchi}}, \bibinfo {author}
  {\bibfnamefont {Guangyu}\ \bibnamefont {Zhang}}, \bibinfo {author}
  {\bibfnamefont {Adrian}\ \bibnamefont {Bachtold}}, \bibinfo {author}
  {\bibfnamefont {Allan~H.}\ \bibnamefont {MacDonald}}, \ and\ \bibinfo
  {author} {\bibfnamefont {Dmitri~K.}\ \bibnamefont {Efetov}},\ }\bibfield
  {title} {\enquote {\bibinfo {title} {Superconductors, orbital magnets and
  correlated states in magic-angle bilayer graphene},}\ }\href {\doibase
  10.1038/s41586-019-1695-0} {\bibfield  {journal} {\bibinfo  {journal}
  {Nature}\ }\textbf {\bibinfo {volume} {574}},\ \bibinfo {pages} {653--657}
  (\bibinfo {year} {2019})}\BibitemShut {NoStop}%
\bibitem [{\citenamefont {Park}\ \emph {et~al.}(2021)\citenamefont {Park},
  \citenamefont {Cao}, \citenamefont {Watanabe}, \citenamefont {Taniguchi},\
  and\ \citenamefont {Jarillo-Herrero}}]{park2021tunable}%
  \BibitemOpen
  \bibfield  {author} {\bibinfo {author} {\bibfnamefont {Jeong~Min}\
  \bibnamefont {Park}}, \bibinfo {author} {\bibfnamefont {Yuan}\ \bibnamefont
  {Cao}}, \bibinfo {author} {\bibfnamefont {Kenji}\ \bibnamefont {Watanabe}},
  \bibinfo {author} {\bibfnamefont {Takashi}\ \bibnamefont {Taniguchi}}, \ and\
  \bibinfo {author} {\bibfnamefont {Pablo}\ \bibnamefont {Jarillo-Herrero}},\
  }\bibfield  {title} {\enquote {\bibinfo {title} {Tunable strongly coupled
  superconductivity in magic-angle twisted trilayer graphene},}\ }\href
  {\doibase 10.1038/s41586-021-03192-0} {\bibfield  {journal} {\bibinfo
  {journal} {Nature}\ }\textbf {\bibinfo {volume} {590}},\ \bibinfo {pages}
  {249--255} (\bibinfo {year} {2021})}\BibitemShut {NoStop}%
\bibitem [{\citenamefont {Hao}\ \emph {et~al.}(2021)\citenamefont {Hao},
  \citenamefont {Zimmerman}, \citenamefont {Ledwith}, \citenamefont {Khalaf},
  \citenamefont {Najafabadi}, \citenamefont {Watanabe}, \citenamefont
  {Taniguchi}, \citenamefont {Vishwanath},\ and\ \citenamefont
  {Kim}}]{hao2021electric}%
  \BibitemOpen
  \bibfield  {author} {\bibinfo {author} {\bibfnamefont {Zeyu}\ \bibnamefont
  {Hao}}, \bibinfo {author} {\bibfnamefont {A.~M.}\ \bibnamefont {Zimmerman}},
  \bibinfo {author} {\bibfnamefont {Patrick}\ \bibnamefont {Ledwith}}, \bibinfo
  {author} {\bibfnamefont {Eslam}\ \bibnamefont {Khalaf}}, \bibinfo {author}
  {\bibfnamefont {Danial~Haie}\ \bibnamefont {Najafabadi}}, \bibinfo {author}
  {\bibfnamefont {Kenji}\ \bibnamefont {Watanabe}}, \bibinfo {author}
  {\bibfnamefont {Takashi}\ \bibnamefont {Taniguchi}}, \bibinfo {author}
  {\bibfnamefont {Ashvin}\ \bibnamefont {Vishwanath}}, \ and\ \bibinfo {author}
  {\bibfnamefont {Philip}\ \bibnamefont {Kim}},\ }\bibfield  {title} {\enquote
  {\bibinfo {title} {Electric field{\textendash}tunable superconductivity in
  alternating-twist magic-angle trilayer graphene},}\ }\href {\doibase
  10.1126/science.abg0399} {\bibfield  {journal} {\bibinfo  {journal}
  {Science}\ }\textbf {\bibinfo {volume} {371}},\ \bibinfo {pages} {1133--1138}
  (\bibinfo {year} {2021})}\BibitemShut {NoStop}%
\bibitem [{\citenamefont {Park}\ \emph {et~al.}(2022)\citenamefont {Park},
  \citenamefont {Cao}, \citenamefont {Xia}, \citenamefont {Sun}, \citenamefont
  {Watanabe}, \citenamefont {Taniguchi},\ and\ \citenamefont
  {Jarillo-Herrero}}]{park2022robust}%
  \BibitemOpen
  \bibfield  {author} {\bibinfo {author} {\bibfnamefont {Jeong~Min}\
  \bibnamefont {Park}}, \bibinfo {author} {\bibfnamefont {Yuan}\ \bibnamefont
  {Cao}}, \bibinfo {author} {\bibfnamefont {Li-Qiao}\ \bibnamefont {Xia}},
  \bibinfo {author} {\bibfnamefont {Shuwen}\ \bibnamefont {Sun}}, \bibinfo
  {author} {\bibfnamefont {Kenji}\ \bibnamefont {Watanabe}}, \bibinfo {author}
  {\bibfnamefont {Takashi}\ \bibnamefont {Taniguchi}}, \ and\ \bibinfo {author}
  {\bibfnamefont {Pablo}\ \bibnamefont {Jarillo-Herrero}},\ }\bibfield  {title}
  {\enquote {\bibinfo {title} {Robust superconductivity in magic-angle
  multilayer graphene family},}\ }\href {\doibase 10.1038/s41563-022-01287-1}
  {\bibfield  {journal} {\bibinfo  {journal} {Nature Materials}\ }\textbf
  {\bibinfo {volume} {21}},\ \bibinfo {pages} {877--883} (\bibinfo {year}
  {2022})}\BibitemShut {NoStop}%
\bibitem [{\citenamefont {Zhang}\ \emph {et~al.}(2022)\citenamefont {Zhang},
  \citenamefont {Polski}, \citenamefont {Lewandowski}, \citenamefont {Thomson},
  \citenamefont {Peng}, \citenamefont {Choi}, \citenamefont {Kim},
  \citenamefont {Watanabe}, \citenamefont {Taniguchi}, \citenamefont {Alicea},
  \citenamefont {von Oppen}, \citenamefont {Refael},\ and\ \citenamefont
  {Nadj-Perge}}]{zhang2022promotion}%
  \BibitemOpen
  \bibfield  {author} {\bibinfo {author} {\bibfnamefont {Yiran}\ \bibnamefont
  {Zhang}}, \bibinfo {author} {\bibfnamefont {Robert}\ \bibnamefont {Polski}},
  \bibinfo {author} {\bibfnamefont {Cyprian}\ \bibnamefont {Lewandowski}},
  \bibinfo {author} {\bibfnamefont {Alex}\ \bibnamefont {Thomson}}, \bibinfo
  {author} {\bibfnamefont {Yang}\ \bibnamefont {Peng}}, \bibinfo {author}
  {\bibfnamefont {Youngjoon}\ \bibnamefont {Choi}}, \bibinfo {author}
  {\bibfnamefont {Hyunjin}\ \bibnamefont {Kim}}, \bibinfo {author}
  {\bibfnamefont {Kenji}\ \bibnamefont {Watanabe}}, \bibinfo {author}
  {\bibfnamefont {Takashi}\ \bibnamefont {Taniguchi}}, \bibinfo {author}
  {\bibfnamefont {Jason}\ \bibnamefont {Alicea}}, \bibinfo {author}
  {\bibfnamefont {Felix}\ \bibnamefont {von Oppen}}, \bibinfo {author}
  {\bibfnamefont {Gil}\ \bibnamefont {Refael}}, \ and\ \bibinfo {author}
  {\bibfnamefont {Stevan}\ \bibnamefont {Nadj-Perge}},\ }\bibfield  {title}
  {\enquote {\bibinfo {title} {Promotion of superconductivity in magic-angle
  graphene multilayers},}\ }\href {\doibase 10.1126/science.abn8585} {\bibfield
   {journal} {\bibinfo  {journal} {Science}\ }\textbf {\bibinfo {volume}
  {377}},\ \bibinfo {pages} {1538--1543} (\bibinfo {year} {2022})}\BibitemShut
  {NoStop}%
\bibitem [{\citenamefont {Zhou}\ \emph {et~al.}(2021)\citenamefont {Zhou},
  \citenamefont {Xie}, \citenamefont {Taniguchi}, \citenamefont {Watanabe},\
  and\ \citenamefont {Young}}]{zhou2021superconductivity}%
  \BibitemOpen
  \bibfield  {author} {\bibinfo {author} {\bibfnamefont {Haoxin}\ \bibnamefont
  {Zhou}}, \bibinfo {author} {\bibfnamefont {Tian}\ \bibnamefont {Xie}},
  \bibinfo {author} {\bibfnamefont {Takashi}\ \bibnamefont {Taniguchi}},
  \bibinfo {author} {\bibfnamefont {Kenji}\ \bibnamefont {Watanabe}}, \ and\
  \bibinfo {author} {\bibfnamefont {Andrea~F.}\ \bibnamefont {Young}},\
  }\bibfield  {title} {\enquote {\bibinfo {title} {Superconductivity in
  rhombohedral trilayer graphene},}\ }\href {\doibase
  10.1038/s41586-021-03926-0} {\bibfield  {journal} {\bibinfo  {journal}
  {Nature}\ }\textbf {\bibinfo {volume} {598}},\ \bibinfo {pages} {434--438}
  (\bibinfo {year} {2021})}\BibitemShut {NoStop}%
\bibitem [{\citenamefont {Zhou}\ \emph {et~al.}(2022)\citenamefont {Zhou},
  \citenamefont {Holleis}, \citenamefont {Saito}, \citenamefont {Cohen},
  \citenamefont {Huynh}, \citenamefont {Patterson}, \citenamefont {Yang},
  \citenamefont {Taniguchi}, \citenamefont {Watanabe},\ and\ \citenamefont
  {Young}}]{zhou2022isospin}%
  \BibitemOpen
  \bibfield  {author} {\bibinfo {author} {\bibfnamefont {Haoxin}\ \bibnamefont
  {Zhou}}, \bibinfo {author} {\bibfnamefont {Ludwig}\ \bibnamefont {Holleis}},
  \bibinfo {author} {\bibfnamefont {Yu}~\bibnamefont {Saito}}, \bibinfo
  {author} {\bibfnamefont {Liam}\ \bibnamefont {Cohen}}, \bibinfo {author}
  {\bibfnamefont {William}\ \bibnamefont {Huynh}}, \bibinfo {author}
  {\bibfnamefont {Caitlin~L.}\ \bibnamefont {Patterson}}, \bibinfo {author}
  {\bibfnamefont {Fangyuan}\ \bibnamefont {Yang}}, \bibinfo {author}
  {\bibfnamefont {Takashi}\ \bibnamefont {Taniguchi}}, \bibinfo {author}
  {\bibfnamefont {Kenji}\ \bibnamefont {Watanabe}}, \ and\ \bibinfo {author}
  {\bibfnamefont {Andrea~F.}\ \bibnamefont {Young}},\ }\bibfield  {title}
  {\enquote {\bibinfo {title} {Isospin magnetism and spin-polarized
  superconductivity in bernal bilayer graphene},}\ }\href {\doibase
  10.1126/science.abm8386} {\bibfield  {journal} {\bibinfo  {journal}
  {Science}\ }\textbf {\bibinfo {volume} {375}},\ \bibinfo {pages} {774--778}
  (\bibinfo {year} {2022})}\BibitemShut {NoStop}%
\bibitem [{\citenamefont {Zhang}\ \emph {et~al.}(2023)\citenamefont {Zhang},
  \citenamefont {Polski}, \citenamefont {Thomson}, \citenamefont
  {Lantagne-Hurtubise}, \citenamefont {Lewandowski}, \citenamefont {Zhou},
  \citenamefont {Watanabe}, \citenamefont {Taniguchi}, \citenamefont {Alicea},\
  and\ \citenamefont {Nadj-Perge}}]{zhang2023spin}%
  \BibitemOpen
  \bibfield  {author} {\bibinfo {author} {\bibfnamefont {Yiran}\ \bibnamefont
  {Zhang}}, \bibinfo {author} {\bibfnamefont {Robert}\ \bibnamefont {Polski}},
  \bibinfo {author} {\bibfnamefont {Alex}\ \bibnamefont {Thomson}}, \bibinfo
  {author} {\bibfnamefont {{\'{E}}tienne}\ \bibnamefont {Lantagne-Hurtubise}},
  \bibinfo {author} {\bibfnamefont {Cyprian}\ \bibnamefont {Lewandowski}},
  \bibinfo {author} {\bibfnamefont {Haoxin}\ \bibnamefont {Zhou}}, \bibinfo
  {author} {\bibfnamefont {Kenji}\ \bibnamefont {Watanabe}}, \bibinfo {author}
  {\bibfnamefont {Takashi}\ \bibnamefont {Taniguchi}}, \bibinfo {author}
  {\bibfnamefont {Jason}\ \bibnamefont {Alicea}}, \ and\ \bibinfo {author}
  {\bibfnamefont {Stevan}\ \bibnamefont {Nadj-Perge}},\ }\bibfield  {title}
  {\enquote {\bibinfo {title} {Enhanced superconductivity in
  spin{\textendash}orbit proximitized bilayer graphene},}\ }\href {\doibase
  10.1038/s41586-022-05446-x} {\bibfield  {journal} {\bibinfo  {journal}
  {Nature}\ }\textbf {\bibinfo {volume} {613}},\ \bibinfo {pages} {268--273}
  (\bibinfo {year} {2023})}\BibitemShut {NoStop}%
\bibitem [{\citenamefont {Holleis}\ \emph {et~al.}(2023)\citenamefont
  {Holleis}, \citenamefont {Patterson}, \citenamefont {Zhang}, \citenamefont
  {Yoo}, \citenamefont {Zhou}, \citenamefont {Taniguchi}, \citenamefont
  {Watanabe}, \citenamefont {Nadj-Perge},\ and\ \citenamefont
  {Young}}]{holleis23Ising}%
  \BibitemOpen
  \bibfield  {author} {\bibinfo {author} {\bibfnamefont {Ludwig}\ \bibnamefont
  {Holleis}}, \bibinfo {author} {\bibfnamefont {Caitlin~L.}\ \bibnamefont
  {Patterson}}, \bibinfo {author} {\bibfnamefont {Yiran}\ \bibnamefont
  {Zhang}}, \bibinfo {author} {\bibfnamefont {Heun~Mo}\ \bibnamefont {Yoo}},
  \bibinfo {author} {\bibfnamefont {Haoxin}\ \bibnamefont {Zhou}}, \bibinfo
  {author} {\bibfnamefont {Takashi}\ \bibnamefont {Taniguchi}}, \bibinfo
  {author} {\bibfnamefont {Kenji}\ \bibnamefont {Watanabe}}, \bibinfo {author}
  {\bibfnamefont {Stevan}\ \bibnamefont {Nadj-Perge}}, \ and\ \bibinfo {author}
  {\bibfnamefont {Andrea~F.}\ \bibnamefont {Young}},\ }\bibfield  {title}
  {\enquote {\bibinfo {title} {Ising superconductivity and nematicity in bernal
  bilayer graphene with strong spin orbit coupling},}\ }\href {\doibase
  10.48550/arXiv.2303.00742} {\  (\bibinfo {year} {2023}),\
  10.48550/arXiv.2303.00742}\BibitemShut {NoStop}%
\bibitem [{\citenamefont {Cao}\ \emph {et~al.}(2021{\natexlab{b}})\citenamefont
  {Cao}, \citenamefont {Park}, \citenamefont {Watanabe}, \citenamefont
  {Taniguchi},\ and\ \citenamefont {Jarillo-Herrero}}]{cao2021pauli}%
  \BibitemOpen
  \bibfield  {author} {\bibinfo {author} {\bibfnamefont {Yuan}\ \bibnamefont
  {Cao}}, \bibinfo {author} {\bibfnamefont {Jeong~Min}\ \bibnamefont {Park}},
  \bibinfo {author} {\bibfnamefont {Kenji}\ \bibnamefont {Watanabe}}, \bibinfo
  {author} {\bibfnamefont {Takashi}\ \bibnamefont {Taniguchi}}, \ and\ \bibinfo
  {author} {\bibfnamefont {Pablo}\ \bibnamefont {Jarillo-Herrero}},\ }\bibfield
   {title} {\enquote {\bibinfo {title} {Pauli-limit violation and re-entrant
  superconductivity in moir{\'{e}} graphene},}\ }\href {\doibase
  10.1038/s41586-021-03685-y} {\bibfield  {journal} {\bibinfo  {journal}
  {Nature}\ }\textbf {\bibinfo {volume} {595}},\ \bibinfo {pages} {526--531}
  (\bibinfo {year} {2021}{\natexlab{b}})}\BibitemShut {NoStop}%
\bibitem [{\citenamefont {Lin}\ \emph {et~al.}(2022)\citenamefont {Lin},
  \citenamefont {Siriviboon}, \citenamefont {Scammell}, \citenamefont {Liu},
  \citenamefont {Rhodes}, \citenamefont {Watanabe}, \citenamefont {Taniguchi},
  \citenamefont {Hone}, \citenamefont {Scheurer},\ and\ \citenamefont
  {Li}}]{lin2022zero}%
  \BibitemOpen
  \bibfield  {author} {\bibinfo {author} {\bibfnamefont {Jiang-Xiazi}\
  \bibnamefont {Lin}}, \bibinfo {author} {\bibfnamefont {Phum}\ \bibnamefont
  {Siriviboon}}, \bibinfo {author} {\bibfnamefont {Harley~D.}\ \bibnamefont
  {Scammell}}, \bibinfo {author} {\bibfnamefont {Song}\ \bibnamefont {Liu}},
  \bibinfo {author} {\bibfnamefont {Daniel}\ \bibnamefont {Rhodes}}, \bibinfo
  {author} {\bibfnamefont {K.}~\bibnamefont {Watanabe}}, \bibinfo {author}
  {\bibfnamefont {T.}~\bibnamefont {Taniguchi}}, \bibinfo {author}
  {\bibfnamefont {James}\ \bibnamefont {Hone}}, \bibinfo {author}
  {\bibfnamefont {Mathias~S.}\ \bibnamefont {Scheurer}}, \ and\ \bibinfo
  {author} {\bibfnamefont {J.I.A.}\ \bibnamefont {Li}},\ }\bibfield  {title}
  {\enquote {\bibinfo {title} {Zero-field superconducting diode effect in
  small-twist-angle trilayer graphene},}\ }\href {\doibase
  10.1038/s41567-022-01700-1} {\bibfield  {journal} {\bibinfo  {journal}
  {Nature Physics}\ }\textbf {\bibinfo {volume} {18}},\ \bibinfo {pages}
  {1221--1227} (\bibinfo {year} {2022})}\BibitemShut {NoStop}%
\bibitem [{\citenamefont {Lake}\ \emph {et~al.}(2022)\citenamefont {Lake},
  \citenamefont {Patri},\ and\ \citenamefont {Senthil}}]{lake2022pairing}%
  \BibitemOpen
  \bibfield  {author} {\bibinfo {author} {\bibfnamefont {Ethan}\ \bibnamefont
  {Lake}}, \bibinfo {author} {\bibfnamefont {Adarsh~S.}\ \bibnamefont {Patri}},
  \ and\ \bibinfo {author} {\bibfnamefont {T.}~\bibnamefont {Senthil}},\
  }\bibfield  {title} {\enquote {\bibinfo {title} {Pairing symmetry of twisted
  bilayer graphene: A phenomenological synthesis},}\ }\href {\doibase
  10.1103/PhysRevB.106.104506} {\bibfield  {journal} {\bibinfo  {journal}
  {Phys. Rev. B}\ }\textbf {\bibinfo {volume} {106}},\ \bibinfo {pages}
  {104506} (\bibinfo {year} {2022})}\BibitemShut {NoStop}%
\bibitem [{\citenamefont {de~Vries}\ \emph {et~al.}(2021)\citenamefont
  {de~Vries}, \citenamefont {Portol{\'{e}}s}, \citenamefont {Zheng},
  \citenamefont {Taniguchi}, \citenamefont {Watanabe}, \citenamefont {Ihn},
  \citenamefont {Ensslin},\ and\ \citenamefont {Rickhaus}}]{de2021gate}%
  \BibitemOpen
  \bibfield  {author} {\bibinfo {author} {\bibfnamefont {Folkert~K.}\
  \bibnamefont {de~Vries}}, \bibinfo {author} {\bibfnamefont {El{\'{\i}}as}\
  \bibnamefont {Portol{\'{e}}s}}, \bibinfo {author} {\bibfnamefont {Giulia}\
  \bibnamefont {Zheng}}, \bibinfo {author} {\bibfnamefont {Takashi}\
  \bibnamefont {Taniguchi}}, \bibinfo {author} {\bibfnamefont {Kenji}\
  \bibnamefont {Watanabe}}, \bibinfo {author} {\bibfnamefont {Thomas}\
  \bibnamefont {Ihn}}, \bibinfo {author} {\bibfnamefont {Klaus}\ \bibnamefont
  {Ensslin}}, \ and\ \bibinfo {author} {\bibfnamefont {Peter}\ \bibnamefont
  {Rickhaus}},\ }\bibfield  {title} {\enquote {\bibinfo {title} {Gate-defined
  josephson junctions in magic-angle twisted bilayer graphene},}\ }\href
  {\doibase 10.1038/s41565-021-00896-2} {\bibfield  {journal} {\bibinfo
  {journal} {Nature Nanotechnology}\ }\textbf {\bibinfo {volume} {16}},\
  \bibinfo {pages} {760--763} (\bibinfo {year} {2021})}\BibitemShut {NoStop}%
\bibitem [{\citenamefont {Rodan-Legrain}\ \emph {et~al.}(2021)\citenamefont
  {Rodan-Legrain}, \citenamefont {Cao}, \citenamefont {Park}, \citenamefont
  {de~la Barrera}, \citenamefont {Randeria}, \citenamefont {Watanabe},
  \citenamefont {Taniguchi},\ and\ \citenamefont
  {Jarillo-Herrero}}]{rodan2021highly}%
  \BibitemOpen
  \bibfield  {author} {\bibinfo {author} {\bibfnamefont {Daniel}\ \bibnamefont
  {Rodan-Legrain}}, \bibinfo {author} {\bibfnamefont {Yuan}\ \bibnamefont
  {Cao}}, \bibinfo {author} {\bibfnamefont {Jeong~Min}\ \bibnamefont {Park}},
  \bibinfo {author} {\bibfnamefont {Sergio~C.}\ \bibnamefont {de~la Barrera}},
  \bibinfo {author} {\bibfnamefont {Mallika~T.}\ \bibnamefont {Randeria}},
  \bibinfo {author} {\bibfnamefont {Kenji}\ \bibnamefont {Watanabe}}, \bibinfo
  {author} {\bibfnamefont {Takashi}\ \bibnamefont {Taniguchi}}, \ and\ \bibinfo
  {author} {\bibfnamefont {Pablo}\ \bibnamefont {Jarillo-Herrero}},\ }\bibfield
   {title} {\enquote {\bibinfo {title} {Highly tunable junctions and non-local
  josephson effect in magic-angle graphene tunnelling devices},}\ }\href
  {\doibase 10.1038/s41565-021-00894-4} {\bibfield  {journal} {\bibinfo
  {journal} {Nature Nanotechnology}\ }\textbf {\bibinfo {volume} {16}},\
  \bibinfo {pages} {769--775} (\bibinfo {year} {2021})}\BibitemShut {NoStop}%
\bibitem [{\citenamefont {Diez-Merida}\ \emph {et~al.}(2021)\citenamefont
  {Diez-Merida}, \citenamefont {Diez-Carlon}, \citenamefont {Yang},
  \citenamefont {Xie}, \citenamefont {Gao}, \citenamefont {Watanabe},
  \citenamefont {Taniguchi}, \citenamefont {Lu}, \citenamefont {Law},\ and\
  \citenamefont {Efetov}}]{diez2021magnetic}%
  \BibitemOpen
  \bibfield  {author} {\bibinfo {author} {\bibfnamefont {J.}~\bibnamefont
  {Diez-Merida}}, \bibinfo {author} {\bibfnamefont {A.}~\bibnamefont
  {Diez-Carlon}}, \bibinfo {author} {\bibfnamefont {S.~Y.}\ \bibnamefont
  {Yang}}, \bibinfo {author} {\bibfnamefont {Y.~M.}\ \bibnamefont {Xie}},
  \bibinfo {author} {\bibfnamefont {X.~J.}\ \bibnamefont {Gao}}, \bibinfo
  {author} {\bibfnamefont {K.}~\bibnamefont {Watanabe}}, \bibinfo {author}
  {\bibfnamefont {T.}~\bibnamefont {Taniguchi}}, \bibinfo {author}
  {\bibfnamefont {X.}~\bibnamefont {Lu}}, \bibinfo {author} {\bibfnamefont
  {K.~T.}\ \bibnamefont {Law}}, \ and\ \bibinfo {author} {\bibfnamefont
  {Dmitri~K.}\ \bibnamefont {Efetov}},\ }\bibfield  {title} {\enquote {\bibinfo
  {title} {Magnetic josephson junctions and superconducting diodes in magic
  angle twisted bilayer graphene},}\ }\href {\doibase
  10.48550/arXiv.2110.01067} {\  (\bibinfo {year} {2021}),\
  10.48550/arXiv.2110.01067}\BibitemShut {NoStop}%
\bibitem [{\citenamefont {Xie}\ \emph {et~al.}(2023)\citenamefont {Xie},
  \citenamefont {Efetov},\ and\ \citenamefont {Law}}]{xie2022valley}%
  \BibitemOpen
  \bibfield  {author} {\bibinfo {author} {\bibfnamefont {Ying-Ming}\
  \bibnamefont {Xie}}, \bibinfo {author} {\bibfnamefont {Dmitri~K.}\
  \bibnamefont {Efetov}}, \ and\ \bibinfo {author} {\bibfnamefont {K.~T.}\
  \bibnamefont {Law}},\ }\bibfield  {title} {\enquote {\bibinfo {title}
  {$\varphi_0$-Josephson junction in twisted bilayer graphene induced by a
  valley-polarized state},}\ }\href {\doibase 10.1103/physrevresearch.5.023029}
  {\bibfield  {journal} {\bibinfo  {journal} {Physical Review Research}\
  }\textbf {\bibinfo {volume} {5}} (\bibinfo {year} {2023}),\
  10.1103/physrevresearch.5.023029}\BibitemShut {NoStop}%
\bibitem [{\citenamefont {Hu}\ \emph {et~al.}(2022)\citenamefont {Hu},
  \citenamefont {Sun}, \citenamefont {Xie},\ and\ \citenamefont
  {Law}}]{hu2022valley}%
  \BibitemOpen
  \bibfield  {author} {\bibinfo {author} {\bibfnamefont {Jin-Xin}\ \bibnamefont
  {Hu}}, \bibinfo {author} {\bibfnamefont {Zi-Ting}\ \bibnamefont {Sun}},
  \bibinfo {author} {\bibfnamefont {Ying-Ming}\ \bibnamefont {Xie}}, \ and\
  \bibinfo {author} {\bibfnamefont {K.~T.}\ \bibnamefont {Law}},\ }\bibfield
  {title} {\enquote {\bibinfo {title} {Valley polarization induced josephson
  diode effect in twisted bilayer graphene},}\ }\href {\doibase
  10.48550/arXiv.2211.14846} {\  (\bibinfo {year} {2022}),\
  10.48550/arXiv.2211.14846}\BibitemShut {NoStop}%
\bibitem [{\citenamefont {Portol{\'{e}}s}\ \emph {et~al.}(2022)\citenamefont
  {Portol{\'{e}}s}, \citenamefont {Iwakiri}, \citenamefont {Zheng},
  \citenamefont {Rickhaus}, \citenamefont {Taniguchi}, \citenamefont
  {Watanabe}, \citenamefont {Ihn}, \citenamefont {Ensslin},\ and\ \citenamefont
  {de~Vries}}]{portoles2022tunable}%
  \BibitemOpen
  \bibfield  {author} {\bibinfo {author} {\bibfnamefont {El{\'{\i}}as}\
  \bibnamefont {Portol{\'{e}}s}}, \bibinfo {author} {\bibfnamefont {Shuichi}\
  \bibnamefont {Iwakiri}}, \bibinfo {author} {\bibfnamefont {Giulia}\
  \bibnamefont {Zheng}}, \bibinfo {author} {\bibfnamefont {Peter}\ \bibnamefont
  {Rickhaus}}, \bibinfo {author} {\bibfnamefont {Takashi}\ \bibnamefont
  {Taniguchi}}, \bibinfo {author} {\bibfnamefont {Kenji}\ \bibnamefont
  {Watanabe}}, \bibinfo {author} {\bibfnamefont {Thomas}\ \bibnamefont {Ihn}},
  \bibinfo {author} {\bibfnamefont {Klaus}\ \bibnamefont {Ensslin}}, \ and\
  \bibinfo {author} {\bibfnamefont {Folkert~K.}\ \bibnamefont {de~Vries}},\
  }\bibfield  {title} {\enquote {\bibinfo {title} {A tunable monolithic {SQUID}
  in twisted bilayer graphene},}\ }\href {\doibase 10.1038/s41565-022-01222-0}
  {\bibfield  {journal} {\bibinfo  {journal} {Nature Nanotechnology}\ }\textbf
  {\bibinfo {volume} {17}},\ \bibinfo {pages} {1159--1164} (\bibinfo {year}
  {2022})}\BibitemShut {NoStop}%
\bibitem [{\citenamefont {Sukhachov}\ \emph {et~al.}(2023)\citenamefont
  {Sukhachov}, \citenamefont {von Oppen},\ and\ \citenamefont
  {Glazman}}]{sukhachov2022andreev}%
  \BibitemOpen
  \bibfield  {author} {\bibinfo {author} {\bibfnamefont
  {P.{\hspace{0.167em}}O.}\ \bibnamefont {Sukhachov}}, \bibinfo {author}
  {\bibfnamefont {Felix}\ \bibnamefont {von Oppen}}, \ and\ \bibinfo {author}
  {\bibfnamefont {L.{\hspace{0.167em}}I.}\ \bibnamefont {Glazman}},\ }\bibfield
   {title} {\enquote {\bibinfo {title} {Andreev reflection in scanning
  tunneling spectroscopy of unconventional superconductors},}\ }\href {\doibase
  10.1103/physrevlett.130.216002} {\bibfield  {journal} {\bibinfo  {journal}
  {Physical Review Letters}\ }\textbf {\bibinfo {volume} {130}} (\bibinfo
  {year} {2023}),\ 10.1103/physrevlett.130.216002}\BibitemShut {NoStop}%
\bibitem [{\citenamefont {Josephson}(1962)}]{josephson1962possible}%
  \BibitemOpen
  \bibfield  {author} {\bibinfo {author} {\bibfnamefont {B.D.}\ \bibnamefont
  {Josephson}},\ }\bibfield  {title} {\enquote {\bibinfo {title} {Possible new
  effects in superconductive tunnelling},}\ }\href {\doibase
  10.1016/0031-9163(62)91369-0} {\bibfield  {journal} {\bibinfo  {journal}
  {Physics Letters}\ }\textbf {\bibinfo {volume} {1}},\ \bibinfo {pages}
  {251--253} (\bibinfo {year} {1962})}\BibitemShut {NoStop}%
\bibitem [{\citenamefont {Sigrist}\ and\ \citenamefont
  {Ueda}(1991)}]{sigrist1991phenomenological}%
  \BibitemOpen
  \bibfield  {author} {\bibinfo {author} {\bibfnamefont {Manfred}\ \bibnamefont
  {Sigrist}}\ and\ \bibinfo {author} {\bibfnamefont {Kazuo}\ \bibnamefont
  {Ueda}},\ }\bibfield  {title} {\enquote {\bibinfo {title} {Phenomenological
  theory of unconventional superconductivity},}\ }\href {\doibase
  10.1103/revmodphys.63.239} {\bibfield  {journal} {\bibinfo  {journal}
  {Reviews of Modern Physics}\ }\textbf {\bibinfo {volume} {63}},\ \bibinfo
  {pages} {239--311} (\bibinfo {year} {1991})}\BibitemShut {NoStop}%
\bibitem [{\citenamefont {Golubov}\ \emph {et~al.}(2004)\citenamefont
  {Golubov}, \citenamefont {Kupriyanov},\ and\ \citenamefont
  {Il'ichev}}]{golubov2004current}%
  \BibitemOpen
  \bibfield  {author} {\bibinfo {author} {\bibfnamefont {A.~A.}\ \bibnamefont
  {Golubov}}, \bibinfo {author} {\bibfnamefont {M.~Yu.}\ \bibnamefont
  {Kupriyanov}}, \ and\ \bibinfo {author} {\bibfnamefont {E.}~\bibnamefont
  {Il'ichev}},\ }\bibfield  {title} {\enquote {\bibinfo {title} {The
  current-phase relation in josephson junctions},}\ }\href {\doibase
  10.1103/revmodphys.76.411} {\bibfield  {journal} {\bibinfo  {journal}
  {Reviews of Modern Physics}\ }\textbf {\bibinfo {volume} {76}},\ \bibinfo
  {pages} {411--469} (\bibinfo {year} {2004})}\BibitemShut {NoStop}%
\bibitem [{\citenamefont {Tsuei}\ \emph {et~al.}(1994)\citenamefont {Tsuei},
  \citenamefont {Kirtley}, \citenamefont {Chi}, \citenamefont {Yu-Jahnes},
  \citenamefont {Gupta}, \citenamefont {Shaw}, \citenamefont {Sun},\ and\
  \citenamefont {Ketchen}}]{Tetal94}%
  \BibitemOpen
  \bibfield  {author} {\bibinfo {author} {\bibfnamefont {C.~C.}\ \bibnamefont
  {Tsuei}}, \bibinfo {author} {\bibfnamefont {J.~R.}\ \bibnamefont {Kirtley}},
  \bibinfo {author} {\bibfnamefont {C.~C.}\ \bibnamefont {Chi}}, \bibinfo
  {author} {\bibfnamefont {Lock~See}\ \bibnamefont {Yu-Jahnes}}, \bibinfo
  {author} {\bibfnamefont {A.}~\bibnamefont {Gupta}}, \bibinfo {author}
  {\bibfnamefont {T.}~\bibnamefont {Shaw}}, \bibinfo {author} {\bibfnamefont
  {J.~Z.}\ \bibnamefont {Sun}}, \ and\ \bibinfo {author} {\bibfnamefont
  {M.~B.}\ \bibnamefont {Ketchen}},\ }\bibfield  {title} {\enquote {\bibinfo
  {title} {Pairing symmetry and flux quantization in a tricrystal
  superconducting ring of
  $\mathrm{Y}{\mathrm{ba}}_{2}{\mathrm{cu}}_{3}{\mathrm{o}}_{7\ensuremath{-}\ensuremath{\delta}}$},}\
  }\href {\doibase 10.1103/PhysRevLett.73.593} {\bibfield  {journal} {\bibinfo
  {journal} {Phys. Rev. Lett.}\ }\textbf {\bibinfo {volume} {73}},\ \bibinfo
  {pages} {593--596} (\bibinfo {year} {1994})}\BibitemShut {NoStop}%
\bibitem [{\citenamefont {Tsuei}\ and\ \citenamefont {Kirtley}(2000)}]{TK00}%
  \BibitemOpen
  \bibfield  {author} {\bibinfo {author} {\bibfnamefont {C.~C.}\ \bibnamefont
  {Tsuei}}\ and\ \bibinfo {author} {\bibfnamefont {J.~R.}\ \bibnamefont
  {Kirtley}},\ }\bibfield  {title} {\enquote {\bibinfo {title} {Pairing
  symmetry in cuprate superconductors},}\ }\href {\doibase
  10.1103/RevModPhys.72.969} {\bibfield  {journal} {\bibinfo  {journal} {Rev.
  Mod. Phys.}\ }\textbf {\bibinfo {volume} {72}},\ \bibinfo {pages} {969--1016}
  (\bibinfo {year} {2000})}\BibitemShut {NoStop}%
\bibitem [{\citenamefont {Cr\'epel}\ \emph {et~al.}(2022)\citenamefont
  {Cr\'epel}, \citenamefont {Cea}, \citenamefont {Fu},\ and\ \citenamefont
  {Guinea}}]{crepel2022unconventional}%
  \BibitemOpen
  \bibfield  {author} {\bibinfo {author} {\bibfnamefont {Valentin}\
  \bibnamefont {Cr\'epel}}, \bibinfo {author} {\bibfnamefont {Tommaso}\
  \bibnamefont {Cea}}, \bibinfo {author} {\bibfnamefont {Liang}\ \bibnamefont
  {Fu}}, \ and\ \bibinfo {author} {\bibfnamefont {Francisco}\ \bibnamefont
  {Guinea}},\ }\bibfield  {title} {\enquote {\bibinfo {title} {Unconventional
  superconductivity due to interband polarization},}\ }\href {\doibase
  10.1103/PhysRevB.105.094506} {\bibfield  {journal} {\bibinfo  {journal}
  {Phys. Rev. B}\ }\textbf {\bibinfo {volume} {105}},\ \bibinfo {pages}
  {094506} (\bibinfo {year} {2022})}\BibitemShut {NoStop}%
\bibitem [{\citenamefont {Cea}\ and\ \citenamefont
  {Guinea}(2021)}]{cea21Coulomb}%
  \BibitemOpen
  \bibfield  {author} {\bibinfo {author} {\bibfnamefont {Tommaso}\ \bibnamefont
  {Cea}}\ and\ \bibinfo {author} {\bibfnamefont {Francisco}\ \bibnamefont
  {Guinea}},\ }\bibfield  {title} {\enquote {\bibinfo {title} {Coulomb
  interaction, phonons, and superconductivity in twisted bilayer graphene},}\
  }\href {\doibase 10.1073/pnas.2107874118} {\bibfield  {journal} {\bibinfo
  {journal} {Proceedings of the National Academy of Sciences}\ }\textbf
  {\bibinfo {volume} {118}} (\bibinfo {year} {2021}),\
  10.1073/pnas.2107874118}\BibitemShut {NoStop}%
\bibitem [{\citenamefont {Blonder}\ \emph {et~al.}(1982)\citenamefont
  {Blonder}, \citenamefont {Tinkham},\ and\ \citenamefont
  {Klapwijk}}]{blonder1982transition}%
  \BibitemOpen
  \bibfield  {author} {\bibinfo {author} {\bibfnamefont {G.~E.}\ \bibnamefont
  {Blonder}}, \bibinfo {author} {\bibfnamefont {M.}~\bibnamefont {Tinkham}}, \
  and\ \bibinfo {author} {\bibfnamefont {T.~M.}\ \bibnamefont {Klapwijk}},\
  }\bibfield  {title} {\enquote {\bibinfo {title} {Transition from metallic to
  tunneling regimes in superconducting microconstrictions: Excess current,
  charge imbalance, and supercurrent conversion},}\ }\href {\doibase
  10.1103/physrevb.25.4515} {\bibfield  {journal} {\bibinfo  {journal}
  {Physical Review B}\ }\textbf {\bibinfo {volume} {25}},\ \bibinfo {pages}
  {4515--4532} (\bibinfo {year} {1982})}\BibitemShut {NoStop}%
\bibitem [{\citenamefont {Setiawan}\ and\ \citenamefont
  {Hofmann}(2022)}]{setiawan2022analytic}%
  \BibitemOpen
  \bibfield  {author} {\bibinfo {author} {\bibfnamefont {F.}~\bibnamefont
  {Setiawan}}\ and\ \bibinfo {author} {\bibfnamefont {Johannes}\ \bibnamefont
  {Hofmann}},\ }\bibfield  {title} {\enquote {\bibinfo {title} {Analytic
  approach to transport in superconducting junctions with arbitrary carrier
  density},}\ }\href {\doibase 10.1103/physrevresearch.4.043087} {\bibfield
  {journal} {\bibinfo  {journal} {Physical Review Research}\ }\textbf {\bibinfo
  {volume} {4}} (\bibinfo {year} {2022}),\
  10.1103/physrevresearch.4.043087}\BibitemShut {NoStop}%
\bibitem [{\citenamefont {Lewandowski}\ \emph {et~al.}(2023)\citenamefont
  {Lewandowski}, \citenamefont {Lantagne-Hurtubise}, \citenamefont {Thomson},
  \citenamefont {Nadj-Perge},\ and\ \citenamefont
  {Alicea}}]{lewandowski2023andreev}%
  \BibitemOpen
  \bibfield  {author} {\bibinfo {author} {\bibfnamefont {Cyprian}\ \bibnamefont
  {Lewandowski}}, \bibinfo {author} {\bibfnamefont {{\'{E}}tienne}\
  \bibnamefont {Lantagne-Hurtubise}}, \bibinfo {author} {\bibfnamefont {Alex}\
  \bibnamefont {Thomson}}, \bibinfo {author} {\bibfnamefont {Stevan}\
  \bibnamefont {Nadj-Perge}}, \ and\ \bibinfo {author} {\bibfnamefont {Jason}\
  \bibnamefont {Alicea}},\ }\bibfield  {title} {\enquote {\bibinfo {title}
  {Andreev reflection spectroscopy in strongly paired superconductors},}\
  }\href {\doibase 10.1103/physrevb.107.l020502} {\bibfield  {journal}
  {\bibinfo  {journal} {Physical Review B}\ }\textbf {\bibinfo {volume} {107}}
  (\bibinfo {year} {2023}),\ 10.1103/physrevb.107.l020502}\BibitemShut
  {NoStop}%
\bibitem [{aMi()}]{aMiscspin}%
  \BibitemOpen
  \href@noop {} {}\bibinfo {note} {In the case of the $f$-wave superconductor,
  there is no restriction on the value of a given component of the spin, which
  can be $s_z = 0, \pm 1$.}\BibitemShut {Stop}%
\bibitem [{SM()}]{SM}%
  \BibitemOpen
  \href@noop {} {}\bibinfo {note} {See supplementary information, which
  includes Refs.\
  \cite{moon2014optical,bistritzer2011moire,perfetto2009equilibrium,kitaev2001unpaired,ishii1970josephson}}\BibitemShut
  {NoStop}%
\bibitem [{\citenamefont {Anderson}(1959)}]{anderson1959theory}%
  \BibitemOpen
  \bibfield  {author} {\bibinfo {author} {\bibfnamefont {P.W.}\ \bibnamefont
  {Anderson}},\ }\bibfield  {title} {\enquote {\bibinfo {title} {Theory of
  dirty superconductors},}\ }\href {\doibase 10.1016/0022-3697(59)90036-8}
  {\bibfield  {journal} {\bibinfo  {journal} {Journal of Physics and Chemistry
  of Solids}\ }\textbf {\bibinfo {volume} {11}},\ \bibinfo {pages} {26--30}
  (\bibinfo {year} {1959})}\BibitemShut {NoStop}%
\bibitem [{\citenamefont {Lin}\ and\ \citenamefont
  {Tom\'anek}(2018)}]{lin2018minimum}%
  \BibitemOpen
  \bibfield  {author} {\bibinfo {author} {\bibfnamefont {Xianqing}\
  \bibnamefont {Lin}}\ and\ \bibinfo {author} {\bibfnamefont {David}\
  \bibnamefont {Tom\'anek}},\ }\bibfield  {title} {\enquote {\bibinfo {title}
  {Minimum model for the electronic structure of twisted bilayer graphene and
  related structures},}\ }\href {\doibase 10.1103/PhysRevB.98.081410}
  {\bibfield  {journal} {\bibinfo  {journal} {Phys. Rev. B}\ }\textbf {\bibinfo
  {volume} {98}},\ \bibinfo {pages} {081410} (\bibinfo {year}
  {2018})}\BibitemShut {NoStop}%
\bibitem [{\citenamefont {Guinea}\ and\ \citenamefont
  {Walet}(2018)}]{GuineaNiels2018}%
  \BibitemOpen
  \bibfield  {author} {\bibinfo {author} {\bibfnamefont {Francisco}\
  \bibnamefont {Guinea}}\ and\ \bibinfo {author} {\bibfnamefont {Niels~R.}\
  \bibnamefont {Walet}},\ }\bibfield  {title} {\enquote {\bibinfo {title}
  {Electrostatic effects, band distortions, and superconductivity in twisted
  graphene bilayers},}\ }\href {\doibase 10.1073/pnas.1810947115} {\bibfield
  {journal} {\bibinfo  {journal} {Proceedings of the National Academy of
  Sciences}\ }\textbf {\bibinfo {volume} {115}},\ \bibinfo {pages}
  {13174--13179} (\bibinfo {year} {2018})}\BibitemShut {NoStop}%
\bibitem [{\citenamefont {Rademaker}\ \emph {et~al.}(2019)\citenamefont
  {Rademaker}, \citenamefont {Abanin},\ and\ \citenamefont
  {Mellado}}]{RademakerHartree2019}%
  \BibitemOpen
  \bibfield  {author} {\bibinfo {author} {\bibfnamefont {Louk}\ \bibnamefont
  {Rademaker}}, \bibinfo {author} {\bibfnamefont {Dmitry~A.}\ \bibnamefont
  {Abanin}}, \ and\ \bibinfo {author} {\bibfnamefont {Paula}\ \bibnamefont
  {Mellado}},\ }\bibfield  {title} {\enquote {\bibinfo {title} {Charge
  smoothening and band flattening due to hartree corrections in twisted bilayer
  graphene},}\ }\href {\doibase 10.1103/PhysRevB.100.205114} {\bibfield
  {journal} {\bibinfo  {journal} {Phys. Rev. B}\ }\textbf {\bibinfo {volume}
  {100}},\ \bibinfo {pages} {205114} (\bibinfo {year} {2019})}\BibitemShut
  {NoStop}%
\bibitem [{\citenamefont {Gonzalez-Arraga}\ \emph {et~al.}(2017)\citenamefont
  {Gonzalez-Arraga}, \citenamefont {Lado}, \citenamefont {Guinea},\ and\
  \citenamefont {San-Jose}}]{gonzalez2017}%
  \BibitemOpen
  \bibfield  {author} {\bibinfo {author} {\bibfnamefont {Luis~A.}\ \bibnamefont
  {Gonzalez-Arraga}}, \bibinfo {author} {\bibfnamefont {J.~L.}\ \bibnamefont
  {Lado}}, \bibinfo {author} {\bibfnamefont {Francisco}\ \bibnamefont
  {Guinea}}, \ and\ \bibinfo {author} {\bibfnamefont {Pablo}\ \bibnamefont
  {San-Jose}},\ }\bibfield  {title} {\enquote {\bibinfo {title} {Electrically
  controllable magnetism in twisted bilayer graphene},}\ }\href {\doibase
  10.1103/PhysRevLett.119.107201} {\bibfield  {journal} {\bibinfo  {journal}
  {Phys. Rev. Lett.}\ }\textbf {\bibinfo {volume} {119}},\ \bibinfo {pages}
  {107201} (\bibinfo {year} {2017})}\BibitemShut {NoStop}%
\bibitem [{\citenamefont {Vahedi}\ \emph {et~al.}(2021)\citenamefont {Vahedi},
  \citenamefont {Peters}, \citenamefont {Missaoui}, \citenamefont {Honecker},\
  and\ \citenamefont {de~Laissardi{\`{e}}re}}]{vahedi2021magnetism}%
  \BibitemOpen
  \bibfield  {author} {\bibinfo {author} {\bibfnamefont {Javad}\ \bibnamefont
  {Vahedi}}, \bibinfo {author} {\bibfnamefont {Robert}\ \bibnamefont {Peters}},
  \bibinfo {author} {\bibfnamefont {Ahmed}\ \bibnamefont {Missaoui}}, \bibinfo
  {author} {\bibfnamefont {Andreas}\ \bibnamefont {Honecker}}, \ and\ \bibinfo
  {author} {\bibfnamefont {Guy~Trambly}\ \bibnamefont
  {de~Laissardi{\`{e}}re}},\ }\bibfield  {title} {\enquote {\bibinfo {title}
  {Magnetism of magic-angle twisted bilayer graphene},}\ }\href {\doibase
  10.21468/scipostphys.11.4.083} {\bibfield  {journal} {\bibinfo  {journal}
  {{SciPost} Physics}\ }\textbf {\bibinfo {volume} {11}} (\bibinfo {year}
  {2021}),\ 10.21468/scipostphys.11.4.083}\BibitemShut {NoStop}%
\bibitem [{\citenamefont {Sainz-Cruz}\ \emph {et~al.}(2021)\citenamefont
  {Sainz-Cruz}, \citenamefont {Cea}, \citenamefont {Pantale\'on},\ and\
  \citenamefont {Guinea}}]{sainzcruz21high}%
  \BibitemOpen
  \bibfield  {author} {\bibinfo {author} {\bibfnamefont {H\'ector}\
  \bibnamefont {Sainz-Cruz}}, \bibinfo {author} {\bibfnamefont {Tommaso}\
  \bibnamefont {Cea}}, \bibinfo {author} {\bibfnamefont {Pierre~A.}\
  \bibnamefont {Pantale\'on}}, \ and\ \bibinfo {author} {\bibfnamefont
  {Francisco}\ \bibnamefont {Guinea}},\ }\bibfield  {title} {\enquote {\bibinfo
  {title} {High transmission in twisted bilayer graphene with angle
  disorder},}\ }\href {\doibase 10.1103/PhysRevB.104.075144} {\bibfield
  {journal} {\bibinfo  {journal} {Phys. Rev. B}\ }\textbf {\bibinfo {volume}
  {104}},\ \bibinfo {pages} {075144} (\bibinfo {year} {2021})}\BibitemShut
  {NoStop}%
\bibitem [{\citenamefont {Haldane}(1988)}]{haldane1988model}%
  \BibitemOpen
  \bibfield  {author} {\bibinfo {author} {\bibfnamefont {F.~D.~M.}\
  \bibnamefont {Haldane}},\ }\bibfield  {title} {\enquote {\bibinfo {title}
  {Model for a quantum hall effect without landau levels: Condensed-matter
  realization of the "parity anomaly"},}\ }\href {\doibase
  10.1103/physrevlett.61.2015} {\bibfield  {journal} {\bibinfo  {journal}
  {Physical Review Letters}\ }\textbf {\bibinfo {volume} {61}},\ \bibinfo
  {pages} {2015--2018} (\bibinfo {year} {1988})}\BibitemShut {NoStop}%
\bibitem [{Pai()}]{PairingAmplitudes}%
  \BibitemOpen
  \href@noop {} {}\bibinfo {note} {To induce a superconducting gap of magnitude
  $\Delta$, the necessary amplitudes are $\Delta_S=\Delta/2$ for $s$-wave,
  $\Delta_K=\Delta/4$ for Kitaev and $\Delta_F\approx\Delta/(6\sqrt3)$ for
  $f$-wave pairing.}\BibitemShut {Stop}%
\bibitem [{Mis()}]{Misc1}%
  \BibitemOpen
  \href@noop {} {}\bibinfo {note} {We use a commensurate lattice, but we
  caution that incommensurability has an impact on transport
  \cite{gonccalves2021incommensurability}}\BibitemShut {NoStop}%
\bibitem [{\citenamefont {Gon{\c{c}}alves}\ \emph {et~al.}(2021)\citenamefont
  {Gon{\c{c}}alves}, \citenamefont {Olyaei}, \citenamefont {Amorim},
  \citenamefont {Mondaini}, \citenamefont {Ribeiro},\ and\ \citenamefont
  {Castro}}]{gonccalves2021incommensurability}%
  \BibitemOpen
  \bibfield  {author} {\bibinfo {author} {\bibfnamefont {Miguel}\ \bibnamefont
  {Gon{\c{c}}alves}}, \bibinfo {author} {\bibfnamefont {Hadi~Z}\ \bibnamefont
  {Olyaei}}, \bibinfo {author} {\bibfnamefont {Bruno}\ \bibnamefont {Amorim}},
  \bibinfo {author} {\bibfnamefont {Rubem}\ \bibnamefont {Mondaini}}, \bibinfo
  {author} {\bibfnamefont {Pedro}\ \bibnamefont {Ribeiro}}, \ and\ \bibinfo
  {author} {\bibfnamefont {Eduardo~V}\ \bibnamefont {Castro}},\ }\bibfield
  {title} {\enquote {\bibinfo {title} {Incommensurability-induced sub-ballistic
  narrow-band-states in twisted bilayer graphene},}\ }\href {\doibase
  10.1088/2053-1583/ac3259} {\bibfield  {journal} {\bibinfo  {journal} {2D
  Materials}\ }\textbf {\bibinfo {volume} {9}},\ \bibinfo {pages} {011001}
  (\bibinfo {year} {2021})}\BibitemShut {NoStop}%
\bibitem [{bMi()}]{bMiscSIS}%
  \BibitemOpen
  \href@noop {} {}\bibinfo {note} {In SIS junctions with the leads at $n=+2.4$
  ($-2.4$), the chemical potential of the link is placed in the middle of the
  gap between the flat bands and the hole-like (electron-like) remote
  bands.}\BibitemShut {Stop}%
\bibitem [{\citenamefont {Ambegaokar}\ and\ \citenamefont
  {Baratoff}(1963)}]{ambegaokar1963tunneling}%
  \BibitemOpen
  \bibfield  {author} {\bibinfo {author} {\bibfnamefont {Vinay}\ \bibnamefont
  {Ambegaokar}}\ and\ \bibinfo {author} {\bibfnamefont {Alexis}\ \bibnamefont
  {Baratoff}},\ }\bibfield  {title} {\enquote {\bibinfo {title} {Tunneling
  between superconductors},}\ }\href {\doibase 10.1103/physrevlett.10.486}
  {\bibfield  {journal} {\bibinfo  {journal} {Physical Review Letters}\
  }\textbf {\bibinfo {volume} {10}},\ \bibinfo {pages} {486--489} (\bibinfo
  {year} {1963})}\BibitemShut {NoStop}%
\bibitem [{\citenamefont {Alvarado}\ and\ \citenamefont
  {Yeyati}(2021)}]{alvarado2021transport}%
  \BibitemOpen
  \bibfield  {author} {\bibinfo {author} {\bibfnamefont {M.}~\bibnamefont
  {Alvarado}}\ and\ \bibinfo {author} {\bibfnamefont {A.~Levy}\ \bibnamefont
  {Yeyati}},\ }\bibfield  {title} {\enquote {\bibinfo {title} {Transport and
  spectral properties of magic-angle twisted bilayer graphene junctions based
  on local orbital models},}\ }\href {\doibase 10.1103/PhysRevB.104.075406}
  {\bibfield  {journal} {\bibinfo  {journal} {Phys. Rev. B}\ }\textbf {\bibinfo
  {volume} {104}},\ \bibinfo {pages} {075406} (\bibinfo {year}
  {2021})}\BibitemShut {NoStop}%
\bibitem [{\citenamefont {Zagoskin}(1997)}]{zagoskin1997half}%
  \BibitemOpen
  \bibfield  {author} {\bibinfo {author} {\bibfnamefont {Alexandre~M}\
  \bibnamefont {Zagoskin}},\ }\bibfield  {title} {\enquote {\bibinfo {title}
  {The half-periodic josephson effect in an s-wave superconductor -
  normal-metal - d-wave superconductor junction},}\ }\href {\doibase
  10.1088/0953-8984/9/31/001} {\bibfield  {journal} {\bibinfo  {journal}
  {Journal of Physics: Condensed Matter}\ }\textbf {\bibinfo {volume} {9}},\
  \bibinfo {pages} {L419--L426} (\bibinfo {year} {1997})}\BibitemShut {NoStop}%
\bibitem [{\citenamefont {Bulaevskii}\ \emph {et~al.}(1977)\citenamefont
  {Bulaevskii}, \citenamefont {Kuzii},\ and\ \citenamefont
  {Sobyanin}}]{bulaevskii77superconducting}%
  \BibitemOpen
  \bibfield  {author} {\bibinfo {author} {\bibfnamefont {L~N}\ \bibnamefont
  {Bulaevskii}}, \bibinfo {author} {\bibfnamefont {V~V}\ \bibnamefont {Kuzii}},
  \ and\ \bibinfo {author} {\bibfnamefont {A~A}\ \bibnamefont {Sobyanin}},\
  }\bibfield  {title} {\enquote {\bibinfo {title} {Superconducting system with
  weak coupling to the current in the ground state},}\ }\href
  {https://www.osti.gov/biblio/7316063} {\bibfield  {journal} {\bibinfo
  {journal} {JETP Lett.}\ }\textbf {\bibinfo {volume} {25}} (\bibinfo {year}
  {1977})}\BibitemShut {NoStop}%
\bibitem [{\citenamefont {Zazunov}\ and\ \citenamefont
  {Egger}(2012)}]{zazunov2012supercurrent}%
  \BibitemOpen
  \bibfield  {author} {\bibinfo {author} {\bibfnamefont {Alex}\ \bibnamefont
  {Zazunov}}\ and\ \bibinfo {author} {\bibfnamefont {Reinhold}\ \bibnamefont
  {Egger}},\ }\bibfield  {title} {\enquote {\bibinfo {title} {Supercurrent
  blockade in josephson junctions with a majorana wire},}\ }\href {\doibase
  10.1103/PhysRevB.85.104514} {\bibfield  {journal} {\bibinfo  {journal} {Phys.
  Rev. B}\ }\textbf {\bibinfo {volume} {85}},\ \bibinfo {pages} {104514}
  (\bibinfo {year} {2012})}\BibitemShut {NoStop}%
\bibitem [{\citenamefont {Scheurer}\ and\ \citenamefont
  {Samajdar}(2020)}]{SS20}%
  \BibitemOpen
  \bibfield  {author} {\bibinfo {author} {\bibfnamefont {Mathias~S.}\
  \bibnamefont {Scheurer}}\ and\ \bibinfo {author} {\bibfnamefont {Rhine}\
  \bibnamefont {Samajdar}},\ }\bibfield  {title} {\enquote {\bibinfo {title}
  {Pairing in graphene-based moir\'e superlattices},}\ }\href {\doibase
  10.1103/PhysRevResearch.2.033062} {\bibfield  {journal} {\bibinfo  {journal}
  {Phys. Rev. Research}\ }\textbf {\bibinfo {volume} {2}},\ \bibinfo {pages}
  {033062} (\bibinfo {year} {2020})}\BibitemShut {NoStop}%
\bibitem [{\citenamefont {Linder}\ \emph {et~al.}(2009)\citenamefont {Linder},
  \citenamefont {Black-Schaffer}, \citenamefont {Yokoyama}, \citenamefont
  {Doniach},\ and\ \citenamefont {Sudb\o{}}}]{linder2009josephson}%
  \BibitemOpen
  \bibfield  {author} {\bibinfo {author} {\bibfnamefont {Jacob}\ \bibnamefont
  {Linder}}, \bibinfo {author} {\bibfnamefont {Annica~M.}\ \bibnamefont
  {Black-Schaffer}}, \bibinfo {author} {\bibfnamefont {Takehito}\ \bibnamefont
  {Yokoyama}}, \bibinfo {author} {\bibfnamefont {Sebastian}\ \bibnamefont
  {Doniach}}, \ and\ \bibinfo {author} {\bibfnamefont {Asle}\ \bibnamefont
  {Sudb\o{}}},\ }\bibfield  {title} {\enquote {\bibinfo {title} {Josephson
  current in graphene: Role of unconventional pairing symmetries},}\ }\href
  {\doibase 10.1103/PhysRevB.80.094522} {\bibfield  {journal} {\bibinfo
  {journal} {Phys. Rev. B}\ }\textbf {\bibinfo {volume} {80}},\ \bibinfo
  {pages} {094522} (\bibinfo {year} {2009})}\BibitemShut {NoStop}%
\bibitem [{\citenamefont {Heersche}\ \emph {et~al.}(2007)\citenamefont
  {Heersche}, \citenamefont {Jarillo-Herrero}, \citenamefont {Oostinga},
  \citenamefont {Vandersypen},\ and\ \citenamefont {Morpurgo}}]{Heersche_2007}%
  \BibitemOpen
  \bibfield  {author} {\bibinfo {author} {\bibfnamefont {Hubert~B.}\
  \bibnamefont {Heersche}}, \bibinfo {author} {\bibfnamefont {Pablo}\
  \bibnamefont {Jarillo-Herrero}}, \bibinfo {author} {\bibfnamefont
  {Jeroen~B.}\ \bibnamefont {Oostinga}}, \bibinfo {author} {\bibfnamefont
  {Lieven M.~K.}\ \bibnamefont {Vandersypen}}, \ and\ \bibinfo {author}
  {\bibfnamefont {Alberto~F.}\ \bibnamefont {Morpurgo}},\ }\bibfield  {title}
  {\enquote {\bibinfo {title} {Bipolar supercurrent in graphene},}\ }\href
  {\doibase 10.1038/nature05555} {\bibfield  {journal} {\bibinfo  {journal}
  {Nature}\ }\textbf {\bibinfo {volume} {446}},\ \bibinfo {pages} {56--59}
  (\bibinfo {year} {2007})}\BibitemShut {NoStop}%
\bibitem [{\citenamefont {Rui}\ \emph {et~al.}(2020)\citenamefont {Rui},
  \citenamefont {Sun}, \citenamefont {Kang}, \citenamefont {Peng},
  \citenamefont {Liu},\ and\ \citenamefont {Xu}}]{rui2020superconductivity}%
  \BibitemOpen
  \bibfield  {author} {\bibinfo {author} {\bibfnamefont {Dingran}\ \bibnamefont
  {Rui}}, \bibinfo {author} {\bibfnamefont {Luzhao}\ \bibnamefont {Sun}},
  \bibinfo {author} {\bibfnamefont {N.}~\bibnamefont {Kang}}, \bibinfo {author}
  {\bibfnamefont {Hailin}\ \bibnamefont {Peng}}, \bibinfo {author}
  {\bibfnamefont {Zhongfan}\ \bibnamefont {Liu}}, \ and\ \bibinfo {author}
  {\bibfnamefont {H.~Q.}\ \bibnamefont {Xu}},\ }\bibfield  {title} {\enquote
  {\bibinfo {title} {Superconductivity in an al-twisted bilayer graphene-al
  junction device},}\ }\href {\doibase 10.7567/1347-4065/ab641d} {\bibfield
  {journal} {\bibinfo  {journal} {Japanese Journal of Applied Physics}\
  }\textbf {\bibinfo {volume} {59}},\ \bibinfo {pages} {SGGI07} (\bibinfo
  {year} {2020})}\BibitemShut {NoStop}%
\bibitem [{\citenamefont {Poduval}\ and\ \citenamefont
  {Scheurer}(2023)}]{poduval2023vestigial}%
  \BibitemOpen
  \bibfield  {author} {\bibinfo {author} {\bibfnamefont {Prathyush~P.}\
  \bibnamefont {Poduval}}\ and\ \bibinfo {author} {\bibfnamefont {Mathias~S.}\
  \bibnamefont {Scheurer}},\ }\bibfield  {title} {\enquote {\bibinfo {title}
  {Vestigial singlet pairing in a fluctuating magnetic triplet superconductor:
  Applications to graphene moiré systems},}\ }\href {\doibase
  10.48550/arXiv.2301.01344} {\  (\bibinfo {year} {2023}),\
  10.48550/arXiv.2301.01344}\BibitemShut {NoStop}%
\bibitem [{\citenamefont {Moon}\ \emph {et~al.}(2014)\citenamefont {Moon},
  \citenamefont {Son},\ and\ \citenamefont {Koshino}}]{moon2014optical}%
  \BibitemOpen
  \bibfield  {author} {\bibinfo {author} {\bibfnamefont {Pilkyung}\
  \bibnamefont {Moon}}, \bibinfo {author} {\bibfnamefont {Young-Woo}\
  \bibnamefont {Son}}, \ and\ \bibinfo {author} {\bibfnamefont {Mikito}\
  \bibnamefont {Koshino}},\ }\bibfield  {title} {\enquote {\bibinfo {title}
  {Optical absorption of twisted bilayer graphene with interlayer potential
  asymmetry},}\ }\href {\doibase 10.1103/PhysRevB.90.155427} {\bibfield
  {journal} {\bibinfo  {journal} {Phys. Rev. B}\ }\textbf {\bibinfo {volume}
  {90}},\ \bibinfo {pages} {155427} (\bibinfo {year} {2014})}\BibitemShut
  {NoStop}%
\bibitem [{\citenamefont {Bistritzer}\ and\ \citenamefont
  {MacDonald}(2011)}]{bistritzer2011moire}%
  \BibitemOpen
  \bibfield  {author} {\bibinfo {author} {\bibfnamefont {Rafi}\ \bibnamefont
  {Bistritzer}}\ and\ \bibinfo {author} {\bibfnamefont {Allan~H.}\ \bibnamefont
  {MacDonald}},\ }\bibfield  {title} {\enquote {\bibinfo {title} {Moir{\'{e}}
  bands in twisted double-layer graphene},}\ }\href {\doibase
  10.1073/pnas.1108174108} {\bibfield  {journal} {\bibinfo  {journal}
  {Proceedings of the National Academy of Sciences}\ }\textbf {\bibinfo
  {volume} {108}},\ \bibinfo {pages} {12233--12237} (\bibinfo {year}
  {2011})}\BibitemShut {NoStop}%
\bibitem [{\citenamefont {Perfetto}\ \emph {et~al.}(2009)\citenamefont
  {Perfetto}, \citenamefont {Stefanucci},\ and\ \citenamefont
  {Cini}}]{perfetto2009equilibrium}%
  \BibitemOpen
  \bibfield  {author} {\bibinfo {author} {\bibfnamefont {Enrico}\ \bibnamefont
  {Perfetto}}, \bibinfo {author} {\bibfnamefont {Gianluca}\ \bibnamefont
  {Stefanucci}}, \ and\ \bibinfo {author} {\bibfnamefont {Michele}\
  \bibnamefont {Cini}},\ }\bibfield  {title} {\enquote {\bibinfo {title}
  {Equilibrium and time-dependent josephson current in one-dimensional
  superconducting junctions},}\ }\href {\doibase 10.1103/PhysRevB.80.205408}
  {\bibfield  {journal} {\bibinfo  {journal} {Phys. Rev. B}\ }\textbf {\bibinfo
  {volume} {80}},\ \bibinfo {pages} {205408} (\bibinfo {year}
  {2009})}\BibitemShut {NoStop}%
\bibitem [{\citenamefont {Kitaev}(2001)}]{kitaev2001unpaired}%
  \BibitemOpen
  \bibfield  {author} {\bibinfo {author} {\bibfnamefont {A~Yu}\ \bibnamefont
  {Kitaev}},\ }\bibfield  {title} {\enquote {\bibinfo {title} {Unpaired
  majorana fermions in quantum wires},}\ }\href {\doibase
  10.1070/1063-7869/44/10s/s29} {\bibfield  {journal} {\bibinfo  {journal}
  {Physics-Uspekhi}\ }\textbf {\bibinfo {volume} {44}},\ \bibinfo {pages}
  {131--136} (\bibinfo {year} {2001})}\BibitemShut {NoStop}%
\bibitem [{\citenamefont {Ishii}(1970)}]{ishii1970josephson}%
  \BibitemOpen
  \bibfield  {author} {\bibinfo {author} {\bibfnamefont {Chikara}\ \bibnamefont
  {Ishii}},\ }\bibfield  {title} {\enquote {\bibinfo {title} {Josephson
  currents through junctions with normal metal barriers},}\ }\href {\doibase
  10.1143/ptp.44.1525} {\bibfield  {journal} {\bibinfo  {journal} {Progress of
  Theoretical Physics}\ }\textbf {\bibinfo {volume} {44}},\ \bibinfo {pages}
  {1525--1547} (\bibinfo {year} {1970})}\BibitemShut {NoStop}%
\end{thebibliography}
\end{document}